\documentclass[12pt,english]{article}
\usepackage{lscape}
\usepackage[english]{babel}
\usepackage{comment}
\usepackage{bbm} 

\usepackage{eurosym} 

\usepackage{ragged2e} 

\usepackage{soul}
\usepackage{float}

\usepackage[a4paper,
left=1in,
right=1in,
top=1in,
bottom=1in]{geometry}

\usepackage{amsmath}
\allowdisplaybreaks

\usepackage{graphicx}
\usepackage[colorlinks=true, allcolors=blue,hyperfootnotes=true]{hyperref}
\hypersetup{hypertexnames=false}

\usepackage{tabularx}
\usepackage{booktabs}
\usepackage{longtable}
\usepackage{adjustbox}
\usepackage{amssymb}
\usepackage{amsfonts}
\usepackage{placeins}
\usepackage{dsfont}
\usepackage{xcolor}
\usepackage{soul}
\usepackage[comma,authoryear]{natbib}
\usepackage{multibib}
\newcites{sec}{References in the Online Appendix}

\usepackage{siunitx}

\setcitestyle{citesep={,}}

\usepackage{caption}
\usepackage{subcaption}
\usepackage[flushleft]{threeparttable}
\usepackage{ragged2e}

\usepackage{setspace}

\usepackage[toc,page,header]{appendix}
\usepackage{minitoc}
\noptcrule 

\allowdisplaybreaks

\newtheorem{assumption}{Assumption}

\makeatletter
\let\oldsection\section
\renewcommand{\section}{\vspace{-\parskip}\oldsection}

\let\oldsubsection\subsection
\renewcommand{\subsection}{\vspace{-\parskip}\oldsubsection}

\let\oldparagraph\paragraph
\renewcommand{\paragraph}{\vspace{-\parskip}\oldparagraph}
\makeatother

\begin{document}
	
	\title{ {\textbf{Sovereign Hold-Up and Technology Adoption: Evidence from the North Sea}\thanks{{\scriptsize This paper benefited from comments and discussion with Philippe Aghion, Lassi Ahlvik, Antonin Bergeaud, Estelle Cantillon, Micael Castanheira, Federico Ciliberto, Pierre-Philippe Combes, John Fernald, Nicola Gennaioli, Rachel Griffith, Eric Mengus, Petter 
					Osmundsen, Fabiano Schivardi, Mara Squicciarini, Guido Tabellini, Edoardo Teso, Otto Toivanen, Filipp Ushchev, Nicholas Vreugdenhil, and Riccardo Zago, and seminar and conference participants at Bank of Italy, Bocconi, Bristol, College de France, CREST, ECARES, HEC, LSE, NUS, Sciences Po, and SMU. We would like to thank Banque de France and the Singapore Ministry of Education  (Grant: 20-C244-SMU-002) for financial support. Andrea Stringhetti provided excellent research assistance. All errors are ours.}}
		}
	}
	\author{\textbf{Michele Fioretti}\thanks{{\scriptsize Bocconi University, Department of Economics, IGIER, and CEPR  \href{fioretti.m@unibocconi.it}{fioretti.m@unibocconi.it} }} \and \textbf{Alessandro Iaria}\thanks{{\scriptsize University of Bristol and CEPR, \href{alessandro.iaria@bristol.ac.uk}{alessandro.iaria@bristol.ac.uk}}} \and
		\textbf{Aljoscha Janssen}\thanks{{\scriptsize Singapore Management University, \href{ajanssen@smu.edu.sg}{ajanssen@smu.edu.sg}}} \and
		\textbf{Clément Mazet-Sonilhac}\thanks{{\scriptsize Bocconi University, Department of Finance and IGIER, \href{clement.mazetsonilhac@unibocconi.it}{clement.mazetsonilhac@unibocconi.it}}} \and
		\textbf{Robert K. Perrons}\thanks{{\scriptsize Queensland University of Technology, \href{robert.perrons@qut.edu.au}{
					robert.perrons@qut.edu.au}}}}
	\date{February, 2026}
	
	\maketitle
	\thispagestyle{empty}
	\vspace{-2.8em}
	\begin{abstract} \noindent {\footnotesize 
			Contractual relationships between the state and private firms involving large irreversible investments are vulnerable to \textit{sovereign hold-up risk}: anticipating that the state can unilaterally revise terms once capital is sunk, firms may underinvest. Causal evidence on this mechanism is scarce because sovereign commitment is typically bundled with broader institutional quality. We overcome this identification challenge by exploiting a natural experiment in the North Sea oil and gas industry. In 1985, a Norwegian Supreme Court ruling declared retroactive changes to petroleum licenses unconstitutional, while the UK retained the discretion to revise contracts. Using granular data on the universe of fields and firms from 1975 to 1995, we estimate the impact of this strengthening of sovereign commitment on the adoption of Enhanced Oil Recovery (EOR), a major extraction technology requiring large irreversible investments. Firms exposed to the ruling sharply increased EOR adoption and productivity, gaining market share through aggressive portfolio expansion. We find that private firms with preexisting EOR expertise---rather than state-owned enterprises---drove this transformation, leveraging this expertise to diversify into riskier geologies and adopt complementary technologies. These findings establish sovereign commitment as a primary determinant of investment and technology adoption. By tying the state's hands, the ruling transformed promises into credible commitments, effectively functioning as an industrial policy that unlocked a trajectory of technological deepening. While such constitutional protections are critical for investment, a global survey of constitutions reveals that only 30.6\% of countries prohibit retroactive legislation beyond criminal law.}

		\vfill \noindent {\footnotesize\textit{JEL classifications:}  O33, D23, P48}
		
		\noindent {\footnotesize\textit{Keywords:} Sovereign Commitment; Hold-up Risk; Enforcement; Institutional Credibility; Technology Adoption; Oil and Gas Industry; Enhanced Oil Recovery; Productivity} 
	\end{abstract} 
	

	\thispagestyle{empty}
	\newpage
	\pagenumbering{arabic}
	\onehalfspacing

	\newpage
	\section{Introduction}\label{s:intro}
	
	Contractual relationships between the state and private firms involving large irreversible investments are subject to \textit{sovereign hold-up risk}. Contracts with a sovereign state---the supreme legal authority within its jurisdiction---are inherently unenforceable: absent a higher authority, the state retains discretion to unilaterally and retroactively revise contractual terms. Once a firm sinks capital based on an initial contract, its bargaining position is severely eroded. Although the state may initially offer favorable conditions to attract investment, it faces a time-inconsistent incentive to alter the contractual terms ex post and capture a larger share of surplus.  This phenomenon generates a fundamental commitment problem central to the investment theory of hold-up \citep{williamson1975markets,grout1984investment, thomas1994foreign}: anticipating the risk of ex post expropriation, firms may choose to underinvest in valuable technologies or forgo adoption altogether. 
	
	Sovereign hold-up risk differs from policy uncertainty. While uncertainty regarding economic fundamentals or policy---such as fiscal or regulatory unpredictability---depresses investment by increasing the option value of waiting \citep{bernanke1983irreversibility,dixitpindyck1994,bloom2007uncertainty,bloom2009impact,baker2016measuring}, sovereign hold-up risk concerns the credibility of the state's commitments. It stems from a structural time-inconsistency problem \citep{kydland1977rules}: even a benevolent state faces incentives to renege on promises once capital is sunk. Despite the theoretical prominence of this mechanism, causal evidence isolating the impact of sovereign commitment on technology adoption remains rare.\footnote{We discuss the existing literature in detail below.} The central identification challenge is that sovereign commitment is endogenous and typically ``bundled'' with other dimensions of institutional quality \citep{acemoglu2005unbundling}. Because states that credibly limit their power usually preside over strong legal and economic systems, isolating the consequences of sovereign commitment from the general benefits of good institutions is empirically difficult. 
	
	We overcome this identification challenge in the context of the North Sea upstream oil \& gas (O\&G) industry, comparing Norway and the UK from 1975 to 1995.\footnote{Norway and the UK accounted for 96\% of North Sea oil production in this period \citep{Kemp2012}.} We isolate the role of commitment credibility from broader institutional quality by exploiting a natural experiment rooted in divergent legal foundations. While the UK operates under \textit{parliamentary sovereignty}---whereby the legislature cannot bind its successors, leaving contracts vulnerable to statutory revision---Article 97 of the Norwegian constitution explicitly prohibits retroactive laws. Despite this, Norwegian courts had historically applied Article 97 with flexibility in economic matters, balancing the degree of retroactivity against public interest. The state had thus altered licensing terms retroactively, effectively mirroring the UK's discretion to intervene ex post.\footnote{As discussed in detail in Section \ref{s:supreme_court}, the UK repeatedly altered licensing terms retroactively. For example, it reduced capital expenditure deductibility in 1979 and devalued past tax credits in 2011.} This practice was challenged when the Norwegian government attempted to unilaterally impose stricter royalty terms on existing licenses, leading to a landmark 1985 Supreme Court decision (\textit{Phillips Petroleum v. The State}). The Supreme Court ruled any such retroactive revision unconstitutional, anchoring sovereign commitment in the constitution and curtailing the state's ability to renege on sunk contracts through ordinary legislation.
	
	Our analysis relies on granular panel data covering the universe of O\&G fields and licensees in Norway and the UK. The dataset tracks firms' asset portfolios and detailed field-level measures of geology, production, costs, and deployed technologies. We focus on the adoption of Enhanced Oil Recovery (EOR), a major extraction method of the era. By the 1980s, EOR was a mature, standardized technology available through established service providers. Although adoption did not require in-house invention, it necessitated substantial, irreversible investments, including the drilling of injection wells, the retrofitting of offshore platforms, and the installation of specialized facilities for injectants. Beyond these capital investments, field-specific geological characteristics---such as reservoir permeability, pressure, and oil viscosity---largely determined a field’s technical suitability for EOR, creating heterogeneity across fields in EOR eligibility.
	
	\begin{figure}[H]
		\captionsetup[subfigure]{justification=centering}
		\caption{EOR adoption and production in Norway and in the UK \label{fig:eor_prod_country}}	
		\minipage{0.48\textwidth}
		\subcaption{Share of fields adopting EOR}\label{fig:country_eor}
		\includegraphics[width=\linewidth]{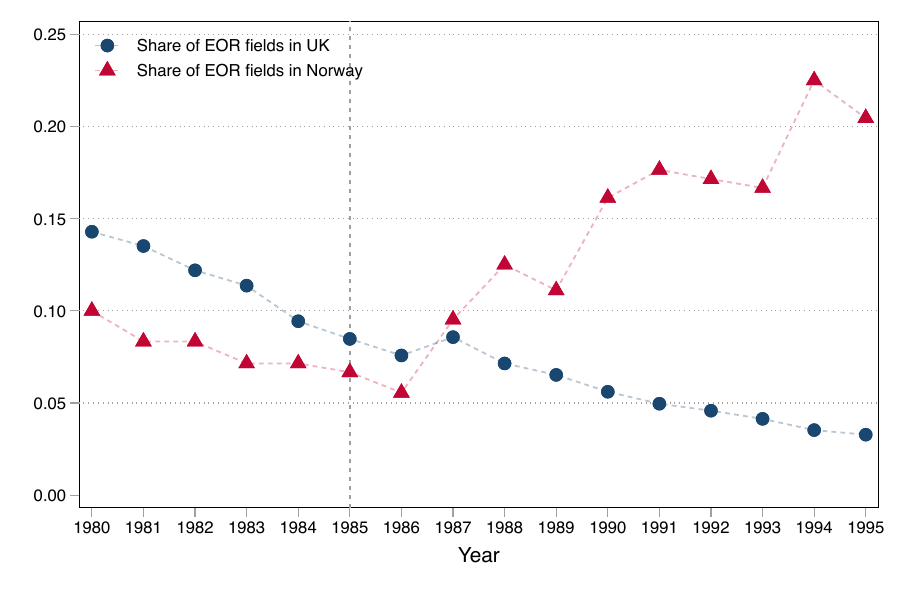}
		\endminipage\hfill
		\minipage{0.48\textwidth}
		\subcaption{Average cumulative production (MMboe)}\label{fig:country_prod}
		\includegraphics[width=\linewidth]{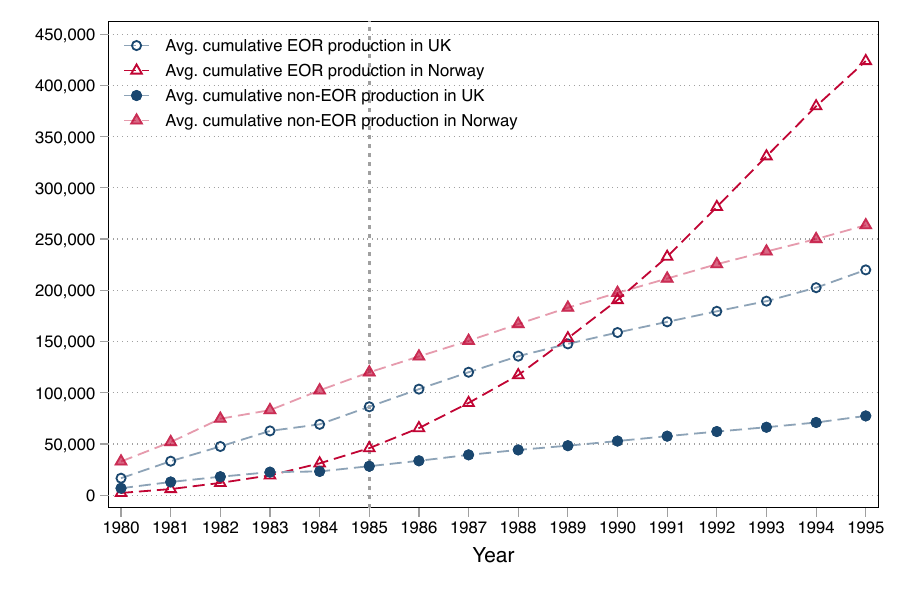}
		\endminipage\hfill
		
		\vspace{-5mm}
		\begin{minipage}{1 \textwidth}
			
			{\footnotesize\singlespacing \textbf{Notes:} Panel (a) plots the share of fields adopting EOR by country and year. Panel (b) plots the average cumulative production of EOR-eligible and non-EOR-eligible fields by country and year. We use cumulative production, rather than annual production, to filter out macro fluctuations. Production is measured in Million Barrels of Oil Equivalent (MMboe). The vertical dotted line marks the year of the Norwegian Supreme Court decision. 
				\par}
		\end{minipage}
	\end{figure}
	
	Our empirical strategy leverages the 1985 Norwegian Supreme Court ruling together with variation in EOR eligibility across the fields included in firms' \textit{pre}-1985 portfolios. We compare firms with similar portfolios of fields but varying exposure to the ruling, as determined by the location of their EOR-eligible fields across the Norway-UK border prior to 1985. For example, consider two otherwise similar firms with comparable pre-1985 portfolios, except that one firm's EOR-eligible fields were concentrated in Norway and the other's in the UK. We expect the 1985 ruling to have increased adoption incentives for the first firm, while leaving incentives for the second unchanged. Figure \ref{fig:eor_prod_country} provides striking evidence consistent with this intuition. Panel (a) shows that, immediately following the 1985 ruling, EOR adoption surged among Norwegian fields, while adoption in the UK continued its pre-1985 decline. Panel (b) documents the resulting divergence in field-level output: after 1985, production from Norwegian EOR-eligible fields accelerated sharply, significantly outperforming the production of comparable British fields.
	
	We formalize the intuition behind our empirical strategy with a dynamic model of technology adoption in the spirit of \citet{jovanovicnyarko1996}. In the North Sea O\&G industry, multiple firms held equity stakes in each field, but only the designated operator made technological decisions and managed daily operations. Through this hands-on involvement, operators accumulated know-how that was partly transferable across fields and technologies. In the model, a firm's unobserved state variable is its stock of know-how, which rises with successful adoptions but whose relevance diminishes with technological distance. Technology adoption occurs when know-how exceeds a threshold that depends on expected returns. By reducing sovereign hold-up risk, the 1985 ruling raised expected returns from EOR adoption in Norway, lowering the adoption threshold for Norwegian EOR-eligible fields while leaving it unchanged for British fields. Through the lens of our model, this asymmetric shock yields a difference-in-differences design for the estimation of the effect of EOR adoption---as induced by a reduction in sovereign hold-up risk---on productivity, market shares, portfolio reallocation, and subsequent technology adoption.
	
	Consistent with Figure \ref{fig:eor_prod_country}, our estimates show that firms holding a 10 percentage point (p.p.) larger share of Norwegian EOR-eligible fields in their pre-1985 portfolios (hereafter ``exposed firms'') were 3.5 p.p. more likely to adopt EOR, nearly doubling their adoption rate. 
	This major increase in EOR adoption translated into substantial productivity gains: exposed firms increased production by 38\% while reducing average variable costs. Through these large productivity gains, the reduction in sovereign hold-up risk induced a profound reallocation of activity within and across firms. Following the 1985 ruling, exposed firms expanded their presence in the North Sea, increasing their market share by 39\%. These market-share gains were accompanied by portfolio expansion along two margins. At the intensive margin, ownership in Norwegian EOR-eligible fields became more concentrated relative to comparable UK fields. At the extensive margin, exposed firms acquired on average one additional field (against a baseline portfolio size of 7.8 fields), with 56.5\% of this expansion in EOR-eligible fields.
	
	We demonstrate that the alleviation of sovereign hold-up risk enabled firms with preexisting EOR-specific know-how to leverage their expertise, thereby correcting a market failure in which specialized technical knowledge had remained underutilized. To investigate this channel, we exploit heterogeneity in firms' preexisting EOR-specific know-how and ownership status (private or state owned). Defining preexisting EOR-specific know-how as having operated a field in the North Sea that adopted EOR prior to 1985, we find that private firms with this specialized expertise were the primary drivers of the industry's transformation. While exposed state-owned firms also gained market share, exposed private firms with preexisting EOR-specific know-how expanded their asset portfolios more aggressively---acquiring an average of 4.7 new fields as opposed to 2.8. Compared to exposed state-owned firms, these private firms also leveraged their expertise to diversify more extensively into ``riskier'' geologies, such as fields with sour oil or deep wells.

	Finally, we examine whether this surge in investment generated spillovers that broadened the direction of technological change. We explore whether EOR-specific technical knowledge facilitated the adoption of technologies other than EOR. We find that private firms with preexisting EOR-specific know-how significantly diversified their technological portfolios, expanding into exploration and production technologies as well as extraction technologies beyond EOR. Consistent with the notion that the relevance of know-how decays with technological distance, technological portfolios expanded significantly more in the direction of related extraction technologies. Moreover, our results highlight that preexisting EOR-specific know-how, rather than state ownership, was the main driver of this diversification. This path dependence indicates that sovereign hold-up risk had arrested a cumulative learning process. By unlocking the adoption of EOR and the accumulation of transferable know-how, the 1985 ruling set the Norwegian shelf on a trajectory of technological diversification and experimentation.
	
	These findings offer a novel perspective on the relationship between institutions and development. We show that the credibility of commitment---isolated from the broader bundle of institutional quality---is a distinct and powerful determinant of economic performance. The Norwegian experience demonstrates that even in a high-income, stable democracy, the residual risk of sovereign hold-up can depress technology adoption and distort the development and deployment of know-how. However, our results also show that this distortion can be corrected by ``tying the state's hands.'' By enforcing the constitutional prohibition on retroactive laws, the Norwegian Supreme Court did not merely resolve a royalty dispute; it addressed a fundamental time inconsistency, transforming the state's promises into credible commitments. This effectively served as a potent industrial policy, unlocking the deployment of know-how and technological deepening.
	
	To evaluate the global prevalence of sovereign hold-up risk, we benchmark constitutional protections against retroactive legislation through a comprehensive survey of national constitutions, using an LLM-assisted and fully auditable coding procedure. We find that explicit prohibitions on non-criminal retroactivity are rare: only 30.6\% of constitutions---including those of just 11 OECD members---contain such protections.
	
	\paragraph{Related literature.}
	Our work builds on the literature on hold-up and incomplete contracts. We focus on sovereign hold-up risk, a time-inconsistency problem in which the state can revise contractual terms once capital is sunk \citep{kydland1977rules}. In theories of relationship-specific investment and residual control rights \citep{klein1978vertical, grossman1986costs, hart1990property}, the prospect of ex post renegotiation shapes ex ante incentives and can distort firms' investment and technology choices. Despite the theoretical prominence of this mechanism, it is difficult to empirically isolate its effects from the broader institutional environment in which the state operates. States with more credible commitments also tend to be embedded in strong legal systems, effective bureaucracies, and stable political environments \citep{acemoglu2005unbundling}. This institutional ``bundling'' makes it difficult to separate the effects of commitment credibility from the broader benefits of high-quality institutions.
	
	This identification problem is compounded by the well-established importance of institutional quality for economic performance. Cross-country evidence links institutions to long-run growth \citep{hall1999play, acemoglu2001colonial, rodrik2004institutions}, while micro-evidence connects secure property rights to investment incentives \citep{besley1995property, goldstein2008institutions} and documents persistent effects of historical institutions \citep{dell2010persistent, banerjee2005history}. Related work on expropriation risk and institutional constraints examines the determinants of nationalization \citep{guriev2011determinants}, how expropriation risk shapes contract terms and extraction behavior \citep{stroebel2013resource, bohn2000ownership}, and how institutional differences affect investment location \citep{cust2020institutions}. \citet{north1989constitutions} argue that the strengthening of Parliament relative to the Crown after the Glorious Revolution transformed government promises into credible commitments, enabling the development of English capital markets. However, this before-after comparison within a single country cannot separate the effects of commitment from concurrent changes. We overcome this identification challenge by comparing Norway and the UK, two countries with similar institutions, within the North Sea O\&G industry and exploiting a Supreme Court ruling that strengthened sovereign commitment in Norway. This allows us to identify the effects of commitment credibility separately from the broader institutional bundle.
	
	Our work also contributes to the literature investigating how legal systems shape economic outcomes. \citet{laporta1997legal,laporta1998law} show that legal origins affect investor protection and financial development, while \citet{djankov2003courts} emphasize how court quality influences contract enforcement. Work on judicial independence shows that courts can transform formal rights into credible protections \citep{feld2003economic,klerman2005value}. We provide evidence on the role of constitutional entrenchment as a commitment device---a feature that distinguishes systems of constitutional supremacy, like Norway's, from systems of parliamentary sovereignty, like the UK's. By anchoring contractual protections in constitutional law---a higher authority than ordinary legislation---courts raise the cost of reneging and limit retroactive interference. Consistent with historical evidence that judicial independence reduced expropriation risk \citep{klerman2005value} and with recent evidence that stronger courts increase corporate innovation by limiting political interference \citep{lai2023judicial}, our findings highlight an implication for technological adoption: constraining the sovereign unlocks the deployment of know-how and facilitates adoption of irreversible, productivity-enhancing technologies.
	
	Our results speak to the literature on industrial policy, which emphasizes that effective policy must solve coordination failures and shape the direction of technological change \citep{rodrik2004industrial,juhasz2023new}. A central limitation of such policies is credibility: when governments cannot commit to stable long-run terms, private firms may refrain from making irreversible and learning-intensive investments \citep{brunner2012credible}. Tying the state's hands through constitutional enforcement operates as a potent form of industrial policy: reducing the retroactive hold-up threat unlocks the deployment of valuable, otherwise idle know-how and accelerates the diffusion of technology in the industry.\footnote{Our results also contribute to the literature on resource misallocation and fiscal incentives in O\&G \citep[e.g.,][]{asker2019mis,stefanski2025extracting,vreugdenhil2020booms,ahlvik2025resource}.}
	
	Our work also contributes to the literature on technology adoption and firm capabilities. Learning-by-doing generates path dependence in technology choices \citep{arrow1962learning,kellogg2011learning} and externalities that shape technology diffusion across firms \citep{griliches1957hybrid,foster1995learning,comin2010exploration,juhasz2024technology}. This literature has also shown that technology diffusion often requires complementary investments in absorptive capacity \citep{cohen1990absorptive, griffith2003r, griffith2004mapping, aghion2015knowledge}. We identify sovereign hold-up risk as an important barrier to this cumulative learning process. By discouraging irreversible investments, sovereign hold-up risk arrests the initiation of learning trajectories. The 1985 ruling unlocked capability development and technological diversification by making initial EOR adoption more attractive---illustrating how institutional constraints can initiate cumulative learning.

	\paragraph{Roadmap.} The remainder of the paper is organized as follows. Section \ref{s:northsea} provides background on the North Sea O\&G sector, the 1985 Norwegian Supreme Court decision, and EOR. Section \ref{s:identification} sets out the theoretical framework and empirical strategy, and Section \ref{s:data} describes the data. Section \ref{s:did} reports estimates on the effect of sovereign hold-up risk on EOR adoption, production, productivity, and various robustness checks. Section \ref{s:mkt_shares} analyzes how EOR adoption shaped market shares and asset portfolios, while Section \ref{s:knowhow} investigates the roles of preexisting EOR-specific know-how and state ownership. Section \ref{s:tech} explores the implications of transferable know-how for technological diversification. Section \ref{s:conclusion} concludes and the Appendix reports additional results and robustness checks.
	
	\section{The Upstream O\&G Sector in the North Sea}\label{s:northsea}
	The North Sea O\&G sector grew until the early 2000s but has since matured, falling from about 8\% of global production, at its peak, to 4\% in 2020 \citep{craig2018history,bpdata}. In our analysis, we focus on the two dominant players of the North Sea O\&G sector, Norway and the UK, which together account for 96\% of the North Sea oil production (Denmark and the Netherlands account for the remaining 4\%) \citep{Kemp2012}. In the following, we outline the core institutions and legal events that shaped the early evolution of the sector in Norway and the UK.
	
	\paragraph{Licenses.} Both governments retain sub-sea ownership and allocate exploration and production rights via competitive bidding rounds.  Norway issues exploration, production, and facilities licenses through annual (mature areas) or biennial (frontiers) rounds. The UK operates similar rounds under the Petroleum Act \citep{gordon2011petroleum,mace2017oil}.  In both jurisdictions, regulators can block bids if the bids do not meet undisclosed technical, financial, or strategic criteria, and ministries often negotiate operational plans prior to award \citep{kemp2013official,kardel2019comparative}.
	
	\paragraph{Ownership and state participation.} 
	The North Sea regulatory framework utilizes a joint-venture structure where license equity is distributed among a consortium of firms, with a designated ``operator'' managing daily production and technological strategy. Although revenues, capital and operating expenditures are shared pro-rata, the choice of operator is a critical determinant of field-level efficiency. In the 1970s, both countries established national oil companies to internalize technical expertise and maximize rent capture. However, their subsequent trajectories define a sharp institutional divergence. 
	
	Norway adopted a model of persistent state-led participation; Statoil (established 1972) was granted statutory preferences in licensing and maintained 50-80\% equity stakes in core fields throughout the 1970s and 1980s. Although partially privatized in 2001, the Norwegian state retains a majority share, ensuring a permanent state presence in the sector’s operational core.\footnote{On January 1, 1985, the State's Direct Financial Interest (SDFI) was created, with Statoil retaining only 20\% of its original portfolio and transferring the remainder to the SDFI, though the state remained the ultimate beneficiary.} Conversely, the UK pursued a rapid de-nationalization strategy. The British National Oil Corporation (BNOC), established in 1975, initially served as a major operator and partner with participation rights in all licenses. Under the Thatcher administration, BNOC’s upstream assets were transferred to a new entity, Britoil, in 1982. Crucially, Britoil continued to act as a primary operator in the North Sea until it was acquired by British Petroleum (BP) in 1988 following a hostile takeover. Parallel to this, the state exited from direct involvement in O\&G production and liquidated its remaining 31.5\% stake in BP in October 1987.
	
	\paragraph{Fiscal Architectures.} Following the 1973 oil shock, the UK and Norway developed nearly parallel fiscal regimes designed to extract Ricardian rents from the North Sea. Both jurisdictions initially employed a ``hybrid'' model: production-based royalties (12.5\% in the UK; 8-16\% in Norway) ensuring immediate state revenue, and profit-based corporate taxes. In 1975, both countries introduced sector-specific rent taxes---the Petroleum Revenue Tax (PRT) in the UK and the Special Petroleum Tax (SPT) in Norway---to capture windfall gains. These regimes targeted a high marginal ``government take,'' often reaching aggregate rates of 75-80\%, yet were designed to be investment-neutral through two key structural mechanisms: ring-fencing and uplift \citep{UK1975PRT, NorwayPTA1975}.
	
	The ``ring-fence'' served as a critical institutional boundary, isolating petroleum income from mainland activities to prevent firms from eroding the resource tax base with unrelated industrial losses. To maintain investment incentives under these high marginal rates, both countries introduced an ``uplift'' allowance to shield the normal return on capital in the absence of interest deductibility. Following initial recalibrations at the turn of the decade (1979 in the UK; 1980 in Norway), these parameters remained remarkably static throughout the 1980s. Although the nominal magnitudes may seem different---the UK provided an immediate 35\% uplift whereas Norway offered a 100\% allowance distributed over 15 years---the NPV of the resulting tax shields was approximately aligned. Specifically, once we account for the interaction with different corporate tax rates and depreciation schedules, the effective relief provided to EOR investments was comparable under the prevailing industry discount rates of 7–10\%. \citep{HMRC_OT00190, valhallRoyalties}. 
	
	This period of institutional similarity between the UK and Norway was further reinforced by synchronous policy adjustments during the price volatility of the 1980s. Both countries responded to the softening of oil prices by phasing out royalties for new developing fields (UK in 1983; Norway in 1986) to eliminate the marginal distortions inherent in gross-revenue taxation \citep{UK1983Royalty}. The fiscal regimes only structurally diverged in the early 1990s through differing approaches to license ``vintaging.'' Although Norway maintained a uniform, high-tax environment across all producing fields, the UK's \textit{Finance Act 1993} aggressively liberalized the regime by abolishing both the PRT and the uplift for all new developments while maintaining the legacy system for pre-1993 assets \citep{UK1993PRT}. This shared trajectory from 1975 to 1993---characterized by identical instruments, stable investment allowances, and common responses to exogenous shocks---motivates our use of British fields as control group for Norwegian fields \citep{NorwaySDFI,nrgi2014}.

	\subsection{1985 Norwegian Supreme Court Decision}\label{s:supreme_court}
	
	A critical institutional divergence between the UK and Norway emerged in the 1980s when the Norwegian Supreme Court curtailed the state's ability to retroactively alter petroleum licensing terms. The dispute originated in December 1972, when the Norwegian state amended the \textit{Continental Shelf Act 1963} to require quarterly rather than biannual royalty payments and introduced a graduated royalty rate (8-16\%). Crucially, the state did not exempt pre-1972 licenses from these changes \citep{dalgaard1987exploitation}. Phillips and its partners in the Ekofisk concession protested by continuing to pay every six months until 1977, when the state demanded compliance with the new regulation \citep{frihagen1985ekofisk}. 
	
	The Phillips group eventually complied but filed suit in 1982 to recover interest losses dating back to 1977. The legal battle progressed from the Oslo City Court (ruling in 1982) to the Eidsivating Court of Appeal (appealed in 1982 with ruling on April 7, 1984), both of which ruled in favor of Phillips \citep{valhallRoyalties, frihagen1985ekofisk}. On December 19, 1985, the Supreme Court upheld these judgments, declaring any unilateral retroactive change to licensing terms unconstitutional if it violated protections against arbitrary deprivation of property. The Court awarded Phillips NOK 140 million (approximately USD 18.2 million) plus legal costs and established that, absent an urgent public-interest mandate, existing licenses could no longer be amended without compensation \citep{mestad1987ekofisk, ulfbeck2016responsibilities}.\footnote{Quoting the Court: ``it must be rather clear that the state could not, for example, change [unilaterally] the time limit for a petroleum license without compensation [...], at least not unless it was based on a very urgent need for regulation'' \citep[][p. 183]{hunter2020character}.} Legal actions from other firms followed, securing additional settlements for NOK 246 million (approximately USD 32 million) plus legal costs.
	
	This ruling resolved a long-standing ambiguity in Norwegian legal doctrine. While Article 97 of the Constitution stated that ``no law may be given retroactive effect,'' its application in civil matters had historically been subject to a ``reasonableness'' test. The landmark \textit{Kløfta} case (1976) had established a tripartite ``sliding scale'' of judicial review, placing economic rights---such as protection against retroactive tax and civil laws---in an intermediate category \citep[][Chapter 5]{kierulf2018judicial}. Under this scheme, the Court balanced the degree of retroactivity against the urgency of the public interest. This framework meant that prior to 1985, O\&G firms faced sovereign hold-up risk in Norway, as the terms of petroleum licenses remained subject to legislative interference that the Court might have deemed reasonable under the prevailing social democratic consensus.\footnote{As analyzed by \cite{hunter2020character}, while petroleum licenses are---in nature---administrative acts both in the UK and in Norway, the 1985 ruling entrenched a Norwegian legal culture that treats them as having a ``protected contractual character'' under the constitution. This differs from the British administrative tradition, where regulatory and fiscal discretion remains paramount.} Following the ruling, a ``consensual model'' of fiscal stability emerged as the institutional norm; as documented by \cite{al2006managing}, this was characterized by a ``free and constructive dialogue'' between the state and licensees, prioritizing negotiated alignments and project-specific adjustments over the unilateral, unheralded legislative amendments common in the British system.
	
	In contrast, the British doctrine of \textit{parliamentary sovereignty} allows unilateral retroactive amendments through ordinary legislation without requiring any compensation \citep{mestad1987ekofisk}.\footnote{The British legal system has allowed retroactive rulings since 1793 \citep{ukparliament, gobbi2010begin}. However, retroactive criminal rulings are prohibited under the European Convention on Human Rights.} Most notably, the \textit{Oil Taxation Act 1975} introduced the \textit{Petroleum Revenue Tax} (PRT) and modified transfer pricing rules retroactively for fields already under development. At the same time, the \textit{Petroleum and Submarine Pipelines Act 1975} fundamentally altered the ``Model Clauses'' of existing licenses, granting the state retroactive control over production levels (depletion strategy) and mandating state participation via the newly formed BNOC. Subsequent legislation, such as the \textit{Oil and Gas (Enterprise) Act 1982} and the \textit{Petroleum Act 1987}, continued the practice of unilateral amendments to regulatory and decommissioning obligations. Also in the following years, the UK has continued to exercise this power.\footnote{For instance, the UK has frequently used ``straddling periods'' to apply tax hikes to profits earned months before the law was announced, as seen with the 2011 Supplementary Charge (increased from 20\% to 32\%) and the 2022 Energy Profits Levy. In addition, in 2011 it capped decommissioning relief at 20\% retroactively, devaluing tax credits that companies had earned by paying into the system at rates as high as 62\% for decades.} Thus, while the 1985 ruling effectively eliminated the possibility of retroactive interventions in Norway, the UK maintained its ability to do so, leaving the risk of sovereign hold-up unmitigated.

	\subsection{Enhanced Oil Recovery (EOR)}\label{s:EOR}
	
	Through a reduction in sovereign hold-up risk, the Supreme Court decision boosted incentives to adopt technologies that required large irreversible investments in Norwegian O\&G fields. At that time, in the North Sea, one of these technologies was Enhanced Oil Recovery (EOR) \citep[e.g.,][]{awan2008survey,gbadamosi2018review}.
	
	As reservoir pressure declines, water or gas injection yields only 20-40\% recovery of a reservoir's resources. Operators can then resort to EOR, which increases recovery up to 50-70\% through the injection of substances that alter fluid properties \citep{toole2003oil,alvarado2010enhanced}.  EOR requires on-platform storage, injection wells, and separators, increasing both capital costs (CAPEX) and operating costs (OPEX).
	
	Only reservoirs that meet specific geological (e.g., adequate size, depth, porosity, permeability, and sulfur content) and technological (e.g., suitable platform design and drive mechanisms) requirements can adopt EOR (see Appendix Table \ref{tab:eor_restrictions} for the detailed criteria), irrespective of the operator's willingness or managerial skills \citep{al2011analysis,nwidee2016eor}. If a field is \textit{EOR-eligible}, reservoir engineers model pressure decline and collaborate with petroleum economists to compare incremental recovery against the increases in CAPEX and OPEX. By the 1980s, EOR was a mature technology and field operators did not need to develop it in-house; adoption could be initiated by contracting an established service provider (e.g., Baker Hughes, Halliburton, Schlumberger, and Weatherford).\footnote{By 1985, this technology was mature and widely available. The earliest recorded EOR attempt was by J.W. Goff in 1901 \citep{rintoul1976spudding}, who injected steam and air into Kern River wells, California.} Importantly, however, the decision to adopt EOR and then its daily use and maintenance were the responsibility of field operators.\footnote{Field operators secured development approvals and directed execution, while service providers, acting at the operators' request, provided design, chemical supply, installation, commissioning, and surveillance. Executive power and accountability remained in the hands of field operators according to the laws of all countries participating to the North Sea O\&G industry.}
	
	
	\section{Empirical Approach}\label{s:identification}
	This section presents a regression framework to estimate the effect of technology adoption on firm outcomes through the exogenous reduction in sovereign hold-up risk induced by the 1985 Supreme Court decision. Section \ref{s:conceptual_framework} introduces a dynamic model of technology adoption characterized by a threshold rule. Section \ref{sec:firm_regressions} then demonstrates how this threshold rule, when combined with an exogenous shock to the adoption threshold, such as the 1985 Supreme Court decision, enables estimation of the effect of technology adoption on firm outcomes, such as production or market shares.
	
	\subsection{Conceptual Framework}\label{s:conceptual_framework}
	We consider technology adoption decisions in O\&G fields. All fields are homogeneous and jointly owned by multiple firms. In each field, one firm acts as the operator, responsible for  decisions such as technology adoption, while the other firms are passive co-owners without decision-making authority. The operator holds share $0< s \leq 1$, and passive co-owners collectively hold the remaining share $1-s$. Profits from field operations accrue proportionally to these ownership shares. To simplify exposition, both the ownership shares and the identity of the operator are treated as exogenous.
	
	We analyze the decision-making of a risk-neutral operator who, in each period, decides whether to adopt a new technology in a different field. Every period, each field generates revenue $R$ and incurs production cost $K$, with $R - K > 0$. Each adoption attempt entails a sunk cost $c>0$, which is borne solely by the operator (e.g., management time, site preparation, etc.). Any successful adoption increases profits by $\Delta \Pi$, with $R - K + \Delta \Pi < M$ for some finite $M$.  Let $A_t$ be the operator's know-how---the accumulated expertise in adopting and managing new technologies. Suppose that $A\in [0,\bar{A}]$ for some finite $\bar{A}$. Successful adoption bears an element of randomness. Adoption given know-how $A$ succeeds with probability $q(A)$, assumed to be continuously differentiable, strictly increasing, and concave. Know-how evolves according to:
	\begin{equation}\label{eq:transition}
		A_{t+1} \;=\; \delta A_t
		\;+\;\gamma\,\mathbbm{1}\{\text{adoption in $t$ succeeds}\},
	\end{equation}
	so that it depreciates by $\delta \in (0,1)$ every period and each successful adoption increases it by $\gamma>0$. The parameter $\gamma$ represents knowledge carryover from current to future technology adoption. A small $\gamma$ means that the knowledge acquired by adopting a  technology in $t$ will not be particularly relevant for the adoption of future technologies (e.g., very different or rapidly evolving technologies). This captures the idea, also in \citet{jovanovicnyarko1996}, that every adoption of a new technology, as technology continues to evolve, requires slightly different know‐how and that past know-how eventually becomes obsolete if not updated. We assume that $\gamma \leq \bar{A}\cdot (1 - \delta)$, which ensures that knowledge remains bounded within $[0, \bar{A}]$ after adoption.
	
	The optimal adoption decision in each period is derived from the Bellman equation:
	\begin{align}\label{eq:bellman}
		V(A)=\max\Bigl\{\,& s\,(R-K)+\beta\,V(\delta A),\\
		& s\bigl(R-K+q(A)\,\Delta\Pi\bigr)-c
		+\beta\bigl[q(A)\,V(\delta A+\gamma)+(1-q(A))\,V(\delta A)\bigr]
		\Bigr\}, \notag
	\end{align}
	where $\beta\in(0,1)$ is the operator's discount factor and $V$ is assumed to be continuously differentiable, strictly increasing, and concave. Given \eqref{eq:bellman}, in each period the operator is indifferent between non-adoption and adoption whenever: 
	\begin{equation}\label{eq:threshold_A}
		A
		\;=\;B(A) \; \equiv \; q^{-1}\left(
		\frac{c}
		{s \cdot \Delta \Pi\;+\;\beta\bigl[V(\delta A+\gamma)-V(\delta A)\bigr]}\right),
	\end{equation}
	where $q^{-1}$ is the inverse of $q$. Appendix \ref{apndx:unique_fx_point} shows that, under mild regularity conditions, equation \eqref{eq:threshold_A} has a unique fixed point $A^*$ and that the adoption rule $A\ge B(A)$ is equivalent to the threshold crossing rule $A\ge A^*$.
	
	\paragraph{Comparative statics.}
	The threshold crossing condition and \eqref{eq:threshold_A} suggest simple comparative statics: lower $c$ or $\delta$, or---symmetrically---higher $s$, $\Delta \Pi$, $\beta$, or $\gamma$ all imply a lower threshold $A^*$ and a higher chance of adoption.\footnote{This holds since $q^{-1}$ is strictly increasing, so $B(A)$ is strictly decreasing in $s$, $\Delta\Pi$, and $\beta$, and strictly increasing in $c$. As $V$ is strictly increasing, $B(A)$ is strictly decreasing in $\gamma$. Moreover, by concavity of $V$, the term $V(\delta A+\gamma)-V(\delta A)$ is strictly decreasing in $\delta$, so $B(A)$ is also strictly increasing in $\delta$.} This also means that, for example, operators with higher know-how $A$ should be more likely to diversify into technologies that are more ``distant'' from their current technology, as distant technologies are likely associated with lower knowledge carryover $\gamma$ and thus higher adoption thresholds $A^*$. 
	
	Based on these simple observations, we next derive difference-in-differences (DiD) regressions that can be estimated when neither $A$ nor threshold $A^*$ is observable. Later, we also rely on these comparative statics to interpret some of our empirical findings, especially when investigating technological diversification in Section \ref{s:tech}. 
	
	\subsection{Derivation of the Regression Model}\label{sec:firm_regressions}
	The model's comparative statics, combined with the 1985 Norwegian Supreme Court decision, can be used to obtain DiD regressions. In what follows, we illustrate this idea, relaying the details to Appendices \ref{a:firm_regressions} and \ref{a:field_regressions}.
	
	\paragraph{Basic setup.} There are two countries (Norway and the UK) and we group years in two periods (pre-1985 and post-1985). We focus on the operator's decision of whether to adopt EOR in the EOR-eligible fields in their portfolio. Prior to the 1985 decision, the adoption threshold in \eqref{eq:threshold_A} is $A^*_{\text{pre}} > 0$ and identical across all fields. The 1985 Norwegian Supreme Court ruling decreased the chance of unfavorable retroactive changes to licensing terms, reducing sovereign hold-up risk in Norway relative to the UK. In our model, this can be seen as an increase in the expected returns $\Delta \Pi$ of successful EOR adoption in Norway relative to the UK. As a result, the adoption threshold in Norwegian EOR-eligible fields fell from $ A^*_{\text{pre,Norway}} = A^*_{\text{pre}}$ to $A^*_{\text{post,Norway}} < A^*_{\text{pre}}$, while the UK threshold remained unchanged at $A^*_{\text{pre,UK}} = A^*_{\text{post,UK}} = A^*_{\text{pre}} $.
	
	We consider firm $i$'s total production as outcome variable; however, the same logic applies to any other outcome affected by changes in $A^*$. For notational simplicity, we assume that each firm $i$ owns $N$ fields in total, with potentially different compositions: $N_{i,\text{Norway EOR}}$ EOR-eligible fields in Norway, $N_{i,\text{UK EOR}}$ EOR-eligible fields in the UK, and $N_{i,\text{Other}} \equiv N - N_{i,\text{Norway EOR}} - N_{i,\text{UK EOR}}$ other fields.\footnote{A DiD regression equation similar to \eqref{eq:did-firm} below can be obtained also when the number of fields $N_i$ varies across firms, but it requires a more involved notation without adding any insight.} We denote $i$'s portfolio of fields by $\mathbf{N}_i=(N_{i,\text{Other}},N_{i,\text{Norway EOR}},N_{i,\text{UK EOR}})$.
	
	Let $y_{a, \text{z}, t}$ denote production at time $t \in \{\text{pre}, \text{post}\}$ for a field of type $\text{z} \in \{\text{EOR}, \text{Other}\}$ with EOR adoption decision $a \in \{0,1\}$. Then, $y_{0,\text{EOR},t}$ ($y_{0,\text{Other},t}$) represents the baseline production without EOR adoption for EOR-eligible (non-EOR-eligible) fields in period $t$. With EOR adoption, an eligible field produces $y_{1,\text{EOR},t} = y_{0,\text{EOR},t} + \Delta y$, where $\Delta y$ denotes the production gain from adoption. Firm $i$'s total production at $t$ is:
	\begin{align}\label{eq:basic_outcome}
		y_{it} =& \; N_{i,\text{Norway EOR}} \cdot [y_{0,\text{EOR},t} + \Delta y \cdot \mathbbm{1}\{A_{it} \geq A^*_{t,\text{Norway}}\}] \notag \\
		& + N_{i,\text{UK EOR}} \cdot [y_{0,\text{EOR},t} + \Delta y \cdot \mathbbm{1}\{A_{it} \geq A^*_{t,\text{UK}}\}] + N_{i,\text{Other}} \cdot y_{0,\text{Other},t} ,
	\end{align}
	where $\mathbbm{1}\{\cdot\}$ is the indicator function for the threshold condition in \eqref{eq:threshold_A}. 
	
	\paragraph{Identifying assumption.} We make the following identifying assumption:
	
	\begin{assumption}\label{indep}
		Know-how $A_{it}$ is independent of firm $i$'s portfolio of fields $\mathbf{N}_i$, so that, for any given threshold $A^*$, $\Pr_t(A_{it} \geq A^* \mid \mathbf{N}_i) = \Pr_t(A_{it} \geq A^*) = 1- F_t(A^*)$.
	\end{assumption}
	This assumption rules out (dis-)economies of scale in EOR adoption \textit{within} each period $t$: $i$'s EOR adoption decisions are made independently for each EOR-eligible field in the portfolio and $A_{it}$ does not evolve within $t$ according to $i$'s adoption decisions in $t$. $F_{\text{pre}}$ denotes the distribution of know-how $A_{i,\text{pre}}$ across operators in $t=\text{pre}$ and $F_{\text{post}}$ the distribution of $A_{i,\text{post}}$ in $t=\text{post}$. Importantly, while constant within each $t$, these distributions may differ between $t=\text{pre}$ and $t=\text{post}$ as firms' know-how evolves according to adoption decisions, so that $F_{\text{pre}} \neq F_{\text{post}}$.
	
	Assumption \ref{indep} is best understood as a parallel-trends condition: within each period, firms with different portfolio compositions aree assumed to have the same distribution of know-how. The effect of the 1985 Norwegian Supreme Court decision is then entirely captured by the reduction in the adoption threshold $A^*_{\text{post,Norway}}$, while $F_{\text{post}}$ reflects aggregate capability evolution common to all firms.
	
	\paragraph{DiD regression.} To cast the problem in a DiD framework, rewrite firm $i$'s production in $t$ as:
	\begin{equation}
		\begin{aligned}\label{eq:identity_prod}
			y_{it} &= \mathbb{E}[y_{it}] + (y_{it} - \mathbb{E}[y_{it}]) \\ 
			&= \mathbb{E}[y_{i,\text{pre}}] + \mathbb{E}[y_{i,\text{post}} - y_{i,\text{pre}} ]\cdot \text{Post}_t + \epsilon_{it},
		\end{aligned}
	\end{equation}
	where the second line decomposes $i$'s expected production at time $t$ ($\mathbb{E}[y_{it}]$) into pre- and post-period averages using indicator $\text{Post}_t$ for $t=\text{post}$ and it captures deviations around these averages ($y_{it} - \mathbb{E}[y_{it}]$) in the error term $\epsilon_{it}$. 
	
	Using the distributions $F_t(\cdot)$ and equation \eqref{eq:basic_outcome}, the expected difference between post- and pre-period outcomes in the second line of \eqref{eq:identity_prod} can be written as:
	\begin{equation}
		\begin{aligned}\label{eq:prod_change_1}
			\mathbb{E}[y_{i,\text{post}}-y_{i,\text{pre}}] &= N_{i,\text{Other}} \cdot [y_{0,\text{Other},\text{post}} - y_{0,\text{Other},\text{pre}}]  \\
			&\quad + (N_{i,\text{Norway EOR}} + N_{i,\text{UK EOR}}) \cdot [y_{0,\text{EOR},\text{post}} - y_{0,\text{EOR},\text{pre}}]  \\
			&\quad + N_{i,\text{Norway EOR}} \cdot \Delta y \cdot [F_{\text{pre}}(A^*_{\text{pre}}) - F_{\text{post}}(A^*_{\text{post,Norway}})]  \\
			&\quad + N_{i,\text{UK EOR}} \cdot \Delta y \cdot [F_{\text{pre}}(A^*_{\text{pre}}) - F_{\text{post}}(A^*_{\text{pre}})].
		\end{aligned}
	\end{equation} 
	Define $\Delta_{\text{Other}} = y_{0,\text{Other},\text{post}} - y_{0,\text{Other},\text{pre}}$ and $\Delta_{\text{EOR}} = y_{0,\text{EOR},\text{post}} - y_{0,\text{EOR},\text{pre}}$ as the baseline production changes for each field type, and let firm $i$'s shares of Norwegian and North Sea EOR-eligible fields be:\footnote{It follows that firm $i$'s UK share of EOR-eligible fields is $\text{Share EOR}_i - \text{Share EOR Norway}_i$.}
	\begin{equation*}
		\text{Share EOR Norway}_i = \frac{N_{i,\text{Norway EOR}}}{N}, \quad \text{Share EOR}_i = \frac{N_{i,\text{Norway EOR}}+N_{i,\text{UK EOR}}}{N}.
	\end{equation*}
	With this notation, Appendix \ref{a:firm_regressions} shows that equation \eqref{eq:prod_change_1} simplifies to:
	\begin{align}\label{eq:prod_change_2}
		\mathbb{E}[y_{i,\text{post}}-y_{i,\text{pre}}] &=N \cdot \Delta_{\text{Other}} \\
		&\quad + \left\{\Delta_{\text{EOR}} - \Delta_{\text{Other}} + \Delta y \cdot [F_{\text{pre}}(A^*_{\text{pre}}) - F_{\text{post}}(A^*_{\text{pre}})]\right\} \cdot N \cdot  \text{Share EOR}_i \notag \\
		&\quad + \Delta y \cdot [F_{\text{post}}(A^*_{\text{pre}}) - F_{\text{post}}(A^*_{\text{post,Norway}})] \cdot N \cdot \text{Share EOR Norway}_i. \notag
	\end{align}
	The first line of \eqref{eq:prod_change_2} represents the baseline trend in production for non-EOR-eligible fields. The second line captures both the differential trend in baseline production between EOR-eligible and non-EOR-eligible fields ($\Delta_{\text{EOR}} - \Delta_{\text{Other}}$) and any difference in production due to changes in EOR adoption that are independent of the Norwegian Supreme Court decision (changes in the probability of EOR adoption from $1-F_{\text{pre}}$ to $1-F_{\text{post}}$). Finally, the third line represents the increase in production due to the \textit{extra} EOR adoption induced by the reduction in sovereign hold-up risk in Norway (the lower adoption threshold $A^*_{\text{post,Norway}}$). 
	
	Combining equation \eqref{eq:prod_change_2} with the expectation of equation \eqref{eq:basic_outcome}  in equation \eqref{eq:identity_prod} yields the following DiD regression with firm and time fixed effects:
	\begin{equation}\label{eq:did-firm}
		y_{it} = \gamma_{\text{treat}} \cdot \text{Share EOR Norway}_i \cdot \text{Post}_t + \gamma_{1} \cdot \text{Share EOR}_i \cdot \text{Post}_t + \alpha_i + \psi_t + \epsilon_{it},
	\end{equation}
	where the coefficients correspond to:
	\begin{equation*}
		\begin{aligned}
			\gamma_{\text{treat}} & \equiv N \cdot \Delta y \cdot [F_{\text{post}}(A^*_{\text{pre}}) - F_{\text{post}}(A^*_{\text{post,Norway}})], \\
			\gamma_{1} & \equiv N \cdot \left\{\Delta_{\text{EOR}} - \Delta_{\text{Other}} + \Delta y \cdot [F_{\text{pre}}(A^*_{\text{pre}}) - F_{\text{post}}(A^*_{\text{pre}})]\right\}, \\
			\alpha_i &\equiv N \cdot \bigl\{y_{0,\text{Other},\text{pre}}\cdot (1 - \text{Share EOR}_i) + y_{0,\text{EOR},\text{pre}}\cdot \text{Share EOR}_i \\ 
			&\quad + \Delta y \cdot [1 - F_{\text{pre}}(A^*_{\text{pre}})] \cdot \text{Share EOR}_i\bigr\}, \\
			\psi_t &\equiv N \cdot \Delta_{\text{Other}} \cdot \text{Post}_t.
		\end{aligned}
	\end{equation*}
	The coefficient of interest $\gamma_{\text{treat}}$ captures the effect of a change in the adoption threshold due to the Norwegian Supreme Court decision (as in the third line of \eqref{eq:prod_change_2}), while $\gamma_1$ captures changes affecting all EOR-eligible fields that are not related to the Norwegian Supreme Court decision (as in the second line of \eqref{eq:prod_change_2}). The coefficients $\alpha_i$ and $\psi_t$ instead represent a time-invariant firm-specific effect and a common time effect, respectively.
	
	Crucially, under Assumption \ref{indep}, 
	the residual term $\epsilon_{it}\equiv y_{it} - \mathbb{E}[y_{it}]$ in \eqref{eq:did-firm} satisfies the desired orthogonality condition $\mathbb{E}[\epsilon_{it} \mid \text{Share EOR}_i,$ $\text{Share EOR Norway}_i] = 0$ (see details in Appendix \ref{a:firm_regressions}).\footnote{In the more realistic case in which the number of fields $N_i$ is specific to each firm $i$, the orthogonality condition $\mathbb{E}[\epsilon_{it} \mid \text{Share EOR}_i,$ $\text{Share EOR Norway}_i] = 0$ additionally requires that $N_i$ and portfolio composition are independent: that is, $\mathbb{E}[N_{i} \mid \text{Share EOR}_i,$ $\text{Share EOR Norway}_i] = \mathbb{E}[N_{i}]$.} This ensures that $\gamma_{\text{treat}}$ can be consistently estimated, even though $\epsilon_{it}$ may depend on firm $i$'s portfolio shares. To avoid endogeneity from changes in firms' field portfolios over $t$ that could violate this condition, we evaluate the portfolio shares in equation \eqref{eq:did-firm} using their pre-1985 values. 
	
	In Appendix~\ref{a:field_regressions}, we derive analogous regressions at the field level, which take the form of a triple-differences specification across country (Norway vs.\ UK), field EOR status (eligible vs.\ not), and time (pre vs.\ post).
	
	\paragraph{Discussion.} In practice, identification relies on comparing firms with nearly identical shares of EOR-eligible fields before 1985, where ``treated'' firms had an additional EOR-eligible field in Norway while ``control'' firms had an additional EOR-eligible field in the UK. By controlling for firm and time fixed effects, other policy changes are expected to affect both groups similarly, effectively ``canceling out'' in DiD regression \eqref{eq:did-firm}.\footnote{Field-level regressions also controlling for country-by-year fixed effects confirm this (Section \ref{s:did}).} The observed difference in post-1985 outcomes then identifies $\gamma_{\text{treat}}$.

	\section{Data} \label{s:data}
	Our dataset comes from Wood McKenzie, a consulting firm specializing in resource extraction data. We construct a field-level dataset to analyze field-specific developments and a firm-level dataset that aggregates field-level data based on ownership shares.
	
	\subsection{Field-level Data}\label{s:data_field}
\begin{figure}[h!]
		\captionsetup[subfigure]{justification=centering}
		\centering
		\caption{British and Norwegian O\&G fields in the North Sea}\label{fig:map}
		\includegraphics[width=.5\linewidth]{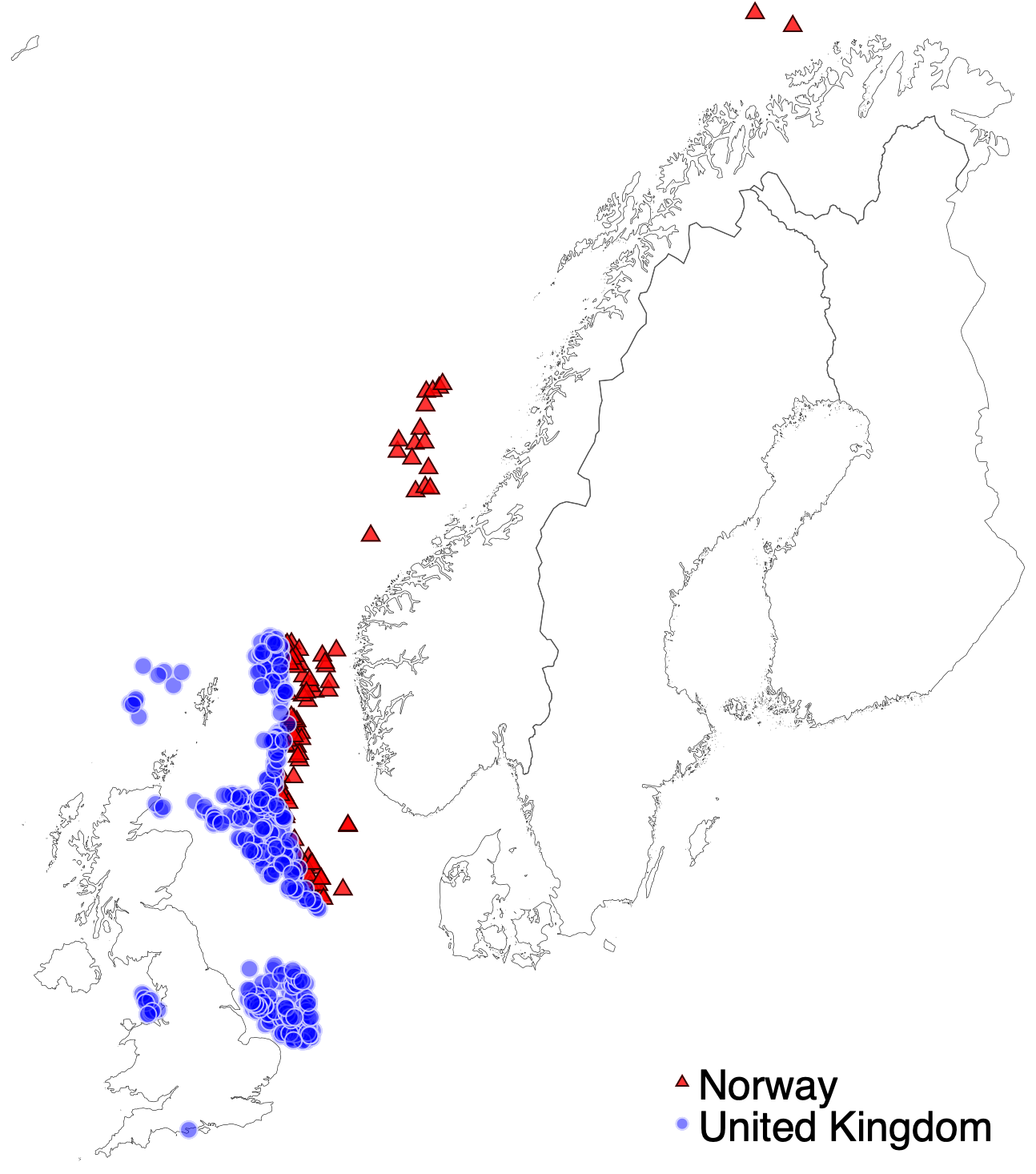}
		\vspace{.5em}
		\begin{minipage}{1 \textwidth}
			{\footnotesize\singlespacing \textbf{Notes:} The figure displays O\&G fields in the southern basin of the North Sea. Blue dots denote the 426 fields under British jurisdiction and red triangles denote the 95 fields under Norwegian jurisdiction.
				\par}
		\end{minipage}
	\end{figure}

	We use data for the North Sea under the jurisdiction of the UK and Norway between 1975 and 1995, as shown in Figure \ref{fig:map}. Blue dots represent UK fields, while red triangles indicate Norwegian fields. Most fields are located on the continental shelf between the two countries, though the UK jurisdiction extends further south. The UK hosts a larger number of O\&G fields, reflecting an earlier industrial development relative to Norway.
	
	Columns 3 and 6 of Table \ref{tab:field-statdes1} show that British and Norwegian O\&G fields are broadly comparable across geology (Panel (a)), oil types (Panel (b)), technology (Panel (c)), and operational practices (Panel (d)). In Panel (a), while Norwegian reservoirs are somewhat larger with more oil in place (as measured in Million Barrels of Oil Equivalent, MMboe), most fields in both countries are similar in size, are located in shallow waters, and have similar geological characteristics. Panel (b) indicates a similar oil-gas mix, with nearly identical API gravity (40.64$^{\circ}$ in the UK vs. 40.30$^{\circ}$ in Norway), suggesting that heavy oil fields are rare in both regions.
	
	\begin{table}[h!]
		\centering
		\caption{Descriptive statistics at the field level}
		\label{tab:field-statdes1}
		\centering
		\begin{adjustbox}{max width={1.1\textwidth},center}
			{
\def\sym#1{\ifmmode^{#1}\else\(^{#1}\)\fi}
\begin{tabular}{l*{6}{c}}
\toprule
          &\multicolumn{3}{c}{UK}                &\multicolumn{3}{c}{Norway}            \\\cmidrule(lr){2-4}\cmidrule(lr){5-7}

          &        Pre-1985&        Post-1985&        Total&         Pre-1985&         Post-1985&        Total\\
          & (1) & (2) & (3) & (4) & (5) & (6)\\
\midrule

\\ \smallskip

\textbf{Panel (a): Field Characteristics types}  & & & & & & \\
\, \, Average initial oil in place (MMboe)&      589.7&      272.6&      355.4&     1366.4&     1235.1&     1270.6\\
\, \, \% fields Large&        0.84&        0.53&        0.61&        0.57  &       0.72  &       0.68\\
\, \, \% fields Moderate&         0.15&        0.38&        0.32&        0.43&        0.25&        0.30\\
\, \, \% fields Small&         0.02  &       0.09   &      0.07     &    0.00    &     0.02      &   0.02\\
\, \, Porosity (\%) &  19.34  &      18.92    &    19.06    &    29.85     &   27.78     &   28.52     \\
\, \, Permeability (Millidarcy) &1,005.56   &    750.62   &    828.34   &    965.30    &  1,602.79    &  1,415.70 \\
\, \, \% shallow and moderate fields ($<$3000m)&        0.82&        0.75&        0.77&        0.50&        0.67&        0.63\\
\, \, \% deep and ultradeep fields ($>$3000m)&        0.18&        0.25&        0.23&        0.50&        0.33&        0.37\\
\, \, Reservoir depth (meters)&     2564&     2630&     2613&     2645&     2619&     2626\\
\, \, Overburden (meters)&          2465&      2542 &     2522 &     2541 &     2476  &    2493     \\
[0.5em]
\textbf{Panel (b): Resource types}  & & & & & & \\
\, \, API Gravity (Fahrenheit)&       40.42&       40.73&       40.64&       41.67&       39.81&       40.30\\
\, \, \% of heavy oil fields&        0.00&        0.02&        0.01&        0.00&        0.00&        0.00\\
\, \, \% of oil (as opposed to gas)&        0.69&        0.57&        0.60&        0.76&        0.72&        0.73\\
\, \, Sulfur content (unit measure)  & 0.48     &    0.49   &      0.48   &      0.23   &      0.22        & 0.22   \\
[0.5em]
\textbf{Panel (c): Technologies}  & & & & & & \\
\, \, Share of active EOR-eligible fields &        0.52&        0.30&        0.35&        0.28&        0.45&        0.40\\
\, \, Share of active EOR-eligible fields (robust definition) &        0.49&        0.29&        0.34&        0.20&        0.40&        0.35\\
\, \, Share of fields adopting EOR&        0.10&        0.05&        0.06&        0.06&        0.16&        0.13\\
\, \, Share of fixed platforms           &  0.86     &    0.72    &     0.76   &      1.00     &    0.74     &    0.81 \\
\, \, Share of fixed subsea            &   0.09    &     0.24   &      0.20   &      0.00    &     0.20    &     0.15 \\
\, \, Share of tension leg platforms    &   0.01    &     0.01    &     0.01    &     0.00    &     0.05    &     0.04 \\ 
[0.5em]
\multicolumn{7}{l}{\textbf{Panel (d): Operations }}  \\
\, \, Average yearly production (MMboe/year)&       20.95&        9.75&       12.68&       27.04&       27.59&       27.44\\
\, \, Average yearly capital expenditure (CAPEX in \$m) &      431.80&       98.89&      185.80&      520.70&      294.36&      355.61\\
\, \, Average yearly operative expenditure (OPEX in \$m) &      144.49&       74.82&       93.01&      303.96&      271.03&      279.94\\
\, \, Average CAPEX per MMboe &      20.61   &     10.14  &      12.87     &   19.26    &    10.67   &     13.00\\
\, \, Average OPEX per MMboe &       6.90     &    7.67    &     7.47    &    11.24    &     9.83    &    10.21\\
[0.5em]
\textbf{Panel (e): Average number of} & & & & & & \\
[0.2em]
\, \,  Active per year (exploration/production)&       40.45&      129.67&      106.38&       11.26&       33.05&       27.15\\
\, \,  Producing fields per year&       30.13&      101.72&       84.61&        6.97&       21.95&       18.62\\
\, \, Firms per year&       66.69&       81.16&       77.38&       18.52&       27.30&       24.93\\
\, \, Operators per year&       12.90&       24.33&       21.50&        6.30&        8.94&        8.21\\
[0.5em]
\midrule
\end{tabular}
}

		\end{adjustbox}
		\vspace{0.2cm}
		\begin{tablenotes}
			\footnotesize \vspace{-0.5em}
			\parbox{\textwidth}{\justifying  \item \hspace{-0.2em}\textbf{Notes:}  Summary statistics are reported for fields in the UK (Columns 1-3) and Norway (Columns 4-6). 
				Columns 1 and 4 present averages for active fields in the exploration or production phase before 1985, Columns 2 and 5 present averages after 1985, and Columns 3 and 6 report averages for all active fields in our 1975-1995 sample. Production is measured in Million Barrels of Oil Equivalent (MMboe). The criteria for a field's EOR eligibility are detailed in Appendix Table \ref{tab:eor_restrictions}. }
		\end{tablenotes}
	\end{table}
	
	Technology adoption is also comparable across countries. We classify fields as EOR-eligible following established engineering criteria detailed in Appendix Table \ref{tab:eor_restrictions}. Panel (c) of Table \ref{tab:field-statdes1} shows that the share of EOR-eligible fields is similar in both countries---35\% in the UK and 40\% in Norway.\footnote{Our main definition of EOR eligibility excludes porosity and permeability. Although there are some differences in porosity and permeability between British and Norwegian fields (Panel (a)), these are largely within the limits of EOR eligibility, see Appendix Table \ref{tab:eor_restrictions}. For completeness, we also repeat our analyses using a ``robust'' definition of eligibility that considers porosity and permeability, which yields similar shares of 34\% and 35\%, respectively. All results in the paper are similar under both measures.} Adoption rates, however, differ markedly: drawing on \cite{awan2008survey} and \cite{gbadamosi2018review}, who list EOR adoptions, we find that Norwegian fields adopted EOR at roughly twice the rate of UK fields by 1995.
	
	
	
	Before 1985, EOR adoption was more common in the UK, whose larger and more technologically advanced O\&G industry hosted nearly four times as many firms and twice as many operators as Norway (Panel (e)). This expertise translated into higher adoption rates and a greater share of active EOR-eligible fields (52\% vs. 28\%). After the 1985 Norwegian Supreme Court ruling, the pattern reversed: in Norway, the share of active EOR-eligible fields rose to 45\% and adoption to 16\%, while in the UK they fell to 30\% and 5\% (Columns 2 and 5, Panel (c)). These shifts also appear in production, with Norway sustaining output through extended field lifespans, while UK production declined sharply, alongside similar trends in CAPEX and OPEX.

	\subsection{Firm-level Data}\label{s:data_firm}
	
	Table \ref{tab:firm-statdes1} compares firms by whether they held shares in Norwegian EOR-eligible fields before the 1985 Supreme Court decision. Panel (a) shows that, on average, 14.8 firms per year owned at least one such field, with state-owned companies like Statoil and BP operating about 25\% of their fields. Firms without Norwegian EOR-eligible holdings expanded from 57 to 73 over the sample, but operated a smaller share (8\%). Global majors such as Amoco and Shell were active in both the UK and Norway. Panel (b) indicates that firms with pre-1985 EOR-eligible fields held larger portfolios, reflecting Statoil's scale and early EOR efforts like Phillips' at Ekofisk.

	\begin{table}[h!]
		\centering
		\caption{Descriptive statistics at the firm level}
		\label{tab:firm-statdes1}
		\centering
		\begin{adjustbox}{max width={1.15\textwidth},center}
			{
\def\sym#1{\ifmmode^{#1}\else\(^{#1}\)\fi}
\begin{tabular}{l*{6}{c}}
\toprule
& \multicolumn{6}{c}{\textbf{Firms that owned EOR-eligible fields in Norway before 1985}} \\
          &\multicolumn{3}{c}{Yes}&\multicolumn{3}{c}{No}\\ 
          \cmidrule(lr){2-4}\cmidrule(lr){5-7}
                   &        Pre-1985&        Post-1985&        Total&         Pre-1985&         Post-1985&        Total\\
                   & (1) & (2) & (3) & (4) & (5) & (6)\\
\midrule 
\\ [0.25em]
\textbf{Panel (a): Firms} & & & & & & \\
[0.5em]
\, \, Average number of firms per year &       13.57&       16.00&       14.84&       57.02&       73.21&       65.82\\
\, \, Average number of state-owned firms per year&        2.64&        3.00&        2.83&        1.00&        1.00&        1.00\\
\, \, Share of firms acting as operators &       0.26&        0.23&        0.25&        0.08&        0.08&        0.08\\
\, \, Average market share&        0.04&        0.05&        0.04&        0.01&        0.00&        0.01\\
[1em]
\textbf{Panel (b): Fields per firm} & & & & & & \\
[0.5em]
\, \, Average number of fields per firm&        7.84&       20.52&       14.45&        1.83&        4.30&        3.17\\
\, \, Average number of producing  fields per firm&        5.38&       14.91&       10.60&        1.72&        3.62&        2.84\\
[1em]
\textbf{Panel (c): Enhanced Oil Recovery} & & & & & & \\
[0.5em]
\, \, Share of EOR-eligible fields (Norway \& UK) &        0.46&        0.50&        0.48&        0.65&        0.44&        0.53\\
\, \, Share of fields adopting EOR (Norway \& UK) &        0.10&        0.18&        0.14&        0.03&        0.05&        0.04\\
\, \, Share of EOR-eligible fields in Norway (Share EOR Norway)&        0.27&        0.38&        0.33&       --&        --&        --\\
[0.5em]
\bottomrule
\end{tabular}
}

		\end{adjustbox}
		\begin{tablenotes}
			\footnotesize \vspace{0.5em}
			\parbox{\textwidth}{\justifying  \item \hspace{-0.2em}\textbf{Notes:}  Summary statistics are reported for firms with (Columns 1-3) and without (Columns 4-6) EOR-eligible fields in Norway before 1985. 
				Columns 1 and 4 present yearly averages for the pre-1985 period, Columns 2 and 5 for the post-1985 period, and Columns 3 and 6 for the entire 1975-1995 sample. The criteria for a field's EOR eligibility are detailed in Appendix Table \ref{tab:eor_restrictions}.}
		\end{tablenotes}
	\end{table}
	
	Panel (c) highlights differences in EOR adoption. Firms without Norwegian EOR-eligible fields before 1985 held a larger share of such fields overall, indicating that ownership patterns did not simply reflect self-selection. Yet far fewer eligible fields actually adopted EOR: post-1985, adoption rose by 66.7\% among firms without Norwegian holdings and by 80\% among those with them. The latter reflects both higher adoption within existing fields (intensive margin) and investment in new EOR-eligible fields (extensive margin). Their share of eligible fields grew from 0.46 to 0.50, while it fell from 0.65 to 0.44 for other firms. As a result, firms with Norwegian EOR-eligible fields before 1985 expanded, raising their share of North Sea production from 4\% to 5\%, while others declined (Panel (a)), potentially connected to drops in global oil prices.

	\section{EOR Adoption and Production}\label{s:did} 
	
	We estimate equation \eqref{eq:did-firm} to assess the impact of sovereign hold-up risk on firms' EOR adoption and production. We define ``Share EOR Norway$_i$'' as the average share of Norwegian EOR-eligible fields in firm $i$'s North Sea portfolio prior to 1985. 
	Similarly, ``Share EOR$_i$'' is the average share of EOR-eligible fields across both Norway and the UK in the same period. To address potential endogeneity, we use field counts rather than ownership-weighted shares, although results are consistent with both definitions. 
	
	\begin{table}[!h]
		\centering
		\caption{Firm-level EOR adoption and production \label{tab:did-firm}}
		\begin{adjustbox}{width=0.7\textwidth,center}
			{
\def\sym#1{\ifmmode^{#1}\else\(^{#1}\)\fi}
\begin{tabular}{l*{3}{c}}
\toprule
& Adoption & \multicolumn{2}{c}{Production}\\
\cmidrule(lr){3-4}
Dependent variable
& Share EOR
& $\ln(1+\text{prod}_{it})$
& $\text{prod}_{it}/\text{OPEX}_{it}$ \\
& (1) & (2) & (3) \\
\midrule
Share EOR Norway $\times$ Post&    0.347\sym{***}&    3.193\sym{***}&    0.257\sym{*}  \\
                &  (0.116)         &  (0.831)             &  (0.140)         \\
[1em]
Share EOR $\times$ Post&    0.021         &   -0.402         &   -0.024         \\
                &  (0.013)         &  (0.266)         &  (0.090)         \\
\midrule
Average dependent variable & 0.053 & 2.562 & 0.328 \\
\midrule
Year FE         &      Yes         &      Yes         &      Yes         \\
Firm FE         &      Yes         &      Yes         &      Yes         \\
Observations    &     1406         &     1406         &     1245         \\
R-squared       &     0.70         &     0.86         &     0.43         \\
\bottomrule
\multicolumn{4}{l}{* -- $p < 0.1$; ** -- $p < 0.05$; *** -- $p < 0.01$}
\end{tabular}
}

		\end{adjustbox}
		\begin{tablenotes}
			\footnotesize 
			\parbox{\textwidth}{\justifying  \item \hspace{-0.1em}\textbf{Notes:} Estimated coefficients from regression \eqref{eq:did-firm}. The dependent variable ``Adoption Share EOR'' in Column 1 is the share of fields that adopted EOR among all the fields in which firm $i$ was active in the North Sea in year $t$. The other dependent variables are production in log (Column 2) and the ratio of production (in Kboe) to OPEX (Column 3). The variable ``Share EOR Norway'' (``Share EOR'') refers to the fraction of Norwegian (North Sea) EOR-eligible fields among all the fields in which firm $i$ was active in the North Sea before 1985. The indicator ``Post'' equals 1 for the years after 1985, the year of the Supreme Court decision, and 0 otherwise. Each regression includes firm and year fixed effects. Standard errors are clustered at the firm level.}
		\end{tablenotes}
	\end{table}
	
	Table \ref{tab:did-firm} reports the coefficient estimates. Column 1 confirms the intuition from Figure \ref{fig:eor_prod_country}: firms with a pre-1985 portfolio of fields that included $10$ percentage points (p.p.) more Norwegian EOR-eligible fields had a $3.5$ p.p. larger share of fields in their portfolios that adopted EOR after 1985.\footnote{As reported in Table \ref{tab:firm-statdes1},  firms with ``Share EOR Norway$_i>0$'' had pre-1985 portfolios including an average of 7.84 fields. Of these, around $0.27\times7.84=2.12$ were, on average, Norwegian EOR-eligible fields. A $10$ p.p. increase in ``Share EOR Norway$_i$'' (from 27\% to 37\%) implies substituting 0.78 of the other fields in the firm's pre-1985 portfolio with 0.78 Norwegian EOR-eligible fields.} In contrast, firms with a pre-1985 portfolio that included British but not Norwegian EOR-eligible fields (with Share EOR$_i>0$ and Share EOR Norway$_i=0$) were not more likely to adopt EOR after 1985. Considering that the average share of fields within a firm's portfolio that adopted EOR was around $4\%$ in the period 1975-1995, these estimates confirm that the Supreme Court decision played a major role in fostering the diffusion of this technology in Norway.
	
	Column 2 shows that firms with a pre-1985 portfolio of fields that included $10$ p.p. more Norwegian EOR-eligible fields had a production of $100\times(\exp(3.193\times0.10)-1)\approx 38\%$ higher in the period 1985-1995. Moreover, for these firms, the increase in production was greater than the increase in operative costs (OPEX), implying a reduction in average variable costs (Column 3). Differently, firms with a pre-1985 portfolio that included British but not Norwegian EOR-eligible fields (with Share EOR$_i>0$ and Share EOR Norway$_i=0$) moved---if at all---in the opposite direction after 1985, slightly decreasing production and increasing average variable costs.
	
\begin{figure}[h!]
		\centering
		\caption{Firm-level event study for production and productivity \label{fig:firm_ES}}
		\captionsetup[subfigure]{justification=centering}
		
		\begin{minipage}[t]{0.48\textwidth}
			\centering
			\subcaption{Production, $\ln(1+\text{prod}_{it})$}
			\includegraphics[width=\linewidth]{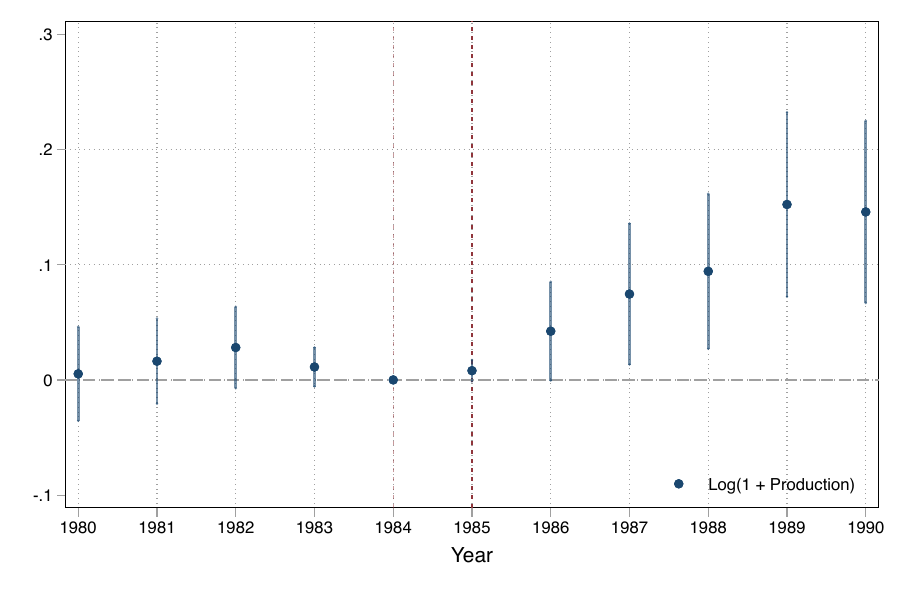}
		\end{minipage}\hfill
		\begin{minipage}[t]{0.48\textwidth}
			\centering
			\subcaption{Inverse Avg. Var. Cost, $\text{prod}_{it}/\text{OPEX}_{it}$}
			\includegraphics[width=\linewidth]{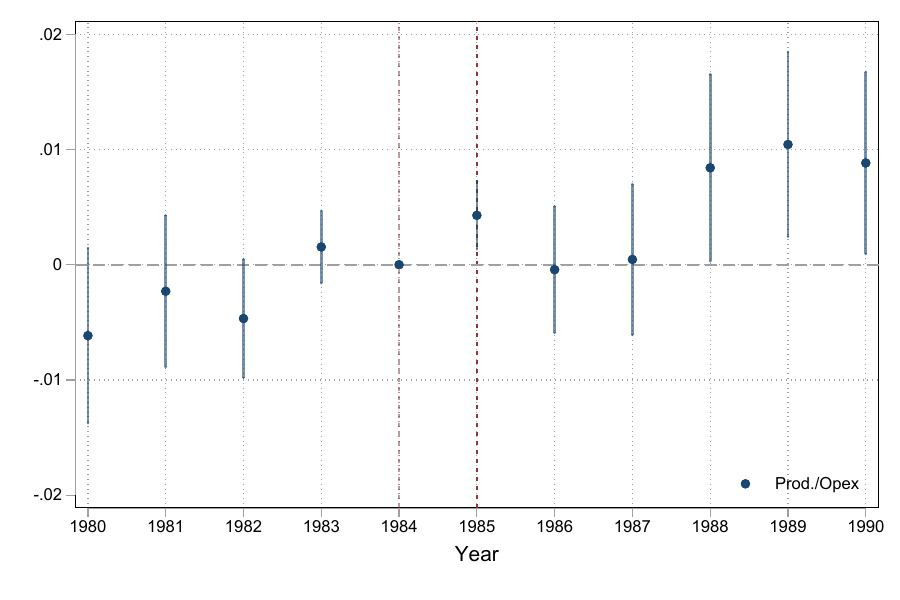}
		\end{minipage}
		
		\vspace{-0.5em}
		
		\parbox{\textwidth}{\footnotesize
			\textbf{Notes:} Estimated coefficients from regression \eqref{eq:event_firm}. Panel (a) shows the estimates of $\widehat{\gamma}_{\text{treat},d}$ using production (in log) as dependent variable, while Panel (b) using production over OPEX (Panel (b)). The variable ``Share EOR Norway'' (``Share EOR'') refers to the fraction of Norwegian (North Sea) EOR-eligible fields among all the fields in which firm $i$ was active in the North Sea before 1985. Each regression includes firm and year fixed effects. Vertical bars around each estimate represent 95\% confidence intervals. Standard errors are clustered at the firm level. Vertical dotted lines mark the years of the rulings of the Eidsivating Court of Appeal (1984) and the Norwegian Supreme Court (1985).}
	\end{figure}

	\sloppy As discussed in Section \ref{s:identification}, the condition $\mathbb{E}[\epsilon_{it} \mid \text{Share EOR}_i, \ \text{Share EOR Norway}_i] = 0$ ensures that the error term is uncorrelated with the explanatory variables, which is crucial for the consistent estimation of DiD regression \eqref{eq:did-firm}. This can be interpreted as a parallel trends assumption, where in the absence of the Supreme Court decision, the changes in outcomes for firms with different pre-1985 shares of EOR-eligible fields would have followed the same trends over time. The following event study examines this assumption:
	\begin{equation}\label{eq:event_firm}
		\begin{aligned}
			y_{it}  &= \sum_{d=-4}^{6} \gamma_{\text{treat},d} \cdot \text{Share EOR Norway}_{i} \cdot \mathbbm{1}\{t-1984=d\} \\
			&+ \sum_{d=-4}^{6} \gamma_{1,d} \cdot \text{Share EOR}_{i} \cdot \mathbbm{1}\{t-1984=d\} + \alpha_i + \psi_{t} + \epsilon_{it},
		\end{aligned}
	\end{equation}
	where $\mathbbm{1}\{t-1984=d\}$ is an indicator that equals 1 when year $t$ happens to be $d$ years before or after 1984. As in regression \eqref{eq:did-firm}, $\alpha_{i}$  and $\psi_{t}$ are firm and year fixed effects. We report the estimates of $\gamma_{\text{treat},d}$ in Figure \ref{fig:firm_ES}, which shows parallel trends in the pre-1985 period for both production and production over OPEX.
	
	\subsection{Robustness Checks}\label{s:rob}
	In Appendix \ref{a:robustness_checks}, we discuss a series of robustness checks, which we summarize here.
	
	\paragraph{Field-level and national policy controls.} 
	To ensure our results are not driven by differences in field characteristics between Norway and the UK and other national-level shocks or divergent energy policies beyond the Norwegian Supreme Court decision of 1985, we estimate a field-level triple-difference specification derived in Appendix \ref{a:field_regressions}. This specification includes field fixed effects and country-by-year fixed effects, which control for all variation occurring across fields (e.g., field maturity) and at the national level in any given year.\footnote{These fixed effects also absorb, for example, any differential fiscal response by Norway and the UK to the 1986 oil price collapse.} As shown in Appendix Table \ref{tab:did-field-appendix}, which reports both OLS and Poisson Pseudo Maximum Likelihood (PPML) estimates (Panels (a) and (b), respectively), Norwegian EOR-eligible fields were 27.6 p.p. more likely to adopt EOR and saw significant production gains across both intensive and extensive margins.\footnote{The PPML estimator is more suitable to account for zeros and heteroskedasticity than standard log-linear OLS specifications.} Consistently, the corresponding event study (Appendix Figure \ref{fig:field_ES}) shows absence of pre-trends and sharp increases immediately after the 1985 decision. 
	
	These findings are mirrored at the firm level in Appendix Table \ref{tab:did-firm-appendix} (Panels (a) and (b)), even when controlling for a firm's general exposure to the Norwegian market through a ``Share Norway$_i \cdot \text{Post}_t$'' interaction (Appendix Table \ref{tab:did-firm_all_treat}).\footnote{The inclusion of this interaction does not qualitatively affect our results but, being highly correlated with other interactions, it inflates the standard errors. Thus, we exclude it from the main text.}
	
	\paragraph{Phasing out Royalties.} 
	A key concern for our findings is that the UK (in 1983) and Norway (in 1986) phased out royalties for new fields, which could have acted as a prospective fiscal incentive. To isolate the Supreme Court's effect on existing assets, we re-estimate our regressions on a restricted sample excluding all fields discovered or licensed after 1983. This ensures all included fields were already operating under sunk legal terms before the ruling. Results in Appendix Table \ref{tab:did-field-appendix} (Panel (d)) and Appendix Table \ref{tab:did-firm-appendix} (Panel (c)) remain unchanged, confirming that the response was driven by the legal protection of existing contracts rather than by prospective tax breaks. Similarly, our results are robust to defining the treatment as starting with the 1984 Court of Appeal ruling (Appendix Table \ref{tab:did-field-appendix}, Panel (c) and Appendix Table \ref{tab:did-firm-appendix}, Panel (b)).\footnote{Our results are also robust to defining the treatment as starting in 1982, the year of Norway's appeal to the Eidsivating Court of Appeal. In fact, Figure \ref{fig:firm_ES} indicates that production and productivity of treated firms did not significantly increase until \textit{after} the 1985 Supreme Court decision.} 

	\paragraph{Mechanisms and displacement.} 
	We validate our conceptual framework through several additional checks. First, we confirm a central prediction of our model in Section \ref{s:conceptual_framework}: EOR adoption was primarily driven by operators with larger ownership stakes ($s$), who capture more of the gains from EOR adoption and thus have higher incentives to overcome sovereign hold-up risks (Appendix Table \ref{tab:did-field_HOS}). Second, results are robust to an alternative EOR-eligibility measure accounting also for permeability and porosity (Appendix Tables \ref{tab:did-field-appendix} and \ref{tab:did-firm-appendix}). Finally, we find no evidence of investment ``crowding-out:'' production in Norwegian non-EOR-eligible fields remained stable, and firms without EOR-eligible Norwegian assets showed no decline in output (Appendix Figure \ref{fig:combined_row}). These results collectively support our interpretation that reductions in sovereign hold-up risk unlocked \textit{new} investment in Norwegian EOR-eligible fields.
	
	\paragraph{Other types of Norwegian fields.}
	
	Although our analysis focuses on EOR-eligible fields, one may wonder whether the Supreme Court decision encouraged, more broadly, investment in all types of Norwegian fields. Columns 1-2 of Table \ref{tab:concentration-firm-other-exposure} present estimates of firm-level regression \eqref{eq:did-firm} using the same dependent variables as Columns 2-3 of Table \ref{tab:did-firm}, but replacing the ownership shares of EOR-eligible fields with the ownership shares of non-EOR-eligible fields. Columns 3-4 of Table \ref{tab:concentration-firm-other-exposure} are discussed in Section \ref{s:mkt_shares}.
	
	Recognizing that significant R\&D efforts in the North Sea have focused on the recovery of deep oil due to the discovery of ultra-deep well fields \citep{doornenbal2022new}, Panel (a) of Table \ref{tab:concentration-firm-other-exposure} examines ownership of deep well fields (reservoir depth greater than 2,000 meters and non EOR eligible). Panel (b) investigates sour-oil fields (fields with sulfur presence), which pose challenges such as corrosion and production losses, although new technologies have improved their viability \citep{burgers2011worldwide}. Panel (c) considers ownership of other non-deep, non-sour, non-EOR-eligible fields, while Panel (d) examines the total ownership of a firm's non-EOR-eligible fields (i.e., the total of the shares from Panels (a)-(c)). These results illustrate that the increases in production and productivity documented for EOR-eligible fields in Table \ref{tab:did-firm} did not occur for other types of field.

	\begin{table}[!h]
		\centering
		\caption{Firm-level production and market shares---non-EOR-eligible fields \label{tab:concentration-firm-other-exposure}}
		\begin{threeparttable}
			{
\def\sym#1{\ifmmode^{#1}\else\(^{#1}\)\fi}
\begin{adjustbox}{width=0.7\textwidth,center}
\begin{tabular}{l*{4}{c}}
\toprule
 &\multicolumn{2}{c}{Production}&\multicolumn{2}{c}{Market Share}\\\cmidrule(lr){2-3}  \cmidrule(lr){4-5}
Dependent variable  &\multicolumn{1}{c}{$\ln(1+\text{prod}_{it})$}&\multicolumn{1}{c}{$\text{prod}_{it}\text{ }/\text{ }\text{OPEX}_{it}$}&\multicolumn{1}{c}{All}&\multicolumn{1}{c}{Positive}\\
                \rule{0pt}{2ex}   
&\multicolumn{1}{c}{(1)}         &\multicolumn{1}{c}{(2)}         &\multicolumn{1}{c}{(3)}         &\multicolumn{1}{c}{(4)}         \\ 
\rule{0pt}{2ex}   \\[-1ex] 
\multicolumn{5}{l}{\textit{Panel (a): Non-EOR, Deep Well Fields (Deep)}}\\
\midrule
Share Deep Norway $\times$ Post&    0.789         &    0.157         &    0.013         &    0.008         \\
                &  (0.841)         &  (0.140)         &  (0.018)         &  (0.023)         \\
[1em]
Share Deep $\times$ Post&   -0.050         &   -0.176         &    0.001         &    0.005\sym{*}  \\
                &  (0.304)         &  (0.130)         &  (0.003)         &  (0.003)         \\
[1em]    
R-squared        &     0.84         &     0.41         &     0.82         &     0.84         \\
\midrule\midrule
\rule{0pt}{4ex}   \\[-3ex] 

\multicolumn{5}{l}{\textit{Panel (b): Non-EOR, Sour Oil Fields (Sour)}}\\
\midrule
Share Sour Norway $\times$ Post&    0.482         &    0.075         &    0.051         &    0.060         \\
                &  (1.313)         &  (0.231)         &  (0.045)         &  (0.046)         \\
[1em]
Share Sour $\times$ Post&   -0.291         &   -0.205         &   -0.007         &   -0.004         \\
                &  (0.338)         &  (0.156)         &  (0.005)         &  (0.005)         \\
[1em]    
R-squared        &     0.84         &     0.42         &     0.82         &     0.85         \\
\midrule\midrule
\rule{0pt}{4ex}   \\[-3ex] 

\multicolumn{5}{l}{\textit{Panel (c): Non-EOR, Non-Deep, Non-Sour Oil Fields (Other)}} \\
\midrule
Share Other Norway $\times$ Post&    0.773         &    0.035         &    0.009         &    0.003         \\
                &  (1.039)         &  (0.112)         &  (0.022)         &  (0.032)         \\
[1em]
Share Other $\times$ Post&    0.214         &    0.079         &   -0.001         &    0.000         \\
                &  (0.343)         &  (0.077)         &  (0.002)         &  (0.002)         \\
[1em]
R-squared       &     0.85         &     0.41         &     0.82         &     0.84         \\
\midrule\midrule
\rule{0pt}{4ex}   \\[-3ex] 

\multicolumn{5}{l}{\textit{Panel (d): Non-EOR Fields}} \\
\midrule
Share non-EOR Norway $\times$ Post&    0.551         &    0.184         &    0.016         &    0.014         \\
                &  (0.739)         &  (0.703)         &  (0.016)         &  (0.020)         \\
[1em]
Share non-EOR $\times$ Post&    0.190         &    0.005         &   -0.004         &   -0.001         \\
                &  (0.304)         &  (0.894)         &  (0.003)         &  (0.002)         \\
[1em]
R-squared       &     0.84         &     0.35         &     0.82         &     0.84         \\
\midrule\midrule
Year FE         &      Yes         &      Yes         &      Yes         &      Yes         \\
Firm FE         &      Yes         &      Yes         &      Yes         &      Yes         \\
Observations    &     1406         &     1406         &     1406         &     1232         \\
\bottomrule
\multicolumn{5}{l}{* -- $p < 0.1$; ** -- $p < 0.05$; *** -- $p < 0.01$}
\end{tabular}
\end{adjustbox}
}
			\begin{tablenotes}
				\footnotesize \vspace{1em}
				\parbox{\textwidth}{\justifying  \item \hspace{-0.2em}\textbf{Notes:}  Estimated coefficients from regression \eqref{eq:did-firm}. In Panel (a), ``Share Deep Norway'' (``Share Deep'') refers to the fraction of Norwegian (North Sea) \textit{ultra-}deep, 
					non-EOR-eligible fields among all the fields in which firm $i$ was active in the North Sea before 1985. Similarly, Panel (b) focuses on sour, non-EOR-eligible fields, Panel (c) on the remaining (other) non-EOR-eligible fields, and Panel (d) on all non-EOR-eligible fields. The dependent variables used in Columns 1-2 are production (in log) and the ratio of production to OPEX, while the dependent variables used in Columns 3-4 are the North Sea production market shares. Column 4 restricts attention to firms with positive market shares. The indicator ``Post'' equals 1 for the years after 1985, the year of the Supreme Court decision, and 0 otherwise. Each regression includes firm and year fixed effects. Standard errors are clustered at the firm level.}
			\end{tablenotes}
		\end{threeparttable}
	\end{table}
	
	These findings indicate that, overall, firms with pre-1985 portfolios that included more Norwegian non-EOR-eligible fields did not experience substantial changes in operations. Although incentivizing investments in EOR-eligible fields was not the explicit objective of the Norwegian Supreme Court decision in 1985, this was nonetheless the primary margin along which firms responded to the ruling.
	
	\section{Market Shares and Portfolio Expansions}\label{s:mkt_shares}
	The results in the previous section show that the Norwegian Supreme Court decision in 1985 encouraged firms to adopt EOR in Norway but not in the UK and that this, in turn, led firms with Norwegian EOR-eligible fields to increase production. We now investigate whether this increase in production allowed EOR adopters in Norway to acquire larger shares of the O\&G market in the North Sea. 
	
	We begin by estimating firm-level regression \eqref{eq:did-firm}, with North Sea production market shares as the dependent variable. Column 1 of Table \ref{tab:concentration-firm} shows that firms whose pre-1985 portfolios included 10 p.p. more Norwegian EOR-eligible fields had 0.39 p.p. higher market shares after 1985. Given that the average firm's share was about 1\%, this is a 39\% increase. Restricting to producing firm-years (Column 2), the effect is 0.73 p.p., or a 73\% increase. By contrast, firms with British but not Norwegian EOR-eligible fields experienced no change.\footnote{Appendix Table \ref{tab:did-firm_all_treat} confirms these results by including also $\text{Share Norway}_i \cdot \text{Post}_t$ as a regressor, to control for possible confounding changes affecting the owners of any Norwegian field.} Columns 3-4 of Table \ref{tab:concentration-firm-other-exposure} further show that this increase in market shares was experienced only by owners of Norwegian EOR-eligible fields, as the market shares of owners of other Norwegian field types did not increase after 1985. 
	
	Appendix Figure \ref{fig:concentration-firm} plots the estimates of event study regression \eqref{eq:event_firm} using North Sea production market shares as the dependent variable and shows parallel trends in the pre-1985 period for firms with Norwegian EOR-eligible fields.\footnote{Appendix Tables \ref{tab:concentration-firm-1983} and \ref{tab:concentration-firm-rob} present robustness checks for different definitions of the post period (since 1984, the year of the ruling of the Eidsivating Court of Appeal) and EOR eligibility (including porosity and permeability, whose values are missing for some fields).}
	
	In addition to the increased productivity of the EOR-eligible fields already included in the portfolio, the market share gains documented in Columns 1-2 of Table \ref{tab:concentration-firm} could also have been driven by portfolio expansions at the extensive margin---through acquisitions of new fields---or at the intensive margin---by increasing the ownership shares of fields already included in the portfolio. We analyze these portfolio expansion strategies in turn.
	
	\paragraph{Extensive margin.}
	Columns 3-5 of Table \ref{tab:concentration-firm} examine whether the Norwegian Supreme Court decision led firms to expand their portfolios of fields. Firms owning 10 p.p. more Norwegian EOR-eligible fields before 1985 expanded their portfolios by one additional field after 1985 ($0.1\times10.42$), with 56.5\% of this growth in EOR-eligible fields (Column 4, $5.88/10.42$) and 43.5\% in non-EOR-eligible fields (Column 5, $4.53/10.42$). In contrast, firms owning 10 p.p. more British but not Norwegian EOR-eligible fields reduced their portfolios by about 0.76 fields ($-24\%$ EOR-eligible, $-76\%$ non-EOR-eligible).

	\begin{table}[!htb]
		\centering
		\caption{Market shares and portfolio expansions}\label{tab:concentration-firm}
		{
\def\sym#1{\ifmmode^{#1}\else\(^{#1}\)\fi}
\begin{adjustbox}{max width={.85\textwidth},center}
\begin{tabular}{l*{5}{c}}
\toprule
                 &\multicolumn{2}{c}{Market Share}&\multicolumn{3}{c}{Number of Fields}\\ \cmidrule(lr){2-3}\cmidrule(lr){4-6}
 Dependent variable               &   All   & Positive     & All  & EOR & Non-EOR   \\
                & \multicolumn{1}{c}{}   &       & Fields  & Eligible & Eligible \\
                &\multicolumn{1}{c}{(1)}         &\multicolumn{1}{c}{(2)}         &\multicolumn{1}{c}{(3)}         &\multicolumn{1}{c}{(4)}         &\multicolumn{1}{c}{(5)}         \\
               \midrule

Share EOR Norway $\times$ Post&    0.039\sym{*}  &    0.073\sym{**} &   10.417\sym{**} &    5.885\sym{***}&    4.532\sym{**} \\
                &  (0.021)         &  (0.033)         &  (4.308)         &  (2.207)         &  (2.253)         \\
[1em]
Share EOR $\times$ Post&   -0.000         &    0.001         &   -7.567\sym{***}&   -1.845\sym{***}&   -5.722\sym{***}\\
                &  (0.003)         &  (0.003)         &  (1.488)         &  (0.614)         &  (1.043)         \\
\midrule
Average dependent variable & 0.015 & 0.017 & 5.858 & 2.410 & 3.449 \\ 
\midrule
Year FE         &      Yes         &      Yes         &      Yes         &      Yes         &      Yes         \\
Firm FE         &      Yes         &      Yes         &      Yes         &      Yes         &      Yes         \\
Observations    &     1406         &     1232         &     1406         &     1406         &     1406         \\
R-squared       &     0.83         &     0.86         &     0.81         &     0.80         &     0.79         \\
\bottomrule
\multicolumn{6}{l}{* -- $p < 0.1$; ** -- $p < 0.05$; *** -- $p < 0.01$}
\end{tabular}
\end{adjustbox}
}

		\begin{tablenotes}
			\footnotesize \vspace{1em}
			\parbox{\textwidth}{\justifying  \item \hspace{-0.2em}\textbf{Notes:}  Estimated coefficients from regression \eqref{eq:did-firm}. ``Market share'' is a firm's market share of production in the North Sea. ``All Fields'' is a firm's total number of fields that were in exploration, development, or production in the North Sea in year $t$. The dependent variables in Columns 4-5 decompose the total number of fields in Column 3 between the number of EOR-eligible fields (Column 4) and the number of non-EOR-eligible fields (Column 5) in which a firm is active in the North Sea. The variable ``Share EOR Norway'' (``Share EOR'') refers to the fraction of Norwegian (North Sea) EOR-eligible fields among all the fields in which firm $i$ was active in the North Sea before 1985. The indicator ``Post'' equals 1 for the years after 1985, the year of the Supreme Court decision, and 0 otherwise. Each regression includes firm and year fixed effects. Standard errors are clustered at the firm level. }
		\end{tablenotes}
	\end{table}
	
	\paragraph{Intensive margin.} Next, we investigate whether the Norwegian Supreme Court decision also led firms to increase their ownership shares of fields already included in their portfolios. We do so by estimating the field-level triple-difference regression derived in Appendix \ref{a:field_regressions}. Columns 1-3 of Table \ref{tab:concentration-field} show that after 1985, ownership shares of Norwegian EOR-eligible fields rose relative to those of UK fields by 27.3 p.p. for operators, 29.1 p.p. for the largest owner, and 40.4 p.p. for the top three owners.  Consistently, Columns 4-5 indicate greater ownership concentration in Norwegian EOR-eligible fields after 1985, as measured by the HHI (with the normalized index in Column 5 accounting for differences in the number of field owners). Column 6 shows that this was partly driven by a lower number of owners per field, which fell by 0.39 on average, nearly a 10\% decline.\footnote{Appendix Table \ref{tab:concentration-field-1983} confirms the robustness of these results when defining ``Post$_t$'' as the years since 1984, the year of the Eidsivating Court of Appeal decision.}

	\begin{table}[!htb]
		\caption{Concentration in field ownership \label{tab:concentration-field}}
		\centering
		\begin{adjustbox}{width=.95\textwidth}{
\def\sym#1{\ifmmode^{#1}\else\(^{#1}\)\fi}
\begin{tabular}{l*{6}{c}}
\toprule
                &\multicolumn{3}{c}{Share of}&\multicolumn{2}{c}{HHI}&\multicolumn{1}{c}{Number of Firms}\\\cmidrule(lr){2-4}\cmidrule(lr){5-6}
                Dependent variable &\multicolumn{1}{c}{Operator}         &\multicolumn{1}{c}{Top 1}         &\multicolumn{1}{c}{Top 3}       &\multicolumn{1}{c}{Standard}         &\multicolumn{1}{c}{Normalized}           &\multicolumn{1}{c}{per Field}         \\
                &\multicolumn{1}{c}{(1)}         &\multicolumn{1}{c}{(2)}         &\multicolumn{1}{c}{(3)}         &\multicolumn{1}{c}{(4)}         &\multicolumn{1}{c}{(5)}         &\multicolumn{1}{c}{(6)}         \\
\midrule
Norway $\times$ EOR $\times$ Post          &    0.273\sym{*}  &    0.291\sym{**} &    0.404\sym{**} &    0.204\sym{*}  &    0.249\sym{*}  &   -0.390\sym{**} \\
                &  (0.141)         &  (0.147)         &  (0.196)         &  (0.115)         &  (0.141)         &  (0.196)         \\
[1em]
EOR $\times$ Post         &   -0.024         &   -0.020         &    0.021         &   -0.019         &    0.023         &    0.208         \\
                &  (0.053)         &  (0.054)         &  (0.089)         &  (0.048)         &  (0.067)         &  (0.167)         \\
\midrule
Country-Year FE &      Yes         &      Yes         &      Yes         &      Yes         &      Yes         &      Yes         \\
Field FE        &      Yes         &      Yes         &      Yes         &      Yes         &      Yes         &      Yes         \\
Observations    &     1988         &     1988         &     1988        &     1988         &     1781          &     1988         \\
R-squared       &     0.59         &     0.47         &     0.46         &     0.64         &     0.52          &     0.96         \\
\bottomrule
\multicolumn{7}{l}{* -- $p < 0.1$; ** -- $p < 0.05$; *** -- $p < 0.01$}
\end{tabular}
}
\end{adjustbox}

		\begin{tablenotes}
			\footnotesize \vspace{1em}
			\parbox{\textwidth}{\justifying \item \hspace{-0.2em}\textbf{Notes:} Estimated coefficients from the triple-difference regression at field level (see equation \eqref{eq:did-firm} and its derivation in Appendix \ref{a:field_regressions}). The dependent variables used in Columns 1-3 are the ownership share of the operator, the largest, and the combined shares of the three largest firms in a field. The dependent variables used in Columns 4-5 are the Herfindahl-Hirschman concentration index (HHI) computed in terms of asset interest shares at the field-level. ``Normalized HHI'' in Column 5 is computed as $\frac{\text{HHI} - 1/N}{1-1/N}$ and helps comparing the HHI across fields in which the number of firms $N$ vary. ``Number of Firms'' in Column 6 is the number of partnering firms in a specific field. The indicator variables ``Norway,'' ``EOR,'' and ``Post'' equal 1 for Norwegian fields, for EOR-eligible fields, and for the years after 1985. Each regression includes country-by-year and field fixed effects. Standard errors are clustered by field.}
		\end{tablenotes}
	\end{table}
	
	\paragraph{Discussion.} These results suggest that the reduction in sovereign hold-up risk due to the Norwegian Supreme Court decision enabled EOR adopters in Norway to acquire larger shares of the O\&G market in the North Sea. Importantly, these market share gains were not only due to the increased productivity of the EOR-eligible fields already included in the portfolios, but were also achieved through two separate portfolio expansion strategies. First, pre-1985 owners of Norwegian EOR-eligible fields expanded their portfolios at the extensive margin, acquiring ownership shares of fields that were not already in their portfolios. Second, they also expanded their portfolios at the intensive margin, increasing their ownership shares of Norwegian EOR-eligible fields already in their portfolios.

	\section{Know-How and State Ownership}\label{s:knowhow}
	
	Motivated by our theoretical framework, in this section we investigate the mechanisms underlying our main results from Section \ref{s:mkt_shares}. Specifically, we test for heterogeneous effects based on firm-level capabilities to understand whether the expansion of market shares and portfolios was driven by the accumulation of technical expertise or by the advantages of state support. To disentangle these channels, we distinguish between firms that possessed EOR-specific know-how in the North Sea and those characterized by state ownership.
	
	We define a firm as possessing EOR-specific know-how in the North Sea if it acted as the operator of at least one North Sea field that adopted EOR prior to 1985. This definition captures the crucial distinction between active management and passive equity ownership. Although passive owners benefit financially from EOR adoption, it is the operator who actively manages the reservoir pressure, coordinates with service providers, and accumulates the learning-by-doing experience central to our model.
	
	State-owned firms---specifically Statoil, BP, Britoil, and Norway State DFI---played a dominant role in the North Sea during this period.\footnote{Appendix Table \ref{tab:north_sea_companies} lists the state-owned firms operating in the North Sea between 1975 and 1995.} Distinguishing these firms is essential because their market expansion may have stemmed not from technical efficiency but from state-backed financial support and preferential regulatory treatment. For instance, to support its early growth, Statoil benefited from ``carried interest'' (where exploration expenses were covered by other licensees) and the ``gliding scale'' (the right to increase license shares up to 80\% at production), terms that were only phased out by 1993 \citep{austvik2014norwegian}. By controlling for state ownership, we aim to isolate the effect of technical know-how from these institutional advantages.

	\subsection{Empirical Strategy}
	To separately isolate the importance of EOR-specific know-how in the North Sea and state ownership, we categorize firms using two indicator variables, ``$\text{K-H}_{i}$'' and ``$\text{S-O}_{i}$.'' The indicator variable ``$\text{K-H}_{i}$'' equals one if firm $i$ was the \textit{operator} of at least a field in the North Sea which adopted EOR before 1985. This definition of know-how ensures an active role for firm $i$ in the adoption and management of EOR prior to the Norwegian Supreme Court decision, rather than only passive ownership of EOR-eligible fields. Differently, the indicator variable ``$\text{S-O}_{i}$'' equals one if firm $i$ was state-owned before 1985.
	
	Because nearly all state-owned firms possessed EOR-specific know-how (in the sense of ``$\text{K-H}_{i}=1$'') but not all firms with EOR-specific know-how were state-owned, we categorize firms into three mutually exclusive groups: firms that prior to 1985 had know-how but were not state-owned---Conoco, Mobil, and Shell (denoted by ``K-H Only$_i$,'' accounting for 263 fields)---, firms that prior to 1985 were state-owned and had know-how, such as Statoil and BP (denoted by ``K-H \& S-O$_i$,'' 157 fields), and other firms (517 fields).\footnote{There are 521 unique fields in total. In our baseline regressions, we exclude two firms that were state-owned but did not have any EOR-eligible field in their pre-1985 portfolios. These are Norway State DFI, which was endowed with shares of Statoil's EOR-eligible fields only in 1984, and Britoil, which was purchased by BP in 1988. Including these two firms in the regressions in the separate category ``S-O Only$_i$'' (stated-owned without know-how) does not affect our results.} We then estimate the heterogeneous effects of the Norwegian Supreme Court decision for each firm category by augmenting the treatment in regression \eqref{eq:did-firm} as follows:
	\begin{equation}\label{eq:know-how-firm}
		\begin{aligned}
			y_{it} &= \theta_{\text{treat}}^{\text{S-O}} \cdot \text{K-H \& S-O}_{i} \cdot \text{Share EOR Norway}_{i} \cdot \text{Post}_t  \\
			&+ \theta_{\text{treat}}^{\text{K-H}} \cdot \text{K-H Only}_{i} \cdot \text{Share EOR Norway}_{i} \cdot \text{Post}_t  \\
			&+ \theta_{1} \cdot \text{Share EOR Norway}_{i} \cdot \text{Post}_t + \theta_{2} \cdot \text{Share EOR}_{i} \cdot \text{Post}_t \\
			&+ \alpha_i + \psi_{t} + \epsilon_{it}.
		\end{aligned}
	\end{equation}
	As in regression \eqref{eq:did-firm}, $\theta_{1}$ captures the effect of a higher pre-1985 ownership of EOR-eligible fields in Norway after the Norwegian Supreme Court decision. $\theta_{\text{treat}}^{\text{S-O}}$ captures the additional effect of the Supreme Court decision on state-owned firms. Because state-owned firms also had direct expertise on EOR adoption in the North Sea by 1985, $\theta_{\text{treat}}^{\text{S-O}}$ combines the additional returns due to the advantageous conditions of state ownership with any extra return related to EOR-specific know-how.\footnote{Statoil's technical learning was driven by strong government support. For instance, the state granted Statoil the Gullfaks field (despite objections from international oil companies), making Statoil an operator in 1981, and in 1987 transferred Statfjord operatorship from Mobil to Statoil—moves whose technical justification has been questioned \citep{thurber2010norway}.} 
	In contrast, $\theta_{\text{treat}}^{\text{K-H}}$ isolates the extra returns of EOR-specific know-how---returns other than those implied by state ownership or the financial benefits shared by all owners of Norwegian EOR-eligible fields.\footnote{We omit the interactions between operator types and ``Post$_t$'' due to the high correlation with their three-way interactions with ``ShareEOR Norway$_i$'' (corr.=0.984 for K-H \& S-O; 0.971 for K-H only).} 
	
	Before proceeding, we note that drilling decisions in a field, due to spatial correlation in well outcomes, usually correlate with drilling decisions in nearby fields \citep{hodgson2024information}. Similarly, EOR adoption in a field could also be influenced by EOR adoptions in nearby fields. However, as we show in Appendix \ref{apndx:dist}, it was operators' know-how---rather than proximity to fields that adopted EOR---the primary driver of EOR adoption. As a consequence, in what follows, we do not further explore the role of geographic proximity.

	\subsection{Estimation Results} \label{s:ms}
	Column 1 of Table \ref{tab:know-firm} presents the estimates of firm-level regression \eqref{eq:know-how-firm} using the market share of production in the North Sea as the dependent variable. Column 1 shows that ``K-H \& S-O'' firms with a pre-1985 portfolio of fields that included 10 p.p. more Norwegian EOR-eligible fields had a market share $1.32$ p.p. higher after 1985, relative to non-state-owned firms without know-how (or 1.32 + 0.1 = 1.42 p.p. higher relative to firms without EOR-eligible fields). As these firms had an average market share of 0.95\% in the period 1975-1985, this corresponded to a 139\% increase. Firms that, prior to 1985, were operators in fields that adopted EOR \textit{and} that were not state-owned (``K-H Only'') also experienced large gains in market shares, with an increase of $0.5$ p.p.. As their average market share was around 0.5\% in the period 1975-1985, this corresponded to a 100\% increase. In addition, $\widehat{\theta}_{1}$ is 12 and 5 times smaller than $\widehat{\theta}_{\text{treat}}^{\text{S-O}}$ and $\widehat{\theta}_{\text{treat}}^{\text{K-H}}$, respectively, suggesting that a large portion of the market share expansion documented in Section \ref{s:mkt_shares} was driven by operators' expertise in adopting and managing EOR during its life cycle, rather than the financial gains due to the mere ownership of EOR-eligible fields.
	
	\begin{table}[!ht]
		\centering
		\caption{Market shares and portfolio expansions (heterogeneity)} \label{tab:know-firm}
		\centering
		\begin{adjustbox}{max width={0.95\textwidth},center}
{
\def\sym#1{\ifmmode^{#1}\else\(^{#1}\)\fi}
\begin{tabular}{l*{4}{c}}
\toprule
        &\multicolumn{1}{c}{Market}&\multicolumn{3}{c}{Number of Fields}\\ \cmidrule(lr){3-5}
      Dependent variable  &   Share      & All  & EOR & Non-EOR   \\
        & \multicolumn{1}{c}{}   & Fields  & Eligible & Eligible \\
        &\multicolumn{1}{c}{(1)}         &\multicolumn{1}{c}{(2)}         &\multicolumn{1}{c}{(3)}         &\multicolumn{1}{c}{(4)}         \\
\midrule
K-H \& S-O $\times$ Share EOR Norway $\times$ Post&    0.132\sym{***}&   28.155\sym{***}&    9.287\sym{***}&   18.868\sym{**} \\
                &  (0.025)         &  (9.825)         &  (2.799)         &  (7.804)         \\
[1em]
K-H Only $\times$ Share EOR Norway $\times$ Post&    0.050\sym{**} &   47.503\sym{***}&   10.545\sym{**} &   36.958\sym{***}\\
                &  (0.023)         & (10.474)         &  (4.858)         &  (6.249)         \\
[1em]
Share EOR Norway $\times$ Post&    0.010\sym{**} &   -1.787         &    2.375         &   -4.162\sym{**} \\
                &  (0.005)         &  (3.452)         &  (2.135)         &  (1.680)         \\
\midrule
Average dependent variable & 0.012 &  5.216 & 2.116 & 3.100 \\
\midrule
Share EOR $\times$ Post &      Yes         &      Yes         &      Yes         &      Yes         \\
Year FE         &      Yes         &      Yes         &      Yes         &      Yes         \\
Firm FE         &      Yes         &      Yes         &      Yes         &      Yes         \\
Observations    &     1377         &     1377         &     1377         &     1377         \\
R-squared        &     0.89         &     0.85         &     0.82         &     0.84         \\
\bottomrule
\multicolumn{5}{l}{* -- $p < 0.1$; ** -- $p < 0.05$; *** -- $p < 0.01$}
\end{tabular}
}
\end{adjustbox}
		\begin{tablenotes}
			\footnotesize \vspace{1em}
			\parbox{\textwidth}{\justifying  \item \hspace{-0.2em}\textbf{Notes:}  Estimated coefficients from regression \eqref{eq:know-how-firm}. The dependent variables are firm $i$'s market share (Column 1), number of fields (Column 2), number of EOR-eligible fields (Column 3), and number of non-EOR-eligible fields (Column 4) in year $t$. Firms are divided into three categories based on the indicator variables ``K-H'' (know-how) and ``S-O'' (state ownership). ``K-H'' equals 1 if firm $i$ was an operator in a field that adopted EOR before 1985, while ``S-O'' equals 1 if firm $i$ was state owned during 1975-1995. Because some firms possessed both know-how and were state owned, we create the variables ``K-H \& S-O,'' an indicator for a state-owned firm with know-how and ``K-H Only,'' an indicator for a non-state-owned firm with know-how. ``Share EOR Norway'' is the pre-1985 share of firm $i$'s fields that were eligible for EOR. The indicator ``Post'' equals 1 for the years after 1985, the year of the Supreme Court decision, and 0 otherwise. All regressions control for ``Share EOR $\cdot$ Post,'' which is firm $i$'s share of EOR-eligible fields in the North Sea before 1985 times the post-1985 indicator, and firm and year fixed effects. The two state-owned firms that did not have know-how are excluded from all regressions (Britoil and Norway State DFI). Standard errors are clustered at the firm level.}
		\end{tablenotes}
	\end{table} 
	
	This is also supported by the estimates in Columns 2-4 of Table \ref{tab:know-firm}, which indicate that know-how was the main force behind the portfolio expansions at the extensive margin documented in Table \ref{tab:concentration-firm}. The estimated effect for ``K-H Only'' firms is substantially larger than that for ``K-H \& S-O'' firms. ``K-O Only'' firms with a pre-1985 portfolio of fields that included 10 p.p. more Norwegian EOR-eligible fields expanded their portfolio by about $4.7$ new fields (Column 2), with 78\% of these being non-EOR-eligible (Column 4). Differently, the portfolios of ``K-H \& S-O'' firms expanded by about $2.8$ fields, with 67\% of these being non-EOR-eligible. This suggests a broader shift in investment strategy, as firms with EOR-specific know-how exhibited a renewed interest in fields with different characteristics suitable for alternative technologies beyond EOR. The effect of simply owning shares of EOR-eligible fields is instead estimated to be negligible.\footnote{Leave-one-out exercises ensure that our estimates are not driven by one of the three ``K-H Only'' firms. The K-H Only effect remains positive and statistically significant when leaving out Conoco/Mobil/Shell one at a time; the S-O estimate is stable (see Appendix Table \ref{tab:loo_know-firm}).}
	
	These results highlight how different types of firms employed different portfolio expansion strategies after the Norwegian Supreme Court decision. First, firms with know-how (both ``K-H \& SO'' and ``K-H Only'') expanded their portfolios toward EOR-eligible fields in a proportion similar to their availability (24\% of the fields in the North Sea are EOR eligible, see Table \ref{tab:field-statdes1}). Second, while the estimated market share increase of ``K-H Only'' firms in Column 1 is smaller than that estimated for ``K-H \& S-O'' firms, their more aggressive entry into new fields (Columns 2-4) suggests that the expected returns---also in terms of market shares---may have materialized over a longer horizon, as these firms may have invested in different fields requiring other, potentially less mature, technologies.
	
	\paragraph{Additional results on portfolio diversification.} In Appendix \ref{a:portfolio}, we further investigate the portfolio diversification strategies of different types of firms in relation to fields' geological characteristics and their implied levels of risk. We estimate regression \eqref{eq:know-how-firm} using as dependent variables the number of EOR-eligible fields and of non-EOR-eligible fields by geological type, distinguishing between features commonly associated with high risk (heavy oil, sour oil, and deep wells) and with low risk.
	
	The results in Appendix Table \ref{tab:know-risk-oil} (discussed in more detail in Appendix \ref{a:portfolio}) indicate that pre-1985 EOR-specific know-how in the North Sea, rather than state support, drove firms' portfolio diversification into high-risk assets after the Supreme Court decision. Firms with this type of expertise not only specialized more in high-risk EOR-eligible fields, but also diversified more into high-risk non-EOR-eligible fields. Importantly, diversification into high-risk fields potentially called for the adoption of risk-mitigating technologies other than EOR. We explore this possibility in the next section.\footnote{Panels (a) and (b) of Appendix Tables \ref{tab:multiple_panels} and \ref{tab:multiple_panels_SO} show that the results discussed in this section and presented in Tables \ref{tab:know-firm} and \ref{tab:know-risk-oil} also hold when controlling for ``$\text{Share Norway}_i \cdot \text{Post}_t$'' and when including Britoil and Norway State DFI, the two state-owned firms without know-how, respectively.}
	
	\paragraph{Oil majors.} Some firms may have acquired EOR-specific expertise outside the North Sea and then leveraged it in Norway after the 1985 Supreme Court decision. In fact, \cite{sheng2013enhanced} documents pervasive EOR adoption by the ``Seven Sisters'' (BP, Chevron, Exxon, Gulf Oil, Mobil, Shell, Texaco) and other global players (Amoco, Conoco, Eni, Phillips, Total). We estimate a regression similar to \eqref{eq:know-how-firm} that replaces the ``K-H$_i$'' and ``S-O$_i$'' indicators with an ``Oil Major$_i$'' indicator. The estimation results in Appendix Table \ref{tab:know-firm-om} suggest that the findings in Table \ref{tab:know-firm} are not driven by EOR-specific expertise potentially acquired anywhere in the world, but rather stress the central role of a firm's expertise within the North Sea. 
	
	Along the same lines, Appendix Table \ref{tab:know-risk-oil-om} replicates the portfolio diversification analysis of Appendix Table \ref{tab:know-risk-oil} replacing the ``K-H$_i$'' and ``S-O$_i$'' indicators with an ``Oil Major$_i$'' indicator. The estimates in Appendix Table \ref{tab:know-risk-oil-om} show that the Supreme Court decision did not alter the composition of oil majors' portfolios in the North Sea. We find no evidence of increased entry into specific fields or diversification across multiple fields, confirming that the results in Appendix Table \ref{tab:know-risk-oil} are not driven by oil majors.

	\section{Technological Diversification}\label{s:tech}
	The 1980s marked a period of significant technological innovation in the O\&G industry in the North Sea. Several key advancements improved extraction efficiency and addressed the harsh offshore conditions of the region. A major development was the use of \textit{concrete gravity base structures} (condeep)---the world's largest man-made movable object of the time---providing stability for offshore platforms, as for example at Gullfaks (Norway). The introduction of \textit{floating production, storage, and offloading} (FPSO) units enabled production in areas unsuitable for fixed platforms, with Petrojarl 1 deployed at Balder (Norway) as an early example. \textit{Subsea production systems} allowed extraction from remote reservoirs, enhancing operations at Gullfaks South (Norway) and Arbroath (UK), for example. \textit{Tension Leg Platforms} (TLPs) also emerged as a key innovation, with Hutton (UK), operated by Conoco, hosting the world's first production TLP, installed in 1984. Finally, \textit{horizontal drilling}, which was pioneered in Brynhild (Norway) and later also deployed in Beatrice (UK), improved reservoir drainage.\footnote{Norway also saw the deployment of TLPs in deeper fields than the Hutton field (150m), such as in Snorre (310m) and Heidrun (350m), with installations in 1992 and 1995, respectively, both operated by Statoil. In the 1980s, several non-Statoil operators in Norway deployed innovative technologies in the North Sea. For example, Aukra (Norway) saw the installation of the Sea Eagle FPSO in 1990 operated by Shell---a ``K-H Only'' firm. In Oseberg (Norway), the Oseberg A condeep was deployed in 1988 by Mobil, which also ``had know-how.'' Additionally, Frigg (Norway) deployed sub-sea production systems and a floating production unit in 1986, operated by BP and Total. These deployments highlight the variety of cutting-edge solutions employed by non-Statoil operators during the period.}
	
	These examples illustrate the technological transformation of the North Sea O\&G industry in the 1980s, which became a laboratory of offshore engineering breakthroughs. In what follows, we investigate whether the reduction in Norwegian sovereign hold-up risk encouraged owners of Norwegian EOR-eligible fields to diversify the range of technologies deployed, and in particular whether pre-1985 adoption of EOR (which we defined as ``know-how'') facilitated the subsequent adoption of other technologies.
	
	\subsection{Definition of Technological Diversification}\label{s:tech_approach}
	For the analysis of technological diversification, we use regressions \eqref{eq:did-firm} and \eqref{eq:know-how-firm}. For each firm $i$, we define the following measure of technological \textit{diversification}:
	\begin{equation}\label{eq:diversification}
		1 - \sum_{\tau=1}^{N} \left( \frac{\text{firm } i\text{'s production through technology } \tau \text{ in year } t}{\text{firm } i\text{'s total production in year } t} \right)^2,    
	\end{equation}
	which is one minus the HHI in terms of firm $i$'s ``technological shares.'' We consider a range of $N$ mutually exclusive technologies, each of which could have been adopted in a given field in a specific year. Then, for each technology $\tau=1,...,N$, we compute the share of firm $i$'s production in year $t$ obtained through $\tau$, relative to $i$'s total production in that year (obtained through any technology). To compute firm $i$'s production from technology $\tau$, we weigh the production of each field in $i$'s portfolio by firm $i$'s ownership share in that field in year $t$, and then standardize the resulting measure (to ease comparisons across technologies).
	Finally, we use \eqref{eq:diversification} as a dependent variable in \eqref{eq:know-how-firm} and interpret $\theta_{\text{treat}}$ as the average dispersion in a firm's portfolio of technologies resulting from its response to the Supreme Court decision.
	
	\paragraph{Technologies.} We classify technologies in two groups (see details in Appendix \ref{apndx:technologies}). The first group is \textit{exploration \& production} technologies, which includes the infrastructures required for field exploration and subsequent O\&G production. For instance, a \textit{fixed platform} is a stationary platform that supports drilling and production activities in shallow waters. A \textit{production semi-submersible platform} is an offshore platform anchored to the seabed for deep-water drilling. A \textit{subsea system} involves wells connected to a platform for remote deep O\&G extraction. The second group is \textit{extraction} technologies, which includes technologies designed to increase recovery rates or improve production by maintaining or increasing reservoir pressure. These include primary and secondary recovery methods, which are used at the early stages of a field's life cycle. In comparison,  EOR is a tertiary recovery technology, used only once primary and secondary recovery methods are no longer effective. 
	
	Returning to the comparative statics of the model in Section \ref{s:conceptual_framework}, we expect that the know-how acquired through EOR adoption transferred more easily to other extraction technologies, which shared common features with EOR (i.e., had higher knowledge carryover $\gamma$), than to exploration \& production technologies (i.e., had lower $\gamma$). As a result, firms with EOR-specific know-how face a relatively lower threshold $A^*$ for the subsequent adoption of technologies in the extraction class than in the exploration \& production class.
	
	\paragraph{Field characteristics and technologies.} Because technological suitability depends on field‐specific geology, we also measure technological diversification along two dimensions---well depth and reservoir size---using \eqref{eq:diversification}. We first discretize each continuous variable into quintiles across the North Sea, then compute the diversification indices over each category, with $\tau$ in \eqref{eq:diversification} denoting a field's quintile of well depth or reservoir size. In this context, larger values of index \eqref{eq:diversification} indicate that a firm obtained its total production through a relatively more diversified portfolio of fields that spanned the range of possible well depths and reservoir sizes in the North Sea.
	
	\paragraph{Technology-related costs.} To further assess technological diversification, we compute index \eqref{eq:diversification} over fields' cumulative tariffs and CAPEX expenditures per MMboe. Tariffs are the charges for using industry infrastructure---production (processing raw resources in a plant), transportation (moving resources via pipelines, ships, or other networks), and storage (tank farms, underground caverns).\footnote{We observe only the total tariffs for each year and field, without further breakdowns.} For each field, we sum the tariffs and CAPEX per MMboe over all years, discretize each cumulative measure into quintiles across the North Sea, and then calculate the diversification index, with $\tau$ indicating the field's tariffs or CAPEX quintile. Lower values of the resulting diversification index indicate that a firm's production was concentrated in fields with similar tariffs or CAPEX profiles, reflecting operational homogeneity. Importantly, a low index does not specify whether these fields were low or high cost, only that they required comparable levels of services or CAPEX.

	\begin{table}[!ht]
		\centering
		\caption{Know-how and technological diversification \label{tab:know-risk}}
		\centering
		\begin{adjustbox}{width=\textwidth,center}
{
\def\sym#1{\ifmmode^{#1}\else\(^{#1}\)\fi}
\begin{tabular}{l*{7}{c}}
\toprule
                &\multicolumn{6}{c}{Diversification in Technological Portfolios ($1 - $HHI)} \\ 
                \cmidrule(lr){2-7}
Dependent variable               &\multicolumn{1}{c}{Exploration}&\multicolumn{1}{c}{Extraction}&\multicolumn{1}{c}{Well }&\multicolumn{1}{c}{Reservoir} &\multicolumn{1}{c}{Tariff} &\multicolumn{1}{c}{CAPEX}\\
                &\multicolumn{1}{c}{\& Prod. Tech.}&\multicolumn{1}{c}{Technologies}&\multicolumn{1}{c}{ Depth}&\multicolumn{1}{c}{Size} &\multicolumn{1}{c}{Intensity} &\multicolumn{1}{c}{Intensity}\\                
                &\multicolumn{1}{c}{(1)}         &\multicolumn{1}{c}{(2)}         &\multicolumn{1}{c}{(3)}         &\multicolumn{1}{c}{(4)}         &\multicolumn{1}{c}{(5)}         &\multicolumn{1}{c}{(6)} \\ 

\multicolumn{7}{l}{\textit{Panel (a): Baseline}}  \\   
\midrule 
Share EOR Norway $\times$ Post&   -1.131\sym{*}  &   -0.944         &   -0.865         &   -0.894         &   -1.240\sym{**} &   -0.937         \\
                &  (0.641)         &  (0.609)         &  (0.683)         &  (0.655)         &  (0.534)         &  (0.643)         \\
[1em]
Share EOR $\times$ Post&   -0.723\sym{**} &   -0.407         &   -0.666\sym{**} &   -0.570\sym{*}  &   -0.490\sym{*}  &   -0.569\sym{**} \\
                &  (0.297)         &  (0.275)         &  (0.281)         &  (0.298)         &  (0.280)         &  (0.283)         \\
[1em] 
Observations    &     1377         &     1377         &     1377         &     1377         &     1377         &     1377         \\
R-squared       &     0.36         &     0.31         &     0.34         &     0.32         &     0.33         &     0.33         \\
\midrule\midrule
\rule{0pt}{4ex}  \\[-3ex] 

\multicolumn{7}{l}{\textit{Panel (b): Know-How (K-H) \& State-Owned (S-O)}}\\   
\midrule 
K-H \& S-O $\times$ Share EOR Norway $\times$ Post&    2.301\sym{***}&    1.419         &    1.928\sym{**} &    2.542\sym{***}&    1.504\sym{**} &    2.050\sym{**} \\
                &  (0.679)         &  (0.861)         &  (0.864)         &  (0.656)         &  (0.701)         &  (0.889)         \\
[1em]
K-H Only $\times$ Share EOR Norway $\times$ Post&    3.546\sym{***}&    5.055\sym{***}&    4.380\sym{***}&    5.855\sym{***}&    4.321\sym{***}&    3.564\sym{***}\\
                &  (0.854)         &  (0.920)         &  (1.134)         &  (0.820)         &  (0.892)         &  (1.176)         \\
[1em]
Share EOR Norway $\times$ Post&   -1.481\sym{**} &   -1.257\sym{**} &   -1.206\sym{*}  &   -1.347\sym{**} &   -1.536\sym{***}&   -1.263\sym{**} \\
                &  (0.595)         &  (0.576)         &  (0.665)         &  (0.561)         &  (0.498)         &  (0.621)         \\
[1em] 
Observations    &     1377         &     1377         &     1377         &     1377         &     1377         &     1377         \\
R-squared       &     0.37         &     0.32         &     0.35         &     0.33         &     0.34         &     0.34         \\
\midrule\midrule
Share EOR $\times$ Post&      Yes         &      Yes         &      Yes         &      Yes         &      Yes         &      Yes         \\
Year FE         &      Yes         &      Yes         &      Yes         &      Yes         &      Yes         &      Yes         \\
Firm FE         &      Yes         &      Yes         &      Yes         &      Yes         &      Yes         &      Yes         \\
\bottomrule
\multicolumn{7}{l}{* -- $p < 0.1$; ** -- $p < 0.05$; *** -- $p < 0.01$}
\end{tabular}
}
\end{adjustbox}

		\begin{tablenotes}
			\footnotesize \vspace{1em}
			\parbox{\textwidth}{\justifying  \item \hspace{-0.2em}\textbf{Notes:}  Panel (a) presents estimated coefficients from regression \eqref{eq:did-firm} and Panel (b) from regression \eqref{eq:know-how-firm}. The dependent variables evaluate technological diversification according to \eqref{eq:diversification} across technology types (Columns 1-2), field characteristics (Columns 3-4), and technology-related expenditures (Column 5-6). See main text for more details. Each dependent variable is standardized. In Panel (b), firms are divided into three categories based on the indicator variables ``K-H'' (know-how) and ``S-O'' (state ownership). ``K-H'' equals 1 if firm $i$ was an operator in a field that adopted EOR before 1985, while ``S-O'' equals 1 if firm $i$ was state owned during 1975-1995. Because some firms possessed both know-how and were state owned, we create the variables ``K-H \& S-O,'' an indicator for a state-owned firm with know-how and ``K-H Only,'' an indicator for a non-state-owned firm with know-how. ``Share EOR Norway'' is the pre-1985 share of firm $i$'s fields that were eligible for EOR. The indicator ``Post'' equals 1 for the years after 1985, the year of the Supreme Court decision, and 0 otherwise. All regressions control for ``Share EOR $\cdot$ Post,'' which is firm $i$'s share of EOR-eligible fields in the North Sea before 1985 times the post-1985 indicator, and firm and year fixed effects. The two state-owned firms that did not have know-how are excluded (Britoil and Norway State DFI). Standard errors are clustered at the firm level.}
		\end{tablenotes}
	\end{table}
	
	\subsection{Estimation Results}
	Panel (a) of Table \ref{tab:know-risk} reports estimates of regression \eqref{eq:did-firm} using the diversification indices described above as dependent variables. The negative and mostly not significant estimates indicate that larger pre-1985 shares of Norwegian EOR-eligible fields were not sufficient, on their own, to spur technological diversification after the Supreme Court decision. 
	
	Panel (b) turns to estimating regression \eqref{eq:know-how-firm}. The second row, which corresponds to firms with EOR-specific know-how that were not state owned, consistently reports the largest increases in technological diversification. The estimated effects are roughly 1.5 to 3.5 times larger for ``K-H Only'' firms compared to those in the first row, which refer to firms with EOR-specific know-how that were state owned. This difference is significant at the 2\% level across all columns except the last, where standard errors are larger.\footnote{The null hypothesis that $\theta^{\text{S-O}}_{\text{treat}}=\theta^{\text{K-H}}_{\text{treat}}$ in \eqref{eq:know-how-firm} is rejected with $p$-value $<$ 0.015 in Columns (1)-(5). For Column (6), we instead find $p$-value = 0.179.} 
	
	Overall, these results---together with those from Table \ref{tab:know-risk-oil}---underscore the central role of know-how in driving technological diversification. 
	
	\noindent As mentioned above, since EOR is a tertiary extraction technology, its characteristics overlap more with those of other extraction technologies in Column 2 (which include primary and secondary methods) than with those of exploration \& production technologies in Column 1. The comparative statics in Section \ref{s:conceptual_framework} predict that pre-1985 EOR-specific know-how should have facilitated the adoption of similar extraction technologies rather than more distant exploration \& production technologies. Consistent with this prediction, the estimated diversification in extraction technologies (Column 2) for ``K-H Only'' firms is substantially larger than that in exploration \& production technologies (Column 1). To test whether this difference is statistically significant, we subtract the dependent variable in Column 1 from that in Column 2 and re-estimate the specification in Panel (b). The coefficient on the interaction with ``K-H Only'' is 1.51 and statistically significant at the 1\% level, implying that a 10 p.p. increase in ``Share EOR Norway'' led to a 0.15 standard deviation larger increase in diversification in extraction relative to exploration \& production technologies. Differently from ``K-H Only'' firms, state-owned firms with direct expertise in EOR did not diversify more in extraction technologies. Their behavior was inconsistent with our theoretical framework, possibly reflecting alternative, politically motivated objectives---such as a prioritization of the exploration of all potentially productive fields for better planning and future allocation.\footnote{Panel (c) of Appendix Table \ref{tab:multiple_panels} shows that controlling for Share Norway$_i$ $\times$ Post$_t$ does not affect our conclusions. Including also state-owned firms without know-how as a separate category, we find that state-owned firms with know-how diversify their technological portfolios more than those without know-how, as documented in Panel (c) of Appendix Table \ref{tab:multiple_panels_SO}.}

	\paragraph{Oil majors.} The results in Appendix Table \ref{tab:know-risk-om} exclude the possibility that oil majors drove technological diversification after the Norwegian Supreme Court decision. In particular, we do not find any significant change in our technological diversification measures for those oil majors that owned more Norwegian EOR-eligible fields before 1985.

	\paragraph{Discussion.} Collectively, our results suggest that firms that owned EOR-eligible fields (rather than any other type of field, Table \ref{tab:concentration-firm-other-exposure}) prior to 1985 expanded their production after the Supreme Court decision both directly through EOR adoption (Table \ref{tab:did-firm}), but also indirectly by expanding their portfolios of assets, especially with other EOR-eligible fields (Tables \ref{tab:concentration-firm} and \ref{tab:concentration-field}). Firms that already possessed EOR-specific know-how in the North Sea before 1985 were those that experienced the largest increases in market shares, and were also those that expanded their portfolios more aggressively after the Supreme Court decision (Table \ref{tab:know-firm}). In particular, these firms not only specialized in other EOR-eligible fields, but also diversified in fields with alternative characteristics (Appendix Table \ref{tab:know-risk-oil}) that required different technologies (Table \ref{tab:know-risk}). 

	\section{Conclusion}\label{s:conclusion}
	We investigate how sovereign hold-up risk affects technology adoption and industry evolution in the upstream North Sea O\&G sector. Exploiting a 1985 Norwegian Supreme Court ruling that constrained unilateral retroactive changes to petroleum licensing terms, we show that stronger sovereign commitment accelerated adoption of EOR, a prominent extraction technology, increased investment and productivity, and enabled firms to gain larger shares of the North Sea O\&G market. Importantly, these gains were largest for firms with prior operational expertise in EOR, not simply those with financial stakes in EOR-eligible assets or those benefiting from state support.
	
	Our findings point to a knowledge-based market failure induced by sovereign hold-up risk. Following the ruling, firms with EOR-specific know-how reshaped their portfolios of assets, both specializing in additional EOR-eligible fields and diversifying into fields with different geological characteristics requiring alternative technologies. Consistent with complementarities in technological knowledge, these firms subsequently adopted a broader range of technologies, particularly other extraction technologies that shared technical features with EOR. This pattern of technological diversification, concentrated among technologies similar to EOR, suggests that know-how accumulated through early adoption is partly transferable, generating economies of scope that facilitate subsequent adoption across related technological domains.
	
	\begin{figure}[!ht]
		\captionsetup[subfigure]{justification=centering}
		\centering
		\caption{Countries with constitutional prohibitions on non-criminal retroactive laws}\label{fig:retroactivity-map} \vspace{-0.4cm}
		\includegraphics[width=\linewidth]{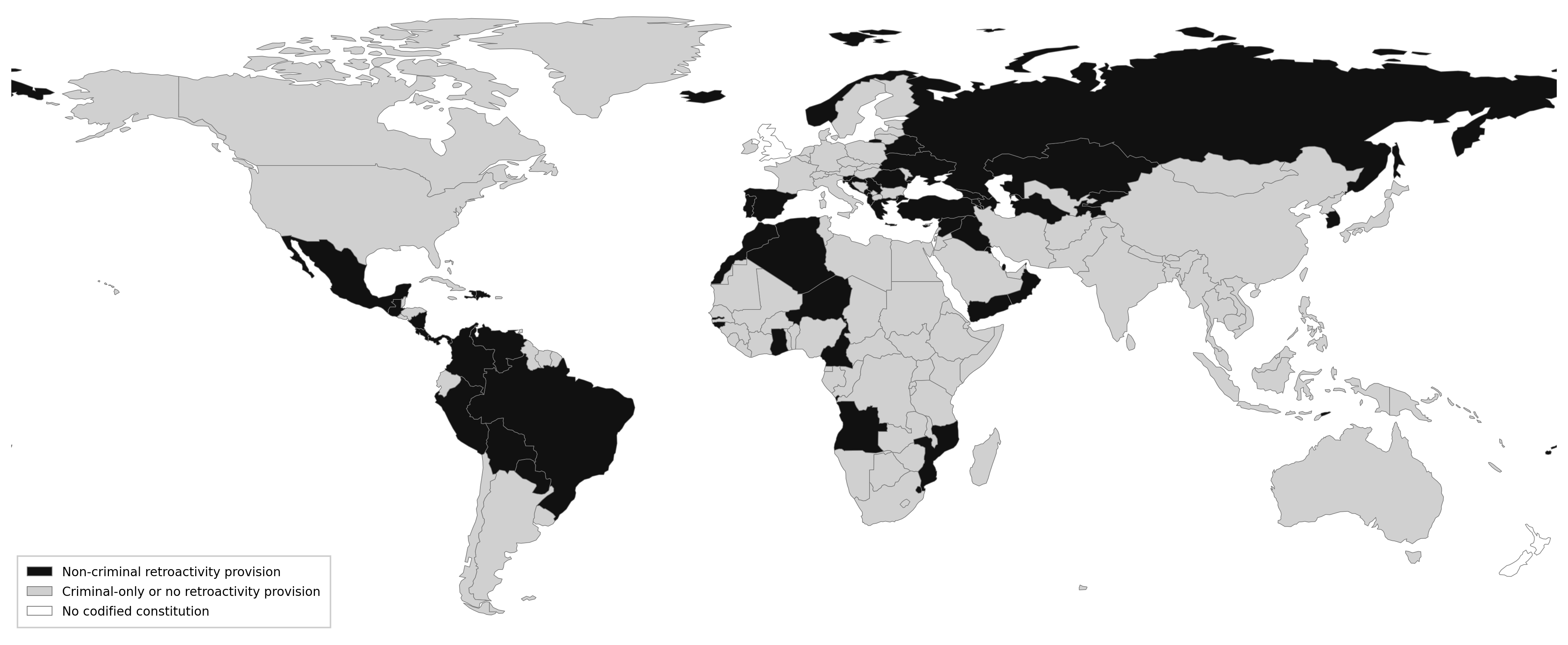}
		\begin{minipage}{1 \textwidth}\vspace{-0.4cm}
			{\footnotesize\singlespacing \textbf{Notes:} Data are constructed from the \textit{Comparative Constitutions Project} (CCP) API using English translations of in-force constitutions. Of the 193 countries included in the CCP database, we consider Israel, New Zealand, and the United Kingdom as not having a codified constitution (shown in white) following \cite{ginsburg2015does}. We identify constitutional prohibitions on retroactivity by retrieving candidate provisions using retroactivity-related cues (e.g., \emph{retroactive}, \emph{retrospective}, \emph{ex post facto}) and then classifying the extracted excerpts with OpenAI \texttt{gpt-5.2} operating under a constrained prompt: the model is required to rely only on the provided constitutional text and to quote verbatim evidence supporting its classification. Countries are assigned to one of two categories: (1) \textit{non-criminal retroactivity provision} (black) if the constitution contains an explicit prohibition on retroactive laws extending beyond criminal punishment (e.g., fiscal, administrative, regulatory), or (2) \textit{criminal-only or no retroactivity provision} (gray) if the constitution only includes retroactivity prohibitions specific to criminal law or does not include explicit retroactivity prohibitions. See Appendix \ref{apndx:retroactivity} for details.
				\par}
		\end{minipage}
	\end{figure}
	The commitment problem we study may extend well beyond the North Sea O\&G sector. Using data from the \textit{Comparative Constitutions Project} (CCP) \citep{ElkinsGinsburg2025CCP}, Figure \ref{fig:retroactivity-map} classifies countries based on whether their constitution prohibits retroactive laws outside the criminal sphere using a large language model (see details in Appendix \ref{apndx:retroactivity}). Of the 190 codified constitutions in the CCP database, 131 (68.9\%) are either silent on non-criminal retroactivity or restrict such protections to criminal punishment. 
	This pattern persists among advanced economies: of the 35 codified constitutions within the OECD, only 11 (31.4\%) explicitly prohibit retroactive non-criminal laws. Yet many long-horizon investments---in infrastructure, the energy transition, or public procurement---depend critically on stable administrative and fiscal regimes, precisely where ex-post interventions are most prevalent. Moreover, even where constitutions explicitly address non-criminal retroactivity, sovereign hold-up risk remains since enforcement ultimately depends on judicial interpretation \citep{kryvoi2021non}. Tax is a canonical example: courts routinely weigh retroactive measures against public-interest and proportionality considerations, sometimes deciding not to enforce constitutional provisions against retroactive taxation.\footnote{Recent international examples of retroactive tax measures include 1980s Australian anti-avoidance laws, French \textit{Conseil Constitutionnel} rulings upholding retroactive taxation in the general interest, and the retroactive 3\% French Digital Services Tax. By contrast, the U.S. Constitution prohibits retroactive criminal laws via the Ex Post Facto Clauses but lacks a blanket ban on civil law retroactivity. See \textit{Calder v. Bull}, 3 U.S. 386 (1798). Consequently, civil retroactive laws are permissible under the Due Process Clause if they serve a legitimate purpose through rational means. The Supreme Court has upheld such statutes regarding black lung compensation (\textit{Usery v. Turner Elkhorn Mining Co.}, 428 U.S. 1 (1976)), pension liabilities (\textit{PBGC v. R.A. Gray \& Co.}, 467 U.S. 717 (1984)), and tax amendments (\textit{United States v. Carlton}, 512 U.S. 26 (1994)). For a comprehensive review of this distinction, see \cite{Hochman1960}.}
	
	The commitment challenges we study may also exist in other settings where governments rely on long-term private expertise. In the case of the North Sea O\&G sector, credibility stemmed from judicial enforcement and constitutional hardening. In other contexts, sovereign commitment can be secured through delegation to independent agencies \citep{rogoff1985optimal,alesina1993central}, international legal frameworks \citep{tomz2012reputation}, or constitutional fiscal rules \citep{badinger2017case}. More broadly, institutional design can help overcome limited state capacity by enabling credible commitments even in environments with weak enforcement, especially when embedded in long-standing legal or cultural norms \citep{north1989constitutions,tabellini2008institutions,mastrorocco2023state}. Absent such credibility, hold-up risk distorts investment not only in physical capital, but also in cumulative knowledge.

	
	\singlespacing
	\bibliographystyle{ecca-mod.bst}
	\bibliography{bibliography.bib}

	\newpage
	\clearpage

	\appendix
	\onehalfspacing
	
	\pagenumbering{arabic}
	\setcounter{page}{1} \renewcommand{\thepage}{OA-\arabic{page}}
	
	\setcounter{footnote}{0}
	\section*{\huge{Online Appendix}}
	
	\setcounter{table}{0} \renewcommand{\thetable}{A\arabic{table}} 
	\setcounter{figure}{0} \renewcommand{\thefigure}{A\arabic{figure}}
	\setcounter{section}{0} \renewcommand{\thesection}{A}
	\setcounter{equation}{0} 
	\renewcommand{\theequation}{A\arabic{equation}}
	
	\section{Theory}\label{apndx:theory}
	
	\subsection{Threshold Crossing Rule for Technology Adoption}\label{apndx:unique_fx_point}
	We supplement the assumptions made in Section \ref{s:conceptual_framework} with the following regularity conditions, which are sufficient for the result: 
	\begin{itemize}
		\item $q$ is $m$-strongly monotone (that is, $q'(A)\ge m>0$),
		\item the boundary conditions $B(0)>0$ and $B(\bar A)<\bar A$, and
		\item the parameter restriction $(s\,\Delta\Pi)^2\,m > c\,\beta\,\delta\,L$. 
	\end{itemize}
	Define $D(A) \equiv s\,\Delta\Pi + \beta\,[V(\delta A+\gamma)-V(\delta A)]$. Since $V$ is strictly increasing, $\gamma > 0$, and $\beta > 0$, we have $V(\delta A + \gamma) > V(\delta A)$, thus $D(A) > s\,\Delta\Pi > 0$. Taking derivatives, $D'(A) = \beta\,\delta\,[V'(\delta A+\gamma)-V'(\delta A)]$. Since $V$ is concave, $V'$ is non-increasing, so $V'(\delta A+\gamma) \leq V'(\delta A)$, implying $D'(A) \leq 0$. Since $V\in C^1$ and concave on the compact interval $[0,\bar A]$, it is Lipschitz continuous with some constant $L<\infty$, meaning $0 < V'(\cdot) \leq L$. Therefore, we have $V'(\delta A) - V'(\delta A+\gamma) \leq L$ (the maximal difference between $V'(\delta A)=L$ and $V'(\delta A+\gamma)$ just above $0$), thus $-D'(A) \leq \beta\,\delta\,L$. 
	
	For $B(A)=q^{-1}(c/D(A))$, differentiation yields $B'(A) = \frac{-c \cdot D'(A)}{D(A)^2 \cdot q'(B(A))} = \frac{c \cdot (-D'(A))}{D(A)^2 \cdot q'(B(A))}$. Since $\beta\,\delta\,L \geq -D'(A) \geq 0$, $D(A) > s\,\Delta\Pi > 0$, and $q'(B(A)) \geq m$, we have $0 \leq B'(A) < \frac{c\,\beta\,\delta\,L}{(s\,\Delta\Pi)^2\,m} < 1$, where the strict inequality in the middle follows from $D(A) > s\,\Delta\Pi$ and the last inequality follows from the parameter restriction. 
	
	Defining $h(A)\equiv B(A) - A$, we then obtain $h'(A) = B'(A) - 1 < 0$, which means that $h$ is strictly decreasing. Moreover, since $V$ and $q$ are continuous and $D(A) > s\,\Delta\Pi > 0$, the map $A \mapsto c/D(A)$ is continuous, and because $q$ is strictly increasing its inverse is continuous. Hence, $B(A) = q^{-1}(c/D(A))$ and $h(A) \equiv B(A) - A$ are continuous. Given $h(0) = B(0) > 0$ and $h(\bar A) = B(\bar A) - \bar A < 0$ (by the boundary conditions), continuity of $h$ guarantees a unique root $A^* \in (0,\bar A)$ of $h(A) = 0$. Finally, because $h(A)$ is continuous and strictly decreasing in $A$, the adoption rule $A\ge B(A)$ is equivalent to $A\ge A^*$

	\subsection{Additional Details on Firm-Level Regressions}\label{a:firm_regressions}
	We provide additional details on the derivations in Section \ref{sec:firm_regressions} for DiD regression \eqref{eq:did-firm}. We start by evaluating firm $i$'s total production in \eqref{eq:basic_outcome} separately for $t\in \{\text{pre, post}\}$. In the pre-period, both Norwegian and UK EOR-eligible fields face the same threshold $A^*_{\text{pre}}$:
	\begin{equation}\label{eq:pre}
		y_{i,\text{pre}} = N_{i,\text{Other}} \cdot y_{0,\text{Other},\text{pre}} + (N_{i,\text{Norway EOR}} + N_{i,\text{UK EOR}}) \cdot [y_{0,\text{EOR},\text{pre}} + \Delta y \cdot \mathbbm{1}\{A_{i,\text{pre}} \geq A^*_{\text{pre}}\}].
	\end{equation}
	In the post-period, the thresholds instead diverge:
	\begin{equation}\label{eq:post}
		\begin{aligned}
			y_{i,\text{post}} = & \; N_{i,\text{Other}} \cdot y_{0,\text{Other},\text{post}} + N_{i,\text{Norway EOR}} \cdot [y_{0,\text{EOR},\text{post}} + \Delta y \cdot \mathbbm{1}\{A_{i,\text{post}} \geq A^*_{\text{post,Norway}}\}] \\
			&+ N_{i,\text{UK EOR}} \cdot [y_{0,\text{EOR},\text{post}} + \Delta y \cdot \mathbbm{1}\{A_{i,\text{post}} \geq A^*_{\text{pre}}\}].
		\end{aligned}
	\end{equation}
	Taking expectations of \eqref{eq:pre} and \eqref{eq:post} over the relevant distributions yields:
	\begin{align}
		\mathbb{E}[y_{i,\text{pre}}] &= N_{i,\text{Other}} \cdot y_{0,\text{Other},\text{pre}} + (N_{i,\text{Norway EOR}} + N_{i,\text{UK EOR}}) \cdot y_{0,\text{EOR},\text{pre}} \notag\\
		&\quad + (N_{i,\text{Norway EOR}} + N_{i,\text{UK EOR}}) \cdot \Delta y \cdot [1 - F_{\text{pre}}(A^*_{\text{pre}})],\label{eq:E_pre} \\ \notag \\
		\mathbb{E}[y_{i,\text{post}}] &= N_{i,\text{Other}} \cdot y_{0,\text{Other},\text{post}} + (N_{i,\text{Norway EOR}} + N_{i,\text{UK EOR}}) \cdot y_{0,\text{EOR},\text{post}} \notag \\
		&\quad + N_{i,\text{Norway EOR}} \cdot \Delta y \cdot [1 - F_{\text{post}}(A^*_{\text{post,Norway}})] \notag \\
		&\quad + N_{i,\text{UK EOR}} \cdot \Delta y \cdot [1 - F_{\text{post}}(A^*_{\text{pre}})].\label{eq:E_post}
	\end{align}
	Consequently, the expected change in production from the pre- to the post-period is:
	\begin{align*}
		\mathbb{E}[y_{i,\text{post}}-y_{i,\text{pre}}] &= N_{i,\text{Other}} \cdot [y_{0,\text{Other},\text{post}} - y_{0,\text{Other},\text{pre}}] \\
		&\quad + (N_{i,\text{Norway EOR}} + N_{i,\text{UK EOR}}) \cdot [y_{0,\text{EOR},\text{post}} - y_{0,\text{EOR},\text{pre}}]\\
		&\quad + N_{i,\text{Norway EOR}} \cdot \Delta y \cdot \{[1 - F_{\text{post}}(A^*_{\text{post,Norway}})] - [1 - F_{\text{pre}}(A^*_{\text{pre}})]\} \\
		&\quad + N_{i,\text{UK EOR}} \cdot \Delta y \cdot \{[1 - F_{\text{post}}(A^*_{\text{pre}})] - [1 - F_{\text{pre}}(A^*_{\text{pre}})]\} \\
		&= N_{i,\text{Other}} \cdot [y_{0,\text{Other},\text{post}} - y_{0,\text{Other},\text{pre}}] \\
		&\quad + (N_{i,\text{Norway EOR}} + N_{i,\text{UK EOR}}) \cdot [y_{0,\text{EOR},\text{post}} - y_{0,\text{EOR},\text{pre}}] \\
		&\quad + N_{i,\text{Norway EOR}} \cdot \Delta y \cdot [F_{\text{pre}}(A^*_{\text{pre}}) - F_{\text{post}}(A^*_{\text{post,Norway}})] \\
		&\quad + N_{i,\text{UK EOR}} \cdot \Delta y \cdot [F_{\text{pre}}(A^*_{\text{pre}}) - F_{\text{post}}(A^*_{\text{pre}})],
	\end{align*}
	which is equation \eqref{eq:prod_change_1} in Section \ref{sec:firm_regressions}. The term $[F_{\text{pre}}(A^*_{\text{pre}}) - F_{\text{post}}(A^*_{\text{post,Norway}})]$ captures both the direct effect of the threshold change and any distributional shifts between periods for Norwegian fields, while $[F_{\text{pre}}(A^*_{\text{pre}}) - F_{\text{post}}(A^*_{\text{pre}})]$ captures only distributional shifts for UK fields. Substituting in equation \eqref{eq:prod_change_1} the expressions for $\Delta_{\text{Other}}$, $\Delta_{\text{EOR}}$, and the portfolio shares, we obtain equation \eqref{eq:prod_change_2} in Section \ref{sec:firm_regressions}:
	\begin{align*}
		\mathbb{E}[y_{i,\text{post}}-y_{i,\text{pre}}] &= N \cdot \Delta_{\text{Other}} + N \cdot [\Delta_{\text{EOR}} - \Delta_{\text{Other}}] \cdot \text{Share EOR}_i \\
		&\quad + N \cdot \Delta y \cdot [F_{\text{pre}}(A^*_{\text{pre}}) - F_{\text{post}}(A^*_{\text{pre}})] \cdot \text{Share EOR}_i  \\
		&\quad + N \cdot \Delta y \cdot [F_{\text{post}}(A^*_{\text{pre}}) - F_{\text{post}}(A^*_{\text{post,Norway}})] \cdot \text{Share EOR Norway}_i \\
		&= N \cdot \Delta_{\text{Other}} + N \cdot \left\{\Delta_{\text{EOR}} - \Delta_{\text{Other}} + \Delta y \cdot [F_{\text{pre}}(A^*_{\text{pre}}) - F_{\text{post}}(A^*_{\text{pre}})]\right\} \cdot \text{Share EOR}_i\\
		&\quad + N \cdot \Delta y \cdot [F_{\text{post}}(A^*_{\text{pre}}) - F_{\text{post}}(A^*_{\text{post,Norway}})] \cdot \text{Share EOR Norway}_i,
	\end{align*}
	where we use the fact that $\text{Share EOR UK}_i = \text{Share EOR}_i - \text{Share EOR Norway}_i$ to re-express the term that multiplies $\text{Share EOR UK}_i$ in equation \eqref{eq:prod_change_1}.
	
	We then combine these expressions to obtain DiD regression \eqref{eq:did-firm} in Section \ref{sec:firm_regressions}. To do this, we use the following identity for firm $i$'s production in period $t$:
	\begin{equation*}
		y_{it} = \mathbb{E}[y_{it}] + (y_{it} - \mathbb{E}[y_{it}]).
	\end{equation*}
	In particular, we use equations \eqref{eq:E_pre} and \eqref{eq:prod_change_2} to decompose $\mathbb{E}[y_{it}]$ into changes induced by the Supreme Court decision (the coefficient of interest $\gamma_{\text{treat}}$), changes unrelated to it (the coefficient $\gamma_1$), a time-invariant firm-specific component (the fixed effect $\alpha_i$), and a common time effect (the fixed effect $\psi_t$). Define $\text{Post}_t$ as an indicator variable equal to 1 when $t=\text{post}$ and 0 otherwise. Then, 
	\begin{equation*}
		\mathbb{E}[y_{it}] = \mathbb{E}[y_{i,\text{pre}}] + \mathbb{E}[y_{i,\text{post}} - y_{i,\text{pre}} ]\cdot \text{Post}_t
	\end{equation*} 
	can be equivalently written as:
	\begin{equation}\label{eq:expected_prod}
		\mathbb{E}[y_{it}] = \gamma_{\mathrm{treat}}\cdot\text{Share EOR Norway}_i\cdot \text{Post}_t +\gamma_{1} \cdot\text{Share EOR}_i\cdot\text{Post}_t + \alpha_i + \psi_t ,
	\end{equation}
	where the coefficients on the right-hand side are:
	\begin{align*}
		\gamma_{\text{treat}} & \equiv N \cdot \Delta y \cdot [F_{\text{post}}(A^*_{\text{pre}}) - F_{\text{post}}(A^*_{\text{post,Norway}})] \\
		\gamma_{1} & \equiv N \cdot \left\{\Delta_{\text{EOR}} - \Delta_{\text{Other}} + \Delta y \cdot [F_{\text{pre}}(A^*_{\text{pre}}) - F_{\text{post}}(A^*_{\text{pre}})]\right\} \\
		\alpha_i &\equiv N \cdot \bigl\{y_{0,\text{Other},\text{pre}}\cdot (1 - \text{Share EOR}_i) + y_{0,\text{EOR},\text{pre}}\cdot \text{Share EOR}_i \\ 
		&\quad + \Delta y \cdot [1 - F_{\text{pre}}(A^*_{\text{pre}})] \cdot \text{Share EOR}_i\bigr\} \\
		\psi_t &\equiv N \cdot \Delta_{\text{Other}} \cdot \text{Post}_t.
	\end{align*}
	Finally, substituting \eqref{eq:basic_outcome} and \eqref{eq:expected_prod} in the above identity, we obtain DiD regression \eqref{eq:did-firm} in Section \ref{sec:firm_regressions}:
	\begin{equation*}
		y_{it} = \gamma_{\mathrm{treat}}\cdot\text{Share EOR Norway}_i\cdot \text{Post}_t + \gamma_{1}\cdot\text{Share EOR}_i\cdot\text{Post}_t + \alpha_i + \psi_t + \epsilon_{it},
	\end{equation*}
	where the residual term $\epsilon_{it}\equiv y_{it} - \mathbb{E}[y_{it}]$ corresponds to:
	\begin{align*}
		\epsilon_{it} &\equiv (1 - \text{Post}_t) \cdot N \cdot \text{Share EOR}_i \cdot \Delta y \cdot \left\{\mathbbm{1}\{A_{i,\text{pre}} \geq A^*_{\text{pre}}\} - [1 - F_{\text{pre}}(A^*_{\text{pre}})]\right\} \\
		&\quad + \text{Post}_t \cdot N \cdot \text{Share EOR Norway}_i \cdot \Delta y \cdot \left\{\mathbbm{1}\{A_{i,\text{post}} \geq A^*_{\text{post,Norway}}\} - [1 - F_{\text{post}}(A^*_{\text{post,Norway}})]\right\} \\
		&\quad + \text{Post}_t \cdot N \cdot \text{Share EOR UK}_i \cdot \Delta y \cdot \left\{\mathbbm{1}\{A_{i,\text{post}} \geq A^*_{\text{pre}}\} - [1 - F_{\text{post}}(A^*_{\text{pre}})]\right\}.
	\end{align*}
	
	\subsection{Derivation of Field-Level Regressions}\label{a:field_regressions}
	
	In this Appendix, we adapt the firm-level derivations of Section \ref{sec:firm_regressions} to obtain field-level triple-difference regression \eqref{eq:did-field}, similar to regression \eqref{eq:did-firm} from the main text.
	
	
	Consider a field $f$ operated by firm $i(f)$ in country $c(f) \in \{\text{Norway}, \text{UK}\}$. Let $\text{EOR}_f$ be an indicator equal to 1 if field $f$ is EOR-eligible and 0 otherwise. Following the framework of Section \ref{sec:firm_regressions}, field $f$'s production in period $t \in \{\text{pre}, \text{post}\}$ is:
	\begin{align*}
		y_{ft} = 
		\begin{cases}
			y_{0,\text{Other},t} & \text{if } \text{EOR}_f = 0 \\
			y_{0,\text{EOR},t} + \Delta y \cdot \mathbbm{1}\{A_{i(f),t} \geq A^*_{t,c(f)}\} & \text{if } \text{EOR}_f = 1,
		\end{cases}
	\end{align*}
	where $A^*_{t,c(f)}$ denotes the adoption threshold in period $t$ for EOR-eligible fields in country $c(f)$. As discussed in Section \ref{sec:firm_regressions}, the 1985 Norwegian Supreme Court decision induced a change in thresholds. Before 1985, all EOR-eligible fields faced the same threshold:
	\begin{align*}
		A^*_{\text{pre,Norway}} = A^*_{\text{pre,UK}} = A^*_{\text{pre}}.
	\end{align*}
	After 1985, the threshold for Norwegian EOR-eligible fields decreased while that for UK fields remained unchanged:
	\begin{align*}
		A^*_{\text{post,Norway}} < A^*_{\text{pre}}, \quad A^*_{\text{post,UK}} = A^*_{\text{pre}}.
	\end{align*}
	The analogous of Assumption \ref{indep} at the field level is:
	\begin{assumption}\label{indep_field}
		know-how $A_{it}$ is distributed across operators independently of the status of EOR eligibility and country of the fields they operate, or $\Pr_t(A_{it} \geq A^* \mid \text{EOR}_f, c(f)) = \Pr_t(A_{it} \geq A^*) = 1 - F_t(A^*).$
	\end{assumption}
	The assumption that the adoption threshold for UK fields remains constant across periods is a normalization that allows us to focus on the Norwegian-specific shift in adoption incentives. In practice, common global shocks—such as the mid-1980s oil price decline—likely affected adoption incentives ($\Delta\Pi$) and thresholds in both countries. In our empirical implementation, these common shocks are captured by the time fixed effects and the baseline EOR trend, ensuring that our treatment effect estimate isolates only the incremental effect of the Norwegian legal ruling relative to the UK benchmark.
	
	Given Assumption \ref{indep_field}, taking expectations:
	\begin{align*}
		\mathbb{E}[y_{ft}] = (1 - \text{EOR}_f) \cdot y_{0,\text{Other},t} + \text{EOR}_f \cdot y_{0,\text{EOR},t} + \text{EOR}_f \cdot \Delta y \cdot [1 - F_t(A^*_{t,c(f)})].
	\end{align*}
	In the pre-period, this is:
	\begin{align}\label{eq:field-pre}
		\mathbb{E}[y_{f,\text{pre}}] = (1 - \text{EOR}_f) \cdot y_{0,\text{Other},\text{pre}} + \text{EOR}_f \cdot y_{0,\text{EOR},\text{pre}} + \text{EOR}_f \cdot \Delta y \cdot [1 - F_{\text{pre}}(A^*_{\text{pre}})].
	\end{align}
	Let $\text{Norway}_f$ be an indicator equal to 1 if $c(f) = \text{Norway}$ and 0 if $c(f) = \text{UK}$. The expected production in the post-period is:
	\begin{align}\label{eq:field-post}
		\mathbb{E}[y_{f,\text{post}}] &= (1 - \text{EOR}_f) \cdot y_{0,\text{Other},\text{post}} + \text{EOR}_f \cdot y_{0,\text{EOR},\text{post}}  \\
		&\quad + \text{EOR}_f \cdot \text{Norway}_f \cdot \Delta y \cdot [1 - F_{\text{post}}(A^*_{\text{post,Norway}})] \notag \\
		&\quad + \text{EOR}_f \cdot (1 - \text{Norway}_f) \cdot \Delta y \cdot [1 - F_{\text{post}}(A^*_{\text{pre}})]. \notag
	\end{align}
	The expected change in production from pre- to post-period is:
	\begin{align*}
		\mathbb{E}[y_{f,\text{post}} - y_{f,\text{pre}}] &= (1 - \text{EOR}_f) \cdot \Delta_{\text{Other}} + \text{EOR}_f \cdot \Delta_{\text{EOR}}  \\
		&\quad + \text{EOR}_f \cdot \text{Norway}_f \cdot \Delta y \cdot [F_{\text{pre}}(A^*_{\text{pre}}) - F_{\text{post}}(A^*_{\text{post,Norway}})] \notag \\
		&\quad + \text{EOR}_f \cdot (1 - \text{Norway}_f) \cdot \Delta y \cdot [F_{\text{pre}}(A^*_{\text{pre}}) - F_{\text{post}}(A^*_{\text{pre}})], \notag
	\end{align*}
	where $\Delta_{\text{Other}} = y_{0,\text{Other},\text{post}} - y_{0,\text{Other},\text{pre}}$ and $\Delta_{\text{EOR}} = y_{0,\text{EOR},\text{post}} - y_{0,\text{EOR},\text{pre}}$. Rearranging, we obtain:
	\begin{align}\label{eq:field-change}
		\mathbb{E}[y_{f,\text{post}} - y_{f,\text{pre}}] &= \Delta_{\text{Other}} + \text{EOR}_f \cdot [\Delta_{\text{EOR}} - \Delta_{\text{Other}}]  \\
		&\quad + \text{EOR}_f \cdot \Delta y \cdot [F_{\text{pre}}(A^*_{\text{pre}}) - F_{\text{post}}(A^*_{\text{pre}})] \notag \\
		&\quad + \text{EOR}_f \cdot \text{Norway}_f \cdot \Delta y \cdot [F_{\text{post}}(A^*_{\text{pre}}) - F_{\text{post}}(A^*_{\text{post,Norway}})]. \notag
	\end{align}
	We now combine expressions \eqref{eq:field-pre}, \eqref{eq:field-post}, and \eqref{eq:field-change} to obtain a field-level triple-difference regression. We start from the following identity for field $f$'s production in period $t$:
	\begin{align*}
		y_{ft} = \mathbb{E}[y_{ft}] + (y_{ft} - \mathbb{E}[y_{ft}]).
	\end{align*}
	We can use our previous expressions to decompose $\mathbb{E}[y_{ft}]$ into the changes induced by the Supreme Court decision (the coefficient of interest $\beta_{\text{treat}}$), changes unrelated to it (the coefficient $\beta_1$), a time-invariant field-specific component (the fixed effect $\tau_f$), and a country-time effect (the fixed effect $\iota_{c(f)t}$). Define $\text{Post}_t$ as an indicator variable equal to 1 when $t = \text{post}$ and 0 otherwise. Then,
	\begin{align*}
		\mathbb{E}[y_{ft}] = \mathbb{E}[y_{f,\text{pre}}] + \mathbb{E}[y_{f,\text{post}} - y_{f,\text{pre}}] \cdot \text{Post}_t
	\end{align*}
	can be decomposed as:
	\begin{align}\label{eq:field-expected}
		\mathbb{E}[y_{ft}] = \beta_{\text{treat}} \cdot \text{Norway}_f \cdot \text{EOR}_f \cdot \text{Post}_t + \beta_1 \cdot \text{EOR}_f \cdot \text{Post}_t + \tau_f + \iota_{c(f)t},
	\end{align}
	where the coefficients on the right-hand side are:\footnote{The expressions for $\tau_f$ and $\iota_{c(f)t}$ are the special case implied by the stylized model. In the empirical analysis, these fixed effects will also absorb any additional field-specific and country$\times$year confounder.}
	\begin{align*}
		\beta_{\text{treat}} &\equiv \Delta y \cdot [F_{\text{post}}(A^*_{\text{pre}}) - F_{\text{post}}(A^*_{\text{post,Norway}})] \\
		\beta_1 &\equiv \Delta_{\text{EOR}} - \Delta_{\text{Other}} + \Delta y \cdot [F_{\text{pre}}(A^*_{\text{pre}}) - F_{\text{post}}(A^*_{\text{pre}})] \\
		\tau_f &\equiv (1 - \text{EOR}_f) \cdot y_{0,\text{Other},\text{pre}} + \text{EOR}_f \cdot [y_{0,\text{EOR},\text{pre}} + \Delta y \cdot (1 - F_{\text{pre}}(A^*_{\text{pre}}))] \\
		\iota_{c(f)t} &\equiv \Delta_{\text{Other}} \cdot \text{Post}_t.
	\end{align*}
	In words, $\beta_{\text{treat}}$ captures the average treatment effect of the Norwegian Supreme Court decision, representing the additional mass of operators in the post-period with know-how between the two thresholds; while $\beta_1$ captures both the differential trend in baseline production between EOR-eligible and non-EOR-eligible fields and any distributional shifts that affect all EOR fields equally. Substituting expression \eqref{eq:field-expected} for $\mathbb{E}[y_{ft}]$ in the above identity, we obtain field-level triple-difference regression:
	\begin{align}\label{eq:did-field}
		y_{ft} = \beta_{\text{treat}} \cdot \text{Norway}_f \cdot \text{EOR}_f \cdot \text{Post}_t + \beta_1 \cdot \text{EOR}_f \cdot \text{Post}_t + \tau_f + \iota_{c(f)t} + \epsilon_{ft},
	\end{align}
	where the the error term is:
	\begin{align*}
		\epsilon_{ft} &\equiv y_{ft} - \mathbb{E}[y_{ft}] \\
		&= (1 - \text{Post}_t) \cdot \text{EOR}_f \cdot \Delta y \cdot \left\{\mathbbm{1}\{A_{i(f),\text{pre}} \geq A^*_{\text{pre}}\} - [1 - F_{\text{pre}}(A^*_{\text{pre}})]\right\} \\
		&\quad + \text{Post}_t \cdot \text{EOR}_f \cdot \text{Norway}_f \cdot \Delta y \cdot \left\{\mathbbm{1}\{A_{i(f),\text{post}} \geq A^*_{\text{post,Norway}}\} - [1 - F_{\text{post}}(A^*_{\text{post,Norway}})]\right\} \\
		&\quad + \text{Post}_t \cdot \text{EOR}_f \cdot (1 - \text{Norway}_f) \cdot \Delta y \cdot \left\{\mathbbm{1}\{A_{i(f),\text{post}} \geq A^*_{\text{pre}}\} - [1 - F_{\text{post}}(A^*_{\text{pre}})]\right\}.
	\end{align*}
	Under Assumption \ref{indep_field}, we obtain the desired orthogonality condition $\mathbb{E}[\epsilon_{ft} \mid \text{EOR}_f,$ $\text{Norway}_f] = 0$. This independence assumption ensures that operators' know-how is not systematically related to the types of fields they operate or their geographic location, allowing for consistent estimation of $\beta_{\text{treat}}$ even though the error term is a function of field characteristics. Note that this field-level specification corresponds to the special case of firm-level DiD regression \eqref{eq:did-firm} where each firm operates exactly one field.

	\newpage\clearpage
	\setcounter{table}{0} \renewcommand{\thetable}{B\arabic{table}} 
	\setcounter{figure}{0} \renewcommand{\thefigure}{B\arabic{figure}}
	\setcounter{section}{0} \renewcommand{\thesection}{B}
	
	\section{EOR Eligibility and State-Owned Firms}\label{apndx:back}
	
	\begin{table}[h]
		\centering
		\caption{Criteria for EOR eligibility }
		\label{tab:eor_restrictions}
		\begin{adjustbox}{width={\textwidth},center}
			\begin{threeparttable}
				\begin{tabular}{llcc}
					\toprule
					{Field Characteristic} & {EOR eligibility requirement} & \multicolumn{2}{c}{EOR eligibility} \\ \cmidrule(lr){3-4}
					&& Standard & Robust \\
					\midrule
					Hydrocarbon & oil, oil \& gas & \checkmark & \checkmark \\
					Development type & fixed platform, subsea, tension leg platform & \checkmark & \checkmark \\
					Field size & giant, grouping, large & \checkmark & \checkmark \\
					Water depth & shallow water & \checkmark & \checkmark \\
					Field primary resource & conventional shelf & \checkmark & \checkmark \\
					API gravity & Between 22.6 and 47.6 &\checkmark & \checkmark \\
					Reservoir depth (m) & Between 909.68 and 4,460.23 &\checkmark & \checkmark \\
					Porosity (\%) & Between 10 and 40 & & \checkmark \\
					Permeability (mD) & Between 0 and 4,000 & & \checkmark \\
					\bottomrule
				\end{tabular}
				\begin{tablenotes}
					\footnotesize 
					\item \hspace{-0.2em} \textbf{Notes:} Technical and geological field's characteristics that define EOR eligibility \cite{al2011analysis,nwidee2016eor}.
					\begin{enumerate}
						\item[-] \textit{Hydrocarbon} refers to the type of hydrocarbons. 
						\item[-]\textit{Development type} refers to the type of infrastructure used for oil drilling. 
						\item[-]\textit{Field size} is corresponds to the size of the reservoir measured during the exploration phase in million barrels of oil equivalent. ``Grouping'' indicates that reserves or production are aggregated across multiple smaller sites. 
						\item[-]\textit{Water depth} categorizes fields into ``shallow water,'' defined as having the seafloor less than 150m from the water surface, and ``deep water.''
						\item[-]\textit{Field primary resource} categorizes the type of resource in the field. ``Conventional shelf'' refers to conventional resources in shallow-water O\&G fields located on the continental shelf, typically in water depths of less than 150m. Examples include the Gulf of Mexico, the North Sea, and the Persian Gulf.
						\item[-]\textit{API gravity} measures the density of crude oil and is measured in ``API degrees.'' 
						\item[-]\textit{Reservoir depth} is the total vertical distance between the surface and the reservoir. Shallow reservoirs are less than 1,500m, intermediate reservoirs range from 1,500m to 3,000m, and deep reservoirs range from 3,000m to 5,000m. 
						\item[-]\textit{Porosity (\%)} refers to the amount of empty space within a rock that can store hydrocarbons. 
						\item[-]\textit{Permeability (mD)} is the ability of a rock to allow fluids to flow through.
					\end{enumerate}
					
				\end{tablenotes}
			\end{threeparttable}
		\end{adjustbox}
	\end{table}

	\begin{table}[ht]
		\caption{List of of state-owned companies in 1975-1995}
		\label{tab:north_sea_companies}
		\centering
		\begin{adjustbox}{width=\textwidth}
			\begin{tabular}{lp{12cm}}
				\toprule
				\textbf{Company Name} & \textbf{Description and North Sea Involvement} \\
				\midrule
				BP & British multinational O\&G company. Originally known as British Petroleum, was a key player in North Sea oil exploration and production during the 1970s and 1980s. The company was state owned until it was fully privatized in 1987. In 1988, BP acquired Britoil, strengthening its North Sea presence. It merged with Amoco in 1998. \\
				\multicolumn{2}{c}{ }\\
				Britoil &  A British oil company active in the North Sea during the 1970s and 1980s, focused on exploration and production. It was acquired by BP in 1988, which enhanced BP's presence in the North Sea and contributed to its leadership in the region.
				\\
				\multicolumn{2}{c}{ }\\
				Norway State DFI & From 1973 to 1985, Statoil was awarded a majority of Norway's petroleum development licenses. In the 1980s, as Statoil's cash flow grew significantly with respect to Norway's gross national product, the government restructured its ownership for political reasons. On January 1, 1985, the State's Direct Financial Interest (SDFI) was created, with Statoil retaining 20\% of its original portfolio and transferring the rest to SDFI.
				\\
				\multicolumn{2}{c}{ }\\
				Statoil & Founded in 1972, Statoil was initially a state-owned company involved in the exploration and production of oil in the North Sea. It was privatized in 2001. It merged with Hydro to form StatoilHydro in 2007. It was later rebranded as Equinor in 2018. \\
				\bottomrule
			\end{tabular}
		\end{adjustbox}
	\end{table}
	
	\newpage\clearpage

	

	\newpage\clearpage
	\setcounter{table}{0} \renewcommand{\thetable}{C\arabic{table}} %
	\setcounter{figure}{0} \renewcommand{\thefigure}{C\arabic{figure}}
	\setcounter{section}{0} \renewcommand{\thesection}{C}
	\section{Robustness Checks on Adoption and Production}\label{a:robustness_checks}

	\paragraph{Field-level regressions.} To account for potential confounding variation across countries and years, such as divergent national policies, which are not directly controlled for by regression \eqref{eq:did-firm}, we estimate field-level triple-difference specification \eqref{eq:did-field} (see Appendix \ref{a:field_regressions} for a derivation along the lines of Section \ref{sec:firm_regressions}):
	\begin{align*}
		y_{ft} = \beta_{\text{treat}} \cdot \text{Norway}_f \cdot \text{EOR}_f \cdot \text{Post}_t + \beta_1 \cdot \text{EOR}_f \cdot \text{Post}_t + \tau_f + \iota_{c(f)t} + \epsilon_{ft},
	\end{align*}
	where $\tau_f$ and $\iota_{c(f)t}$ denote field and country-year fixed effects, respectively. Results are presented in Appendix Table \ref{tab:did-field-appendix}, Panel (a). Column 1 shows that Norwegian EOR-eligible fields were 27.6 percentage points (p.p.) more likely to adopt EOR after 1985 than comparable UK fields.
	
	Columns 2-4 report significant production gains. Column 2 shows a 12.6-fold total increase in production for Norwegian EOR-eligible fields ($\exp(2.610)-1$). Decomposing this growth, Column 3 documents a 5-fold rise at the intensive margin (fields active by 1985), while Column 4 finds a 44 p.p. increase at the extensive margin (the probability that previously inactive fields started production). 
	
	Finally, Column 5 shows that EOR adoption caused substantial efficiency gains. Norwegian EOR-eligible fields produced an additional 60.11 barrels per \$1,000 of OPEX relative to the UK baseline. This represents a 67.5\% increase over the average efficiency of 89.04 bbl/kUSD, with implied average variable costs falling from \$11.23 to \$6.70 per barrel. Panel (b) confirms similar estimates across production and efficiency measures using the Poisson Pseudo Maximum Likelihood (PPML) estimator, which is more suitable to account for zeros and heteroskedasticity than standard log-linear OLS specifications.\footnote{We do not estimate the adoption equation (Column 1) via PPML because the dependent variable is binary; instead, we employ a linear probability model suited for Bernoulli-distributed outcomes. Panel (a) of Appendix Table \ref{tab:did-firm-appendix} confirms the results in Table \ref{tab:did-firm} using a PPML estimator.} 
	
	These results suggest that, even after controlling for country-by-year fixed effects, the reduction in sovereign hold-up risk due to the Norwegian Supreme Court decision boosted EOR adoption and output across both active and previously inactive fields.
	
\begin{figure}[h!]
		\captionsetup[subfigure]{justification=centering}
		\caption{Field-level event study of production  \label{fig:field_ES}}	
		\minipage{0.45\textwidth}
		\subcaption{$\ln{(1+\text{prod}_{ft})}$}\label{fig:prod_ES-field}
		\includegraphics[width=\linewidth]{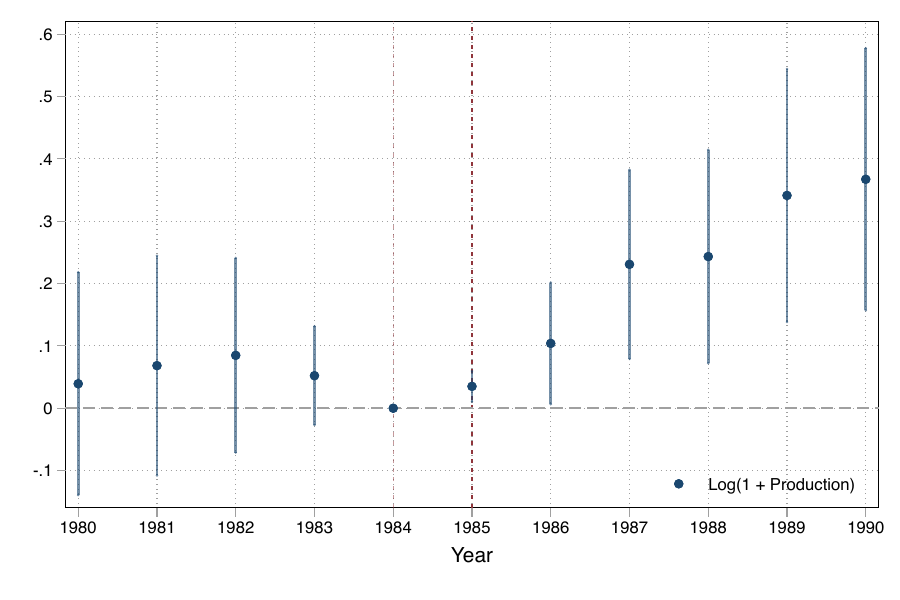}
		\endminipage\hfill
		%
		\minipage{0.45\textwidth}
		\subcaption{prod$_{ft}$ / OPEX$_{ft}$}\label{fig:prod_OPEX_ES-field}
		\includegraphics[width=\linewidth]{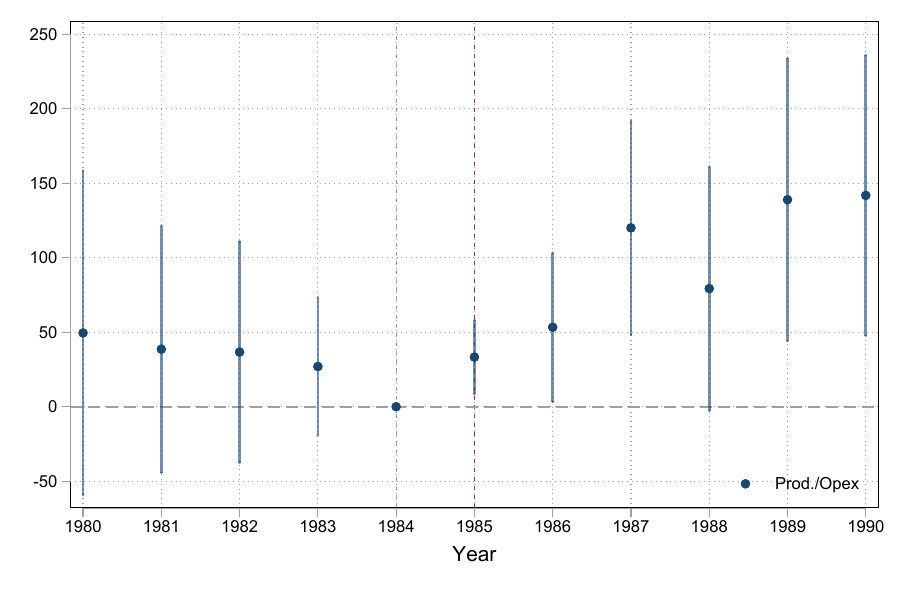}
		\endminipage\hfill
		\begin{minipage}{1 \textwidth}
			{\footnotesize\singlespacing \textbf{Notes:} Estimated coefficients from field-level regression \eqref{eq:event_field}. Each panel shows the estimates of $\widehat{\beta}_{\text{treat},d}$ corresponding to a different dependent variable. The dependent variables are production (in log, Panel (a)),  and the ratio of production to OPEX (in bbl/kUSD, Panel (b)). The indicator variables ``Norway,'' ``EOR,'' and ``Post'' equal 1 for Norwegian fields, for EOR-eligible fields, and for the years after 1985. Each regression includes country-by-year and field fixed effects. Vertical bars around each estimate represent 95\% confidence intervals. Standard errors are clustered at the field level. Vertical dotted lines mark the years when the Eidsivating Court of Appeal (1984) and the Norwegian Supreme Court (1985) delivered their rulings.
				\par}
		\end{minipage}
	\end{figure}

	To track changes in production and costs at the level of the field over time, we estimate the event study regression:
	\begin{equation}\label{eq:event_field}
		\begin{aligned}
			y_{ft} &= \sum_{d=-4}^{6} \beta_{\text{treat},d} \cdot \text{Norway}_{f} \cdot \text{EOR}_f \cdot \mathbbm{1}\{t-1984=d\} \\
			&\quad +  \sum_{d=-4}^{6} \beta_{1,d} \cdot \text{EOR}_f \cdot \mathbbm{1}\{t-1984=d\} + \tau_{f} + \iota_{c(f)t} + \epsilon_{ft}.
		\end{aligned}
	\end{equation}
	Appendix Figure~\ref{fig:field_ES} shows that production and the production-to-OPEX ratio all increased sharply after 1985. The flat pre-trends and the statistically significant increases in 1985 and 1986 suggest that the observed changes cannot be attributed to the phasing out of Norwegian royalties in 1986, which mirrored a similar decision in the UK in 1983, to which we come back in the following paragraphs.
	
	\paragraph{Timing of treatment. } We also check whether firms began responding to the 1984 Eidsivating Court of Appeal ruling, anticipating the Norwegian Supreme Court decision. We re-compute the portfolio shares in \eqref{eq:did-firm} using 1983 as the last year of the pre-period and re-estimate both firm- and field-level regressions, defining $\text{Post}_t$ as 1 for the years since 1984 and 0 otherwise. Both field-level (Appendix Table \ref{tab:did-field-appendix}, Panel (c)) and firm-level (Appendix Table \ref{tab:did-firm-appendix}, Panel (b)) results lead to similar conclusions as those presented above.
	
	\paragraph{Excluding post-1983 licenses.}  
	As noted in Section \ref{s:northsea}, the UK and Norway phased out royalties on new fields in 1983 and 1986, respectively. To ensure these changes do not confound our results, we re‑estimate field-level regression \eqref{eq:did-field} on fields discovered before 1983 and define the treatment as starting with the 1984 Court of Appeal ruling (results are similar if we instead use the 1985 Supreme Court decision). Panel (d) of Appendix Table \ref{tab:did-field-appendix} and Panel (c) of Appendix Table \ref{tab:did-firm-appendix} confirm that our findings for EOR adoption, production, and costs remain unchanged.

	\paragraph{Broader EOR-eligibility criteria.} Panel (e) of Appendix Table \ref{tab:did-field-appendix} and Panel (d) of Appendix Table \ref{tab:did-firm-appendix} confirm that our results are also robust to an alternative measure of EOR eligibility that accounts for field permeability and porosity. The criteria for a field's EOR eligibility are detailed in Appendix Table \ref{tab:eor_restrictions}.

	\paragraph{Triple difference at the firm level.} The first five columns of Appendix Table \ref{tab:did-firm_all_treat} report estimates of an extension of regression \eqref{eq:did-firm} that also controls for ``Share Norway$_i$ $\cdot$ Post$_t$,'' where ``Share Norway$_i$'' represents the average share of Norwegian fields included in firm $i$'s portfolio before 1985. This is the firm-level analogue of the field-level triple-difference regression \eqref{eq:did-field} and controls also for a firm's exposure to changes in the Norwegian market. Overall, the analyses yield results consistent with those of regression \eqref{eq:did-firm}, though with slightly larger standard errors due to the high correlation (0.82) between ``Share EOR Norway$_i$'' and ``Share Norway$_i$.''
	
	\paragraph{Incentives for operators.}  
	A direct implication of the model in Section \ref{s:conceptual_framework} is that the adoption threshold decreases with the operator's ownership share---i.e., $ \partial B(A)/\partial s < 0 $ in equation~\eqref{eq:threshold_A}---as the operator captures a larger portion of the gains from EOR adoption. To test this prediction, we extend the field-level regression in equation~\eqref{eq:did-field} by allowing for heterogeneous treatment effects. Specifically, we interact the treatment with an indicator equal to one if the operator's ownership share in field $ f $ exceeded its North Sea median in $ t{-}1 $. Appendix Table~\ref{tab:did-field_HOS} confirms this model prediction: operators with large ownership share $s$ were the main adopters of EOR.

	\paragraph{Displacement.} To interpret our results, we investigate whether the Supreme Court decision induced crowding-out within Norway---whether the post-1985 expansion of production in Norwegian EOR-eligible fields came at the expense of production in Norwegian non-EOR-eligible fields. Absent crowding-out, we should not observe a decrease in the production of Norwegian non-EOR-eligible after 1985 relative to its own pre-1985 path. Panel (a) of Appendix Figure~\ref{fig:combined_row} shows precisely this: Norwegian non-EOR-eligible output remained broadly stable around 1985, with no visible decline after 1985. 
	
	Panels (b)-(c) further clarify that there were no signs of crowding-out also within cost segments. The production of high-cost non-EOR-eligible fields followed similar trends across countries (Panel (b)), suggesting the absence of any Norway-specific negative shock to the production of these fields around the ruling. Panel (c) shows that the post-1985 rise of production for British non-EOR-eligible fields documented in Panel (a) is mainly driven by low-cost fields. Furthermore, we find stable production levels at control firms without EOR-eligible fields---75 firms---in Norway before 1985 (Panel (d)).
	
	Take together, the relevant implication for crowding-out is that, even as UK low-cost non-EOR output increases, Norwegian low-cost non-EOR output does not exhibit a corresponding decline after 1985. 
	
	
	\begin{table}[!htbp]
		\caption{Robustness checks at the field level\label{tab:did-field-appendix}}
		\centering
		\begin{adjustbox}{width=\textwidth}
			{
\def\sym#1{\ifmmode^{#1}\else\(^{#1}\)\fi}
\begin{adjustbox}{width=0.9\textwidth,center}
\begin{tabular}{l*{5}{c}}
\toprule
 & EOR Adoption &\multicolumn{3}{c}{Production Margins}& Productivity\\\cmidrule(lr){2-2} \cmidrule(lr){3-5} \cmidrule(lr){6-6}
 &\multicolumn{1}{c}{ } &\multicolumn{1}{c}{Intensive and Extensive} & \multicolumn{1}{c}{Only Intensive} & \multicolumn{1}{c}{Only Extensive}&\multicolumn{1}{c}{ }\\
Dependent variable &\multicolumn{1}{c}{$\mathbbm{1}\{\text{Adopted EOR}_{ft}\}$}&\multicolumn{1}{c}{$\ln{(1+\text{prod}_{ft})}$} & \multicolumn{1}{c}{$\ln{(1+\text{prod}_{ft})}$} & \multicolumn{1}{c}{$\mathbbm{1}\{\text{prod}_{ft}>0\}$}&\multicolumn{1}{c}{$\text{prod}_{ft}\text{ }/\text{ } \text{OPEX}_{ft}$}\\
\rule{0pt}{2ex}   
&\multicolumn{1}{c}{(1)}         &\multicolumn{1}{c}{(2)}         &\multicolumn{1}{c}{(3)}         &\multicolumn{1}{c}{(4)}         &\multicolumn{1}{c}{(5)}         \\ 
\rule{0pt}{2ex}   \\[-1ex]
\multicolumn{6}{l}{\textit{Panel (a): Baseline results (OLS)}}\\
\midrule
Norway $\times$ EOR $\times$ Post&0.276\sym{**} & 2.610\sym{***} & 1.787\sym{***} & 0.440\sym{**} & 60.110\sym{**} \\
                &(0.119) & (0.634) & (0.469) & (0.217) & (29.054) \\
[1em]
EOR $\times$ Post&0.065\sym{*} & 0.621\sym{*} & 0.905\sym{**} & 0.037 & 23.552 \\
                &(0.034) & (0.345) & (0.367) & (0.096) & (26.105) \\
[1em]
Field selection &All & All & {Prod. $<$ 1985} & All & All \\
Observations &1989 & 1989 & 924 & 1989 & 1989 \\
R-squared &0.84 & 0.62 & 0.64 & 0.45 & 0.56 \\
\midrule\midrule
\rule{0pt}{4ex}   \\[-3ex]
\multicolumn{6}{l}{\textit{Panel (b): Baseline results (PPML estimator)}}\\
\midrule
Norway $\times$ EOR $\times$ Post&-- & 1.816\sym{***} & 1.346\sym{***} & 0.733\sym{***} & 0.905\sym{*} \\
                &-- & (0.498) & (0.292) & (0.231) & (0.500) \\
[1em]
EOR $\times$ Post&-- & 0.293 & 0.426 & 0.065 & 0.317 \\
                &-- & (0.289) & (0.281) & (0.065) & (0.206) \\
[1em]
Field selection &-- & All & {Prod. $<$ 1985} & All & -- \\
Observations &-- & 1989 & 924 & 1989 & 1905 \\
Pseudo R-squared &-- & 0.76 & 0.79 & 0.05 & 0.48 \\
\midrule\midrule
\rule{0pt}{4ex}   \\[-3ex]
\multicolumn{6}{l}{\textit{Panel (c): Treatment starts in 1984}}\\
\midrule
Norway $\times$ EOR $\times$ Post&0.239\sym{*} & 1.918\sym{***} & 2.426\sym{***} & 0.203 & 40.160 \\
                &(0.122) & (0.537) & (0.457) & (0.193) & (33.257) \\
[1em]
EOR $\times$ Post&0.072\sym{*} & 1.134\sym{***} & 0.977\sym{**} & 0.185\sym{*} & 59.120\sym{*} \\
                &(0.038) & (0.393) & (0.416) & (0.108) & (32.715) \\
[1em]
Field selection &All & All & {Prod. $<$ 1983} & All & Prod. $<$ 1983 \\
Observations &1989 & 1989 & 768 & 1989 & 768 \\
R-squared &0.84 & 0.62 & 0.68 & 0.45 & 0.69 \\
\midrule\midrule
\rule{0pt}{4ex}   \\[-3ex]
\multicolumn{6}{l}{\textit{Panel (d): Excluding fields discovered since 1983}}\\
\midrule
Norway $\times$ EOR $\times$ Post&0.238\sym{**} & 2.354\sym{***} & 2.785\sym{***} & 0.310\sym{*} & 61.493\sym{***} \\
                &(0.116) & (0.430) & (0.358) & (0.167) & (21.827) \\
[1em]
EOR $\times$ Post&0.087\sym{**} & 0.837\sym{**} & 0.742\sym{**} & 0.093 & 31.960 \\
                &(0.035) & (0.333) & (0.364) & (0.092) & (24.112) \\
[1em]
Field selection &Disc. $<$ 1983 & Disc. $<$ 1983 & Disc. $<$ 1983 & Disc. $<$ 1983 & Disc. $<$ 1983 \\
 & & & \& Prod. $<$ 1983 &  &  \\
Observations &1629 & 1629 & 768 & 1629 & 1629 \\
R-squared &0.84 & 0.60 & 0.67 & 0.43 & 0.60 \\
\midrule\midrule
\rule{0pt}{4ex}   \\[-3ex]
\multicolumn{6}{l}{\textit{Panel (e): Different criteria for EOR eligibility}}\\
\midrule
Norway $\times$ EOR $\times$ Post&0.398\sym{***} & 2.777\sym{***} & 1.666\sym{***} & 0.449\sym{*} & 56.002\sym{*} \\
                &(0.088) & (0.667) & (0.546) & (0.269) & (30.258) \\
[1em]
EOR $\times$ Post&0.061\sym{*} & 0.588\sym{*} & 0.831\sym{**} & 0.063 & 22.566 \\
                &(0.037) & (0.346) & (0.369) & (0.095) & (25.303) \\
[1em]
Field selection &All & All & \multicolumn{1}{c}{Prod. $<$ 1985} & All & -- \\
Observations &1989 & 1989 & 924 & 1989 & 1989 \\
R-squared &0.85 & 0.61 & 0.62 & 0.45 & 0.56 \\
\midrule\midrule
Country-Year FE &Yes & Yes & Yes & Yes & Yes \\
Field FE &Yes & Yes & Yes & Yes & Yes \\
\bottomrule
\multicolumn{6}{l}{* -- $p < 0.1$; ** -- $p < 0.05$; *** -- $p < 0.01$}
\end{tabular}
\end{adjustbox}
}
		\end{adjustbox}
		\begin{tablenotes}
			\footnotesize \vspace{0.5em}
			\parbox{\textwidth}{\justifying  \item \hspace{-0.2em}\textbf{Notes:}  Estimated coefficients from field-level regression \eqref{eq:did-field}. Panels (a) and (b) present the baseline results using the OLS or the Poisson Pseudo Maximum Likelihood (PPML) estimators, respectively. In Panel (c), the variable Post is one for years since 1984 (the year of the ruling of the Eidsivating Court of Appeal) and zero otherwise. In Panel (d), fields discovered since 1983 in the North Sea are excluded from the analysis. In Panel (e), EOR eligibility is defined according to the ``robust'' measure in Appendix Table \ref{tab:eor_restrictions}, which also account for porosity and permeability. The dependent variable in Column 1 is an indicator that equals 1 if field $f$ has adopted EOR by year $t$ and 0 otherwise. The dependent variables in Columns 2-4 are functions of field-level production. Columns 2 and 3 use the log of production, with Column 3 restricting the sample to fields that began production before 1985. The dependent variable in Column 4 is an indicator that equals 1 if the field produces in year $t$ and 0 otherwise. The dependent variable in Column 5 is production / opex, measured in barrels per 1k USD. The indicator variables ``Norway,'' ``EOR,'' and ``Post'' equal 1 for Norwegian fields, for EOR-eligible fields, and for the years after 1985 (1983 in Panel (c)). Each regression includes country-by-year and field fixed effects. Standard errors are clustered at the field level.}
		\end{tablenotes}
	\end{table}
	
	\FloatBarrier
	\clearpage\newpage
	
	\begin{table}[!htbp]
		\centering
		\caption{Robustness checks at the firm level \label{tab:did-firm-appendix}}
		\begin{adjustbox}{width=.95\textwidth,center}
			{
\def\sym#1{\ifmmode^{#1}\else\(^{#1}\)\fi}
\begin{adjustbox}{width=0.7\textwidth,center}
\begin{tabular}{l*{3}{c}}
\toprule
& Adoption & \multicolumn{2}{c}{Production}\\
\cmidrule(lr){3-4}
Dependent variable &\multicolumn{1}{c}{Share EOR} & \multicolumn{1}{c}{$\ln(1+\text{prod}_{it})$} & \multicolumn{1}{c}{$\text{prod}_{it}/\text{OPEX}_{it}$} \\
\rule{0pt}{2ex}
&\multicolumn{1}{c}{(1)} &\multicolumn{1}{c}{(2)} &\multicolumn{1}{c}{(3)} \\
\rule{0pt}{2ex}   \\[-1ex]
\multicolumn{4}{l}{\textit{Panel (a): Baseline results (PPML Estimator)}}\\
\midrule
Share EOR Norway $\times$ Post&    3.380\sym{***}&    3.592\sym{***}&    0.875\sym{***}\\
                &  (0.544)         &  (0.635)         &  (0.236)         \\
[1em]
Share EOR $\times$ Post&    0.181         &   -0.164         &   -0.089         \\
                &  (0.391)         &  (0.326)         &  (0.082)         \\
\midrule
Average dependent variable & 0.166 & 62.056 & 0.328 \\
\midrule
Observations    &      449         &     1387         &     1241         \\
Pseudo R-squared       &     0.14         &     0.91         &     0.04         \\
\midrule\midrule
\rule{0pt}{4ex}   \\[-3ex]
\multicolumn{4}{l}{\textit{Panel (b): Treatment starts in 1984}}\\
\midrule
Share EOR Norway $\times$ Post & 0.252\sym{***} & 2.886\sym{***} & 0.270\sym{*} \\
 & (0.066) & (0.838) & (0.158) \\
[1em]
Share EOR $\times$ Post & 0.013 & 0.051 & 0.031 \\
 & (0.010) & (0.278) & (0.105) \\
\midrule
Average dependent variable & 0.054 & 2.591 & 0.330 \\
\midrule
Observations & 1382 & 1382 & 1225 \\
R-squared & 0.66 & 0.85 & 0.43 \\
\midrule\midrule
\rule{0pt}{4ex}   \\[-3ex]
\multicolumn{4}{l}{\textit{Panel (c): Excluding fields discovered since 1983}}\\
\midrule
Share EOR Norway $\times$ Post&    0.397\sym{***}&    3.260\sym{***}&    0.270\sym{*}  \\
                &  (0.111)         &  (0.835)         &  (0.140)         \\
[1em]
Share EOR $\times$ Post&    0.012         &   -0.355         &   -0.025         \\
                &  (0.011)         &  (0.269)         &  (0.090)         \\
\midrule
Average dependent variable & 0.058 &  2.534 & 0.325 \\
\midrule
Observations    &     1406         &     1406         &     1245         \\
R-squared       &     0.73         &     0.86         &     0.43         \\
\midrule\midrule
\rule{0pt}{4ex}   \\[-3ex]
\multicolumn{4}{l}{\textit{Panel (d): Robust definition of EOR eligibility}}\\
\midrule
Share EOR Norway $\times$ Post & 0.335\sym{***} & 2.051\sym{***} & 0.289 \\
 & (0.120) & (0.671) & (0.179) \\
[1em]
Share EOR $\times$ Post & 0.033\sym{**} & -0.289 & -0.011 \\
 & (0.016) & (0.273) & (0.098) \\
\midrule
Average dependent variable & 0.053 & 2.562 & 0.328 \\
\midrule
Observations & 1406 & 1406 & 1245 \\
R-squared & 0.70 & 0.85 & 0.43 \\
\midrule\midrule
Year FE & Yes & Yes & Yes \\
Firm FE & Yes & Yes & Yes \\
\bottomrule
\multicolumn{4}{l}{* -- $p < 0.1$; ** -- $p < 0.05$; *** -- $p < 0.01$}
\end{tabular}
\end{adjustbox}
}
		\end{adjustbox}
		\begin{tablenotes}
			\footnotesize \vspace{1em}
			\parbox{\textwidth}{\justifying  \item \hspace{-0.2em} 
				\textbf{Notes:} Estimated coefficients from regression \eqref{eq:did-firm}. Panel (a) replicates the main text analysis using a PPML estimator. In Panel (b), the variable Post is one for years since 1984 (the year of the ruling of the Eidsivating Court of Appeal) and zero otherwise.   In Panel (d), fields discovered since 1983 in the North Sea are excluded from the analysis. In Panel (e), EOR eligibility is defined according to the ``robust'' measure in Appendix Table \ref{tab:eor_restrictions}, which also account for porosity and permeability. The dependent variable ``Adoption Share EOR'' in Column 1 is the share of fields that adopted EOR among all the fields in which firm $i$ was active in the North Sea in year $t$. The other dependent variables are production in log (Column 2) and the ratio of production (in Kboe) to OPEX (Column 3). The variable ``Share EOR Norway'' (``Share EOR'') refers to the fraction of Norwegian (North Sea) EOR-eligible fields among all the fields in which firm $i$ was active in the North Sea before 1985. In all Panels besides Panel (b), the indicator ``Post'' equals 1 for the years after 1985, the year of the Supreme Court decision, and 0 otherwise. Each regression includes firm and year fixed effects. Standard errors are clustered at the firm level.
			}
		\end{tablenotes}
	\end{table}
	\newpage

	\begin{table}[htb]
		\caption{Field-level adoption---the role of operator ownership } \label{tab:did-field_HOS}
		\begin{adjustbox}{width=.9\textwidth,center}

{
\def\sym#1{\ifmmode^{#1}\else\(^{#1}\)\fi}
\begin{tabular}{l*{4}{c}}
\toprule
 &\multicolumn{4}{c}{EOR Adoption}                                           \\\cmidrule(lr){2-5}
 &\multicolumn{2}{c}{Treatment starts in 1986} 
 &\multicolumn{2}{c}{Treatment starts in 1984}\\ 
 \cmidrule(lr){2-3} \cmidrule(lr){4-5}
 & (1) & (2) & (3) & (4) \\
\midrule

Norway $\times$ EOR $\times$ Post ($\beta_{\text{treat}}$)
& 0.276\sym{**} & 0.076 & 0.239\sym{*} & -0.021 \\
& (0.119) & (0.085) & (0.122) & (0.103) \\
[1em]

EOR $\times$ Post ($\beta_{1}$)
& 0.065\sym{*} & 0.065\sym{**} & 0.072\sym{*} & 0.069\sym{*} \\
& (0.034) & (0.033) & (0.038) & (0.036) \\
[1em]

Norway $\times$ EOR $\times$ Post $\times$ High Ownership Share ($\beta_{\text{het}}$)
&  & 0.260\sym{**} &  & 0.328\sym{**} \\
&  & (0.123) &  & (0.146) \\
\midrule
Post defined as & \multicolumn{2}{c}{years $>$ 1985} & \multicolumn{2}{c}{years $>$ 1983}\\
\midrule 
Country-Year FE & Yes & Yes & Yes & Yes \\
Field FE        & Yes & Yes & Yes & Yes \\
Observations    & 1,989 & 1,771 & 1,989 & 1,771 \\
R-squared       & 0.84 & 0.86 & 0.84 & 0.86 \\
\bottomrule
\multicolumn{5}{l}{\footnotesize * $p < 0.1$; ** $p < 0.05$; *** $p < 0.01$}
\end{tabular}
}

		\end{adjustbox}
		\begin{tablenotes}
			\footnotesize 
			\parbox{\textwidth}{\justifying  \item \hspace{-0.2em}\textbf{Notes:} Estimated coefficients from field-level regression
				\begin{equation*}
					\begin{aligned}
						y_{ft} &= \beta_{\text{het}} \cdot \text{Norway}_{f} \cdot \text{EOR}_f \cdot \text{Post}_t \cdot \text{Operator Above Median Ownership}_{f,t-1} \\
						&\quad + \beta_{\text{treat}} \cdot \text{Norway}_{f} \cdot \text{EOR}_f \cdot \text{Post}_t  + \beta_1 \cdot \text{EOR}_f \cdot \text{Post}_t  + \iota_{f} + \tau_{c(f)t} + \epsilon_{ft},
					\end{aligned}
				\end{equation*}
				where the dependent variable is an indicator that equals 1 if field $f$ has adopted EOR by year $t$ and 0 otherwise. The indicator variables ``Norway,'' ``EOR,'' and ``Post'' equal 1 for Norwegian fields, for EOR-eligible fields, and for the years after 1985 / 1983, and 0 otherwise. The indicator variable $\text{Operator Above Median Ownership}_{f,t-1}$ equals 1 if field $f$'s operator owned more than the median ownership share in the North Sea in $t-1$. The number of observations in Columns (2) and (4) is lower because \text{Operator ``Above Median Ownership}$_{f,t-1}$'' is lagged. Standard errors are clustered at the field level.}
		\end{tablenotes}
	\end{table}

	\FloatBarrier
	
	\begin{table}[!t]
		\centering
		\caption{Firm-level analysis controlling for ``Share Norway $\cdot$ Post'' \label{tab:did-firm_all_treat}}
		\begin{adjustbox}{width=1.1\textwidth,center}

{
\def\sym#1{\ifmmode^{#1}\else\(^{#1}\)\fi}
\begin{adjustbox}{width=.9\textwidth,center}
\begin{tabular}{l*{5}{c}}
\toprule
 & Adoption &\multicolumn{2}{c}{Production and Costs}&\multicolumn{2}{c}{Market Share}\\
 \cmidrule(lr){2-2} \cmidrule(lr){3-4} \cmidrule(lr){5-6}
Dependent variable  
& Share EOR 
& \multicolumn{1}{c}{$\ln(1+\text{prod}_{it})$}
& \multicolumn{1}{c}{$\text{prod}_{it}\text{ }/\text{ }\text{OPEX}_{it}$}
& \multicolumn{1}{c}{All}
& \multicolumn{1}{c}{Positive} \\
\rule{0pt}{2ex}   
& (1) & (2) & (3) & (4) & (5) \\ 
\midrule
Share EOR Norway $\times$ Post
&    0.343\sym{*}  
&    3.812\sym{***}
&    0.208         
&    0.033         
&    0.120\sym{**} \\
&  (0.179)         
&  (1.032)         
&  (0.161)         
&  (0.024)         
&  (0.052)         \\
[1em]

Share EOR $\times$ Post
&    0.021\sym{*}  
&   -0.487\sym{*}  
&   -0.004         
&    0.000         
&   -0.001         \\
&  (0.011)         
&  (0.278)         
&  (0.103)         
&  (0.002)         
&  (0.002)         \\
[1em]

Share Norway $\times$ Post
&    0.003         
&   -0.469         
&   -0.013         
&    0.005         
&   -0.023         \\
&  (0.055)         
&  (0.475)         
&  (0.131)         
&  (0.012)         
&  (0.017)         \\
\midrule
Average dependent variable 
& 0.053 
& 2.562 
& 0.291 
& 0.015 
& 0.017 \\
\midrule
Year FE         & Yes & Yes & Yes & Yes & Yes \\
Firm FE         & Yes & Yes & Yes & Yes & Yes \\
Observations    & 1406 & 1406 & 1406 & 1406 & 1232 \\
R-squared       & 0.70 & 0.86 & 0.41 & 0.83 & 0.86 \\
\bottomrule
\multicolumn{6}{l}{* -- $p < 0.1$; ** -- $p < 0.05$; *** -- $p < 0.01$}
\end{tabular}
\end{adjustbox}
}

		\end{adjustbox}
		\begin{tablenotes}
			\footnotesize \vspace{1em}
			\parbox{\textwidth}{\justifying  \item \hspace{-0.2em}\textbf{Notes:} Estimated coefficients from regression \eqref{eq:did-firm}, additionally controlling for ``Share Norway $\cdot$ Post.'' We exclude this regressor in our main analyses because of multicollinearity with ``Share EOR Norway $\cdot$ Post'' (correlation of 0.8). The table reports estimates for all the key dependent variables analyzed in other tables (a firm's share of EOR adoption, production, production / OPEX, market shares, and positive market shares). The variable ``Share EOR Norway'' (``Share EOR'') refers to the fraction of Norwegian (North Sea) EOR-eligible fields among all the fields in which firm $i$ was active in the North Sea before 1985. ``Share Norway'' is the share of Norwegian fields in which firm $i$ was active in the North Sea before 1985. The indicator ``Post'' equals 1 for the years after 1985, the year of the Supreme Court decision, and 0 otherwise. Each regression includes firm and year fixed effects. Standard errors are clustered at the firm level.}
		\end{tablenotes}
	\end{table}

	\begin{figure}[!b]
		\centering
		\captionsetup[subfigure]{justification=centering}
		\caption{Production trends of non-EOR eligible fields }
		\label{fig:combined_row}
		
		\begin{subfigure}[b]{0.48\textwidth}
			\centering
			\subcaption{All non-EOR-eligible fields}
			\includegraphics[width=\linewidth]{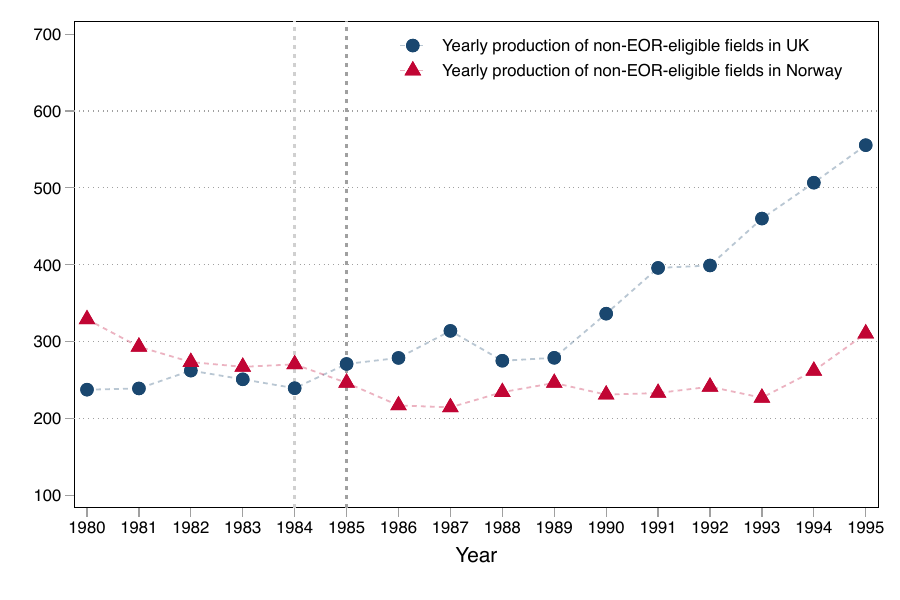}
		\end{subfigure}
		\hfill
		\begin{subfigure}[b]{0.48\textwidth}
			\centering
			\subcaption{High-cost fields}
			\includegraphics[width=\linewidth]{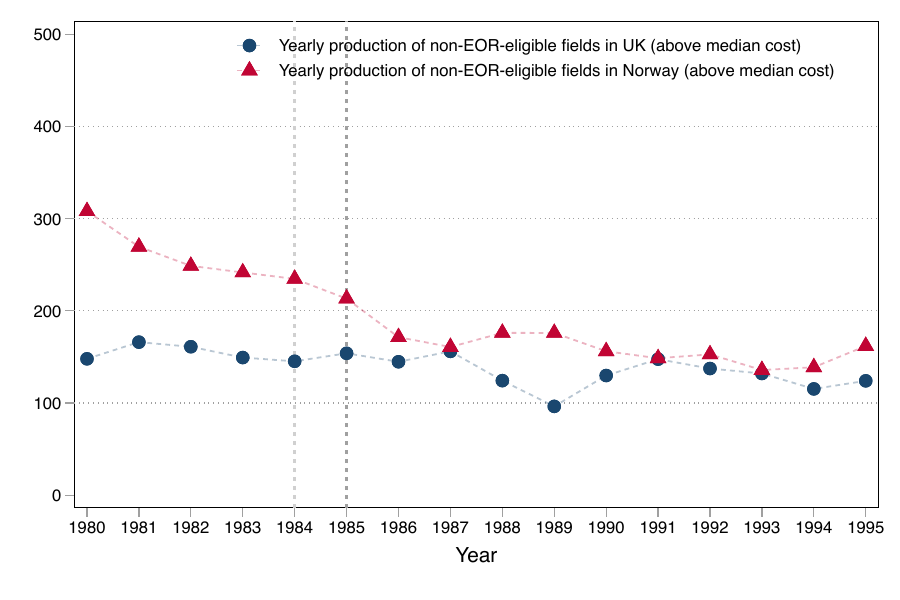}
		\end{subfigure}
		\\
		
		\begin{subfigure}[b]{0.48\textwidth}
			\centering
			\subcaption{Low-cost fields}
			\includegraphics[width=\linewidth]{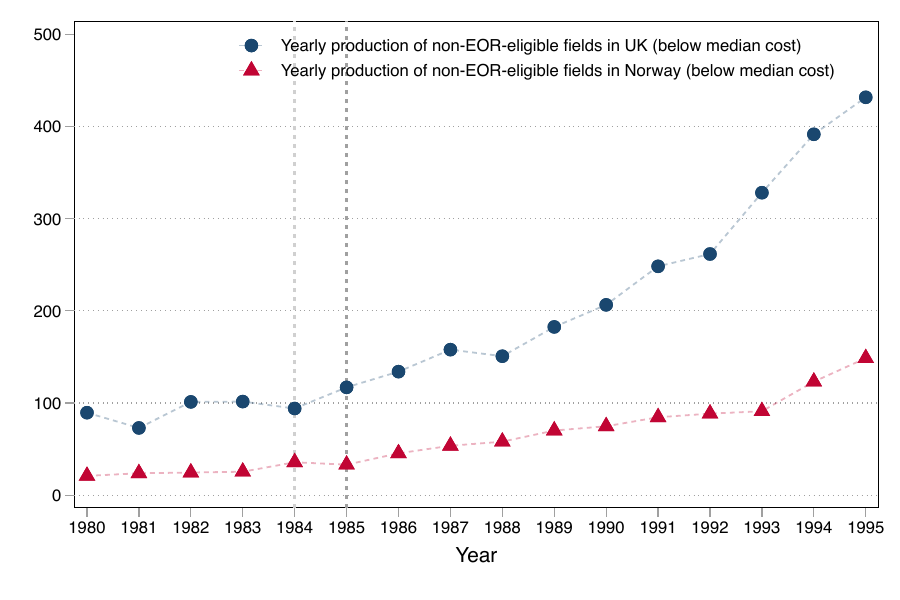}
		\end{subfigure}
		\hfill
		\begin{subfigure}[b]{0.48\textwidth}
			\centering
			\subcaption{Firms w/o Norwegian EOR-eligible fields}
			\includegraphics[width=\linewidth]{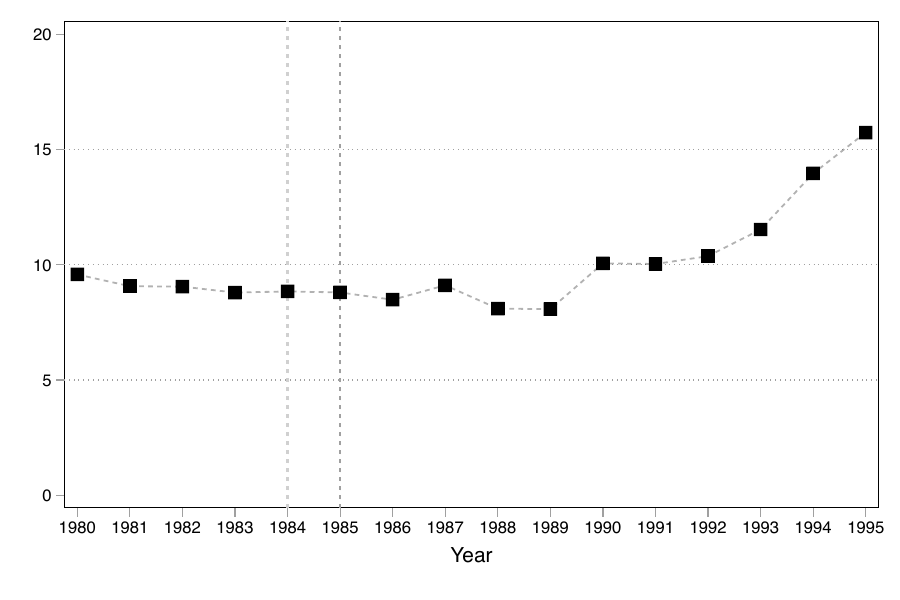}
		\end{subfigure}
		
		\vspace{0.5em}
		\begin{tablenotes}
			\footnotesize \vspace{1em}
			\parbox{\textwidth}{\justifying  \item \hspace{-0.2em}\textbf{Notes:} Panel (a) plots the yearly production in MMboe of non-EOR-eligible fields in the UK (blue dots) and in Norway (red triangles). Panels (b) and (c) split production of non-EOR-eligible fields based on cost per barrel. Panel (b) plots yearly production in MMboe of non-EOR-eligible fields in the UK (blue dots) and in Norway (red triangles) for \textit{high}-cost fields, defined as fields with operative costs of production (dollars per barrel) above median before 1985. Panel (c) plots yearly production in MMboe of non-EOR-eligible fields in the UK (blue dots) and in Norway (red triangles) for \textit{low}-cost fields, defined as fields with operative costs of production (dollars per barrel) below median before 1985.  Panel (d) plots the average yearly production of firms \textit{without} Norwegian EOR-eligible fields as of 1985. Vertical dotted lines mark the years when the Eidsivating Court of Appeal (1984) and the Norwegian Supreme Court (1985) delivered their rulings. }
		\end{tablenotes}
	\end{figure}

	\newpage
	\clearpage
	\FloatBarrier

	\setcounter{table}{0} \renewcommand{\thetable}{D\arabic{table}} %
	\setcounter{figure}{0} \renewcommand{\thefigure}{D\arabic{figure}}
	\setcounter{section}{0} \renewcommand{\thesection}{D}

	\section{Additional Exhibits on Market Shares and Portfolio Expansions}

	\begin{figure}[!h]
		\centering
		\captionsetup[subfigure]{justification=centering}
		\caption{Market shares \label{fig:concentration-firm}}
		\includegraphics[width=0.48 \linewidth]{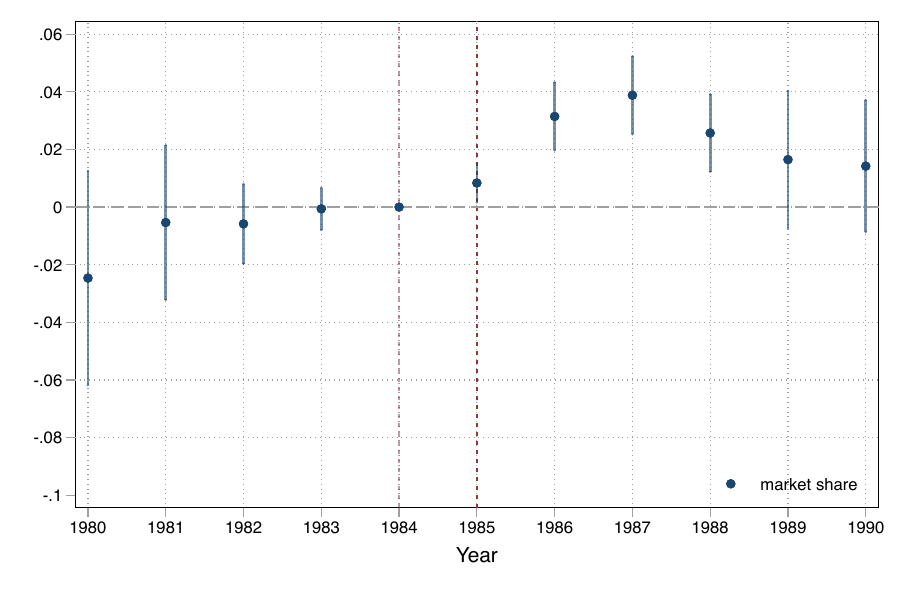}
		\vspace{.5em}
		\begin{minipage}{1 \textwidth}
			{\footnotesize\singlespacing  \textbf{Notes:} Estimated coefficients from regression \eqref{eq:event_firm} with market share as the dependent variable, augmented with controls for ``Share Norway'' interacted with ``Post'' and lagged CAPEX (in logs) to account for time varying difference in firms' size. 
				Fixed effects are at the firm and year level. Vertical bars around each estimate represent 95\% confidence intervals. Standard errors are clustered at the firm level.  Vertical dotted lines mark the years when the Eidsivating Court of Appeal (1984) and the Norwegian Supreme Court (1985) delivered their rulings.
				\par}
		\end{minipage}
	\end{figure}
	
	\begin{table}[!htbp]
		\centering
		\caption{Market shares and portfolio expansions (treatment starts in 1984)\label{tab:concentration-firm-1983}}
		{
\def\sym#1{\ifmmode^{#1}\else\(^{#1}\)\fi}
\begin{adjustbox}{max width={.7\textwidth},center}
\begin{tabular}{l*{5}{c}}
\toprule
                &\multicolumn{2}{c}{Market Share}&\multicolumn{3}{c}{Number of Fields}\\ \cmidrule(lr){2-3}\cmidrule(lr){4-6}
Dependent variable                &   All   & Positive     & All  & EOR & Non-EOR   \\
                & \multicolumn{1}{c}{}   &       & Fields  & Eligible & Eligible \\
                &\multicolumn{1}{c}{(1)}         &\multicolumn{1}{c}{(2)}         &\multicolumn{1}{c}{(3)}         &\multicolumn{1}{c}{(4)}         &\multicolumn{1}{c}{(5)}         \\
\midrule
Share EOR Norway $\times$ Post&    0.051\sym{**} &    0.074\sym{**} &   14.845\sym{***}&    8.335\sym{***}&    6.510\sym{***}\\
                &  (0.022)         &  (0.032)         &  (4.164)         &  (2.053)         &  (2.436)         \\
[1em]
Share EOR $\times$ Post&    0.003         &    0.003         &   -6.130\sym{***}&   -1.319\sym{**} &   -4.811\sym{***}\\
                &  (0.003)         &  (0.003)         &  (1.332)         &  (0.534)         &  (0.958)         \\
\midrule
Average dependent variable & 0.015 & 0.017 & 5.929 & 2.447 & 3.482 \\ 
\midrule
Year FE         &      Yes         &      Yes         &      Yes         &      Yes         &      Yes         \\
Firm FE         &      Yes         &      Yes         &      Yes         &      Yes         &      Yes         \\
Observations    &     1382         &     1212         &     1382         &     1382         &     1382         \\
R-squared       &     0.83         &     0.86         &     0.81         &     0.81         &     0.78         \\
\bottomrule
\multicolumn{6}{l}{* -- $p < 0.1$; ** -- $p < 0.05$; *** -- $p < 0.01$}
\end{tabular}
\end{adjustbox}
}

		\begin{tablenotes}
			\footnotesize 
			\parbox{\textwidth}{\justifying  \item \hspace{-0.2em}\textbf{Notes:} Estimated coefficients from regression \eqref{eq:did-firm}. ``Market share'' is a firm's market share of production in the North Sea. ``All Fields'' is a firm's total number of fields that were in exploration, development, or production in the North Sea in year $t$. The dependent variables in Columns 4-5 decompose the total number of fields in Column 3 between the number of EOR-eligible fields (Column 4) and the number of non-EOR-eligible fields (Column 5). The variable ``Share EOR Norway'' (``Share EOR'') refers to the fraction of Norwegian (North Sea) EOR-eligible fields among all the fields in which firm $i$ was active in the North Sea before 1984. The indicator ``Post'' equals 1 for the years after 1984, the year the Eidsivating Court of Appeal delivered its ruling, and 0 otherwise. Each regression includes firm and year fixed effects. Standard errors are clustered at the firm level.}
		\end{tablenotes}
	\end{table}

	\begin{table}[!htbp]
		\centering
		\caption{Market shares and portfolio expansions (different definition of EOR eligibility) \label{tab:concentration-firm-rob}}
		\centering
		{
\def\sym#1{\ifmmode^{#1}\else\(^{#1}\)\fi}
\begin{adjustbox}{width=.7\textwidth,center}
\begin{tabular}{l*{5}{c}}
\toprule
                &\multicolumn{2}{c}{Market Share}&\multicolumn{3}{c}{Number of Fields}\\ \cmidrule(lr){2-3}\cmidrule(lr){4-6}
Dependent variable                &   All   & Positive     & All  & EOR & Non-EOR   \\
                & \multicolumn{1}{c}{}   &       & Fields  & Eligible & Eligible \\
                &\multicolumn{1}{c}{(1)}         &\multicolumn{1}{c}{(2)}         &\multicolumn{1}{c}{(3)}         &\multicolumn{1}{c}{(4)}         &\multicolumn{1}{c}{(5)}         \\
               \midrule

Share EOR Norway $\times$ Post&    0.014         &    0.016         &    8.267\sym{**} &    4.037\sym{**} &    4.230\sym{*}  \\
                &  (0.017)         &  (0.039)         &  (3.932)         &  (1.838)         &  (2.287)         \\
[1em]
Share EOR $\times$ Post&    0.001         &    0.000         &   -7.203\sym{***}&   -1.637\sym{**} &   -5.566\sym{***}\\
                &  (0.003)         &  (0.002)         &  (1.511)         &  (0.632)         &  (1.049)         \\
\midrule
Average dependent variable & 0.015 & 0.017 & 5.858 & 2.410 & 3.449 \\
\midrule
Year FE         &      Yes         &      Yes         &      Yes         &      Yes         &      Yes         \\
Firm FE         &      Yes         &      Yes         &      Yes         &      Yes         &      Yes         \\
Observations    &     1406         &     1232         &     1406         &     1406         &     1406         \\
R-squared       &     0.82         &     0.84         &     0.81         &     0.79         &     0.79         \\
\bottomrule
\multicolumn{6}{l}{* -- $p < 0.1$; ** -- $p < 0.05$; *** -- $p < 0.01$}
\end{tabular}
\end{adjustbox}
}

		\vspace{-1ex}
		\begin{tablenotes}
			\footnotesize \vspace{1em}
			\parbox{\textwidth}{\justifying  \item \hspace{-0.2em}\textbf{Notes:} Estimated coefficients from regression \eqref{eq:did-firm}. ``Market share'' is a firm's market share of production in the North Sea. ``All Fields'' is a firm's total number of fields that were in exploration, development, or production in the North Sea in year $t$. The dependent variables in Columns 4-5 decompose the total number of fields in Column 3 between the number of EOR-eligible fields (Column 4) and the number of non-EOR-eligible fields (Column 5). The variable ``Share EOR Norway'' (``Share EOR'') refers to the fraction of Norwegian (North Sea) EOR-eligible fields among all the fields in which firm $i$ was active in the North Sea before 1985. EOR eligibility is measured according to the ``robust'' definition in Appendix Table \ref{tab:eor_restrictions}. The indicator ``Post'' equals 1 for the years after 1985, the year of the Supreme Court decision, and 0 otherwise. Each regression includes firm and year fixed effects. Standard errors are clustered at the firm level.}
		\end{tablenotes}
		
	\end{table}
	
	\begin{table}[!htbp]
		\centering
		\caption{Concentration in field ownership (treatment starts in 1984) \label{tab:concentration-field-1983}}
		\centering
		\begin{adjustbox}{width=.85\textwidth, center}
{
\def\sym#1{\ifmmode^{#1}\else\(^{#1}\)\fi}
\begin{tabular}{l*{6}{c}}
\toprule
                &\multicolumn{3}{c}{Share of}&\multicolumn{2}{c}{HHI}&\multicolumn{1}{c}{Number of}\\\cmidrule(lr){2-4}\cmidrule(lr){5-6}
Dependent variable                 &\multicolumn{1}{c}{Operator}         &\multicolumn{1}{c}{Top 1}         &\multicolumn{1}{c}{Top 3}       &\multicolumn{1}{c}{Standard}         &\multicolumn{1}{c}{Normalized}           &\multicolumn{1}{c}{Firms}         \\
                &\multicolumn{1}{c}{(1)}         &\multicolumn{1}{c}{(2)}         &\multicolumn{1}{c}{(3)}         &\multicolumn{1}{c}{(4)}         &\multicolumn{1}{c}{(5)}         &\multicolumn{1}{c}{(6)}         \\
\midrule
Norway $\times$ EOR $\times$ Post&    0.184\sym{*}         &    0.165  &    0.282\sym{*}  &    0.111         &    0.183\sym{*}  &   -0.298\sym{*}  \\
                &  (0.108)         &  (0.112)         &  (0.167)         &  (0.157)         &  (0.108)         &  (0.156)         \\
[1em]
EOR $\times$ Post &    0.014         &    0.013         &    0.076         &    0.060         &    0.047         &    0.173         \\
                &  (0.051)         &  (0.051)         &  (0.089)         &  (0.082)         &  (0.062)         &  (0.153)         \\
\midrule
Country-Year FE &      Yes         &      Yes         &      Yes         &      Yes         &      Yes         &      Yes         \\
Field FE        &      Yes         &      Yes         &      Yes         &      Yes         &      Yes         &      Yes         \\
Observations    &     1988         &     1988         &     1988         &     1988         &     1781         &     1988         \\
R-squared        &     0.59         &     0.47         &     0.45         &     0.64         &     0.52         &     0.96         \\
\bottomrule
\multicolumn{7}{l}{* -- $p < 0.1$; ** -- $p < 0.05$; *** -- $p < 0.01$}
\end{tabular}
}
\end{adjustbox} 
		\begin{tablenotes}
			\footnotesize \vspace{1em}
			\parbox{\textwidth}{\justifying  \item \hspace{-0.2em}\textbf{Notes:} Estimated coefficients from field-level regression \eqref{eq:did-field}. The dependent variables used in Columns 1-3 are the ownership share of the operator, the largest, and the combined shares of the three largest firms in a field. The dependent variables used in Columns 4-5 are the Herfindahl-Hirschman concentration index (HHI) computed in terms of asset interest shares at the field-level. ``Normalized HHI'' in Column 5 is computed as $\frac{\text{HHI} - 1/N}{1-1/N}$ and helps comparing the HHI across fields in which the number of firms $N$ vary. ``Number of Firms'' in Column 6 is the number of partnering firms in a specific field. The indicator variables ``Norway,'' ``EOR,'' and ``Post'' equal 1 for Norwegian fields, for EOR-eligible fields, and for the years after 1984, they year the Eidsivating Court of Appeal delivered its ruling, and 0 otherwise. Each regression includes country-by-year and field fixed effects. Standard errors are clustered at the field level.}
		\end{tablenotes}
	\end{table}

	\clearpage
	\newpage

	\setcounter{table}{0} \renewcommand{\thetable}{E\arabic{table}} %
	\setcounter{figure}{0} \renewcommand{\thefigure}{E\arabic{figure}}
	\setcounter{section}{0} \renewcommand{\thesection}{E}
	
	\section{More on Know-How and State Ownership} \label{apndx:add_tables_KH}

	\begin{table}[!htbp]\centering
		\caption{Leave-one-out robustness for Table \ref{tab:know-firm}}
		\label{tab:loo_know-firm}
		\begin{threeparttable}
			\setlength{\tabcolsep}{6pt}
			\begin{tabular}{lcccc}
				\toprule
				&\multicolumn{1}{c}{Market}&\multicolumn{3}{c}{Number of Fields}\\ \cmidrule(lr){3-5}
				Dependent variable  &   Share      & All  & EOR & Non-EOR   \\
				& \multicolumn{1}{c}{}   & Fields  & Eligible & Eligible \\
				&\multicolumn{1}{c}{(1)}         &\multicolumn{1}{c}{(2)}         &\multicolumn{1}{c}{(3)}         &\multicolumn{1}{c}{(4)}         \\
				\addlinespace[3ex] 
				\multicolumn{5}{l}{\textit{Panel (a): Leave-one-out without Conoco}}\\
				\midrule
				K-H \& S-O $\times$ Share EOR Norway $\times$ Post&       0.132***&      28.246***&       9.323***&      18.924** \\
				&     (0.025)   &     (9.843)   &     (2.809)   &     (7.813)   \\
				K-H Only $\times$ Share EOR Norway $\times$ Post&       0.086***&      62.992***&      15.599   &      47.393***\\
				&     (0.013)   &    (20.850)   &    (10.576)   &    (10.545)   \\[2ex]
				R-squared   &        0.88   &        0.83   &        0.80   &        0.81   \\
				\midrule\midrule
				\addlinespace[3ex]
				\multicolumn{5}{l}{\textit{Panel (b): Leave-one-out without Mobil}}\\
				\midrule
				K-H \& S-O $\times$ Share EOR Norway $\times$ Post&       0.132***&      28.189***&       9.289***&      18.900** \\
				&     (0.025)   &     (9.837)   &     (2.803)   &     (7.811)   \\
				K-H Only $\times$ Share EOR Norway $\times$ Post&       0.026***&      47.820***&      12.394   &      35.426***\\
				&     (0.009)   &    (15.862)   &     (7.701)   &     (8.311)   \\[2ex]
				R-squared   &        0.88   &        0.83   &        0.81   &        0.80   \\
				\midrule\midrule
				\addlinespace[3ex]
				\multicolumn{5}{l}{\textit{Panel (c): Leave-one-out without Shell}}\\
				\midrule
				K-H \& S-O $\times$ Share EOR Norway $\times$ Post&       0.132***&      28.347***&       9.384***&      18.963** \\
				&     (0.025)   &     (9.898)   &     (2.839)   &     (7.836)   \\
				K-H Only $\times$ Share EOR Norway $\times$ Post&       0.051** &      41.070***&       7.407** &      33.662***\\
				&     (0.024)   &     (6.276)   &     (2.847)   &     (4.398)   \\[2ex]
				R-squared   &        0.87   &        0.82   &        0.79   &        0.80   \\
				\midrule
				Year FE     &         Yes   &         Yes   &         Yes   &         Yes   \\
				Firm FE     &         Yes   &         Yes   &         Yes   &         Yes   \\
				Observations&        1356   &        1356   &        1356   &        1356   \\
				\bottomrule
				\multicolumn{5}{l}{* -- $p < 0.1$; ** -- $p < 0.05$; *** -- $p < 0.01$}\\
			\end{tabular}
			\vspace{0.3em}
			\begin{tablenotes}
				\footnotesize \vspace{1em}\footnotesize
				\parbox{\textwidth}{\justifying  \item \hspace{-0.2em}\textbf{Notes:} The table presents robustness checks for Table \ref{tab:know-firm} through a leave-one-out estimation dropping the ``K-H Only'' firm mentioned in each panel header. The bottom panel reports the fixed effects and the number of observations. Standard errors clustered at the firm level.}
			\end{tablenotes}
		\end{threeparttable}
	\end{table}

	\clearpage
	\subsection{Portfolio Diversification}\label{a:portfolio}
	We further investigate the portfolio diversification strategies of different types of firms in relation to fields' geological characteristics and their implied levels of risk. Appendix Table \ref{tab:know-risk-oil} reports estimates of regression \eqref{eq:know-how-firm} using as dependent variables the number of EOR-eligible fields (Columns 1-4) and of non-EOR-eligible fields (Columns 5-8) by geological type. The dependent variables in Columns 1-3 and 5-7 count the number of fields in each firm's portfolio that contain heavy oil, sour oil, or have deep wells, features commonly associated with high risk.\footnote{Column 1 of Table \ref{tab:know-risk-oil} is empty as the geology of EOR-eligible fields is incompatible with heavy oil.} Columns 4 and 8, instead, count the remaining number of fields that do not have these features and are commonly considered low risk.
	
	\begin{table}[!ht]
		\centering
		\caption{Portfolio diversification} \label{tab:know-risk-oil}
		\centering
		\begin{adjustbox}{width=1.1\textwidth,center}
{
\def\sym#1{\ifmmode^{#1}\else\(^{#1}\)\fi}
\begin{tabular}{l*{8}{c}}
\toprule
                &\multicolumn{4}{c}{Number of EOR-Eligible Fields}&\multicolumn{4}{c}{Number of Non-EOR-Eligible Fields}\\ \cmidrule(lr){2-5}\cmidrule(lr){6-9}
               Dependent variable &\multicolumn{1}{c}{Heavy}&\multicolumn{1}{c}{Sour}&\multicolumn{1}{c}{Deep}&\multicolumn{1}{c}{Other}&\multicolumn{1}{c}{Heavy}&\multicolumn{1}{c}{Sour}&\multicolumn{1}{c}{Deep}&\multicolumn{1}{c}{Other}\\
                &\multicolumn{1}{c}{Oil}&\multicolumn{1}{c}{Oil}&\multicolumn{1}{c}{Well}&\multicolumn{1}{c}{(Low Risk)}&\multicolumn{1}{c}{Oil}&\multicolumn{1}{c}{Oil}&\multicolumn{1}{c}{Well}&\multicolumn{1}{c}{(Low Risk)}\\
                &\multicolumn{1}{c}{(1)}         &\multicolumn{1}{c}{(2)}         &\multicolumn{1}{c}{(3)}         &\multicolumn{1}{c}{(4)}           &\multicolumn{1}{c}{(5)}         &\multicolumn{1}{c}{(6)}         &\multicolumn{1}{c}{(7)}         &\multicolumn{1}{c}{(8)}                 \\            
\midrule 
K-H \& S-O $\times$ Share EOR Norway $\times$ Post&    --         &    0.714         &    8.185\sym{***}&    1.140\sym{*}  &    0.856\sym{***}&    8.124         &   14.439\sym{*}  &    1.835\sym{***}\\
                &      --         &  (0.561)         &  (2.496)         &  (0.578)         &  (0.093)         &  (6.391)         &  (7.801)         &  (0.488)         \\
[1em]
K-H Only $\times$ Share EOR Norway $\times$ Post&    --         &    3.009\sym{***}&   10.555\sym{**} &    0.061         &    0.828         &   35.306\sym{***}&   29.809\sym{***}&    3.398         \\
                &      --         &  (0.786)         &  (4.350)         &  (0.634)         &  (0.781)         &  (5.706)         &  (6.851)         &  (2.296)         \\
[1em]
Share EOR Norway $\times$ Post&    --         &   -0.010         &    3.949\sym{*}  &    0.112         &   -0.150\sym{**} &   -0.203         &    0.321         &    0.339         \\
                &      --         &  (0.298)         &  (2.058)         &  (0.130)         &  (0.062)         &  (0.787)         &  (1.045)         &  (0.260)         \\
\midrule
Average dependent variable & -- & 0.580 & 2.015 & 0.097 & 0.060 & 1.864 & 2.356 & 0.156 \\
\midrule
Share EOR $\times$ Post& -- & Yes & Yes & Yes & Yes & Yes & Yes & Yes \\
Year FE         & -- & Yes & Yes & Yes & Yes & Yes & Yes & Yes \\
Firm FE         & -- & Yes & Yes & Yes & Yes & Yes & Yes & Yes \\
Observations    &     --         &     1377         &     1377         &     1377         &     1377         &     1377         &     1377         &     1377         \\
R-squared       &        --         &     0.77         &     0.81         &     0.78         &     0.33         &     0.78         &     0.77         &     0.70         \\
\bottomrule
\multicolumn{9}{l}{* -- $p < 0.1$; ** -- $p < 0.05$; *** -- $p < 0.01$}
\end{tabular}
}
\end{adjustbox}
		\begin{tablenotes}
			\footnotesize \vspace{1em}
			\parbox{\textwidth}{\justifying  \item \hspace{-0.2em}\textbf{Notes:}  Estimated coefficients from regression \eqref{eq:know-how-firm}. The dependent variables are the numbers of different types of EOR-eligible fields (Columns 1-4) and of different types of non-EOR-eligible fields (Columns 5-8) that were in firm $i$'s portfolio in year $t$. Each column focuses on different field types: heavy, sour, deep well, or other non-heavy, non-sour, non-deep. EOR cannot be implemented with heavy oil, hence Column 1 is empty. Firms are divided into three categories based on the indicator variables ``K-H'' (know-how) and ``S-O'' (state ownership). ``K-H'' equals 1 if firm $i$ was an operator in a field that adopted EOR before 1985, while ``S-O'' equals 1 if firm $i$ was state owned during 1975-1995. Because some firms possessed both know-how and were state owned, we create the variables ``K-H \& S-O,'' an indicator for a state-owned firm with know-how and ``K-H Only,'' an indicator for a non-state-owned firm with know-how. ``Share EOR Norway'' is the pre-1985 share of firm $i$'s fields that were eligible for EOR. The indicator ``Post'' equals 1 for the years after 1985, the year of the Supreme Court decision, and 0 otherwise. All regressions control for ``Share EOR $\cdot$ Post,'' which is firm $i$'s share of EOR-eligible fields in the North Sea before 1985 times the post-1985 indicator, and firm and year fixed effects. The two state-owned firms that did not have know-how are excluded from all regressions (Britoil and Norway State DFI). Standard errors are clustered at the firm level.}
		\end{tablenotes}
	\end{table} 
	
	The results of Appendix Table \ref{tab:know-risk-oil} complement those of Table \ref{tab:know-firm}, clarifying that the aggressive portfolio expansion of ``K-H Only'' firms took place primarily toward high-risk fields. Compared to ``K-H \& S-O'' firms, ``K-H Only'' firms not only specialized more in EOR-eligible fields, but also diversified more by entering fields with different geological characteristics, mostly high risk. In particular, Columns 2-3 and 6-7 indicate that ``K-H Only'' firms with a pre-1985 portfolio of fields that included 10 p.p. more Norwegian EOR-eligible fields expanded their portfolios by around $0.3$ EOR-eligible and $3.53$ non-EOR-eligible fields with sour oil, and by around $1.05$ EOR-eligible and $2.98$ non-EOR-eligible fields with deep wells. In contrast, ``K-H \& S-O'' firms with similar pre-1985 portfolios expanded their portfolios moderately in terms of high-risk fields, by around $0.81$ non-EOR-eligible fields with sour oil (not significant), and by around $0.81$ EOR-eligible and $1.44$ non-EOR-eligible fields with deep wells, while they entered more aggressively into fields with a low-risk profile (both EOR-eligible and non-EOR-eligible fields). In contrast, firms without know-how exhibited only a modest expansion into deep‑well (EOR‑eligible) fields while simultaneously reducing their exposure to heavy‑oil fields, which are generally ineligible for EOR, suggesting little diversification.\footnote{Panel (b) of Appendix Table \ref{tab:multiple_panels} shows similar results including ``$\text{Share Norway}_i\cdot \text{Post}_t$'' as an additional regressor. We also obtain similar results when including state-owned firms without know-how as a separate category in Panel (b) of Appendix table \ref{tab:multiple_panels_SO}.}
	
	\newpage
	\subsection{Other Results}
	
	\begin{table}[!htbp]\vspace{-0.6cm}
		\centering
		\caption{Analyses in Tables \ref{tab:know-firm}, \ref{tab:know-risk-oil}, and \ref{tab:know-risk} controlling for ``Share Norway $\cdot$ Post''}
		\label{tab:multiple_panels}
		\centering
		\begin{adjustbox}{width=.92\textwidth,center}
{
\def\sym#1{\ifmmode^{#1}\else\(^{#1}\)\fi}
\begin{tabular}{c}

    \vspace{0.5em}
    \begin{tabular}{l*{4}{c}}
    \multicolumn{5}{c}{\textbf{Panel (a): Replication of Table \ref{tab:know-firm}}}\\
    \toprule\toprule
        &\multicolumn{1}{c}{Market}&\multicolumn{3}{c}{Number of Fields}\\ \cmidrule(lr){3-5}
Dependent variable        &   Share      & All  & EOR & Non-EOR   \\
        & \multicolumn{1}{c}{}   & Fields  & Eligible & Eligible \\
        &\multicolumn{1}{c}{(1)}         &\multicolumn{1}{c}{(2)}         &\multicolumn{1}{c}{(3)}         &\multicolumn{1}{c}{(4)}         \\
\midrule
    K-H \& S-O $\times$ Share EOR Norway $\times$ Post&    0.136\sym{***}&   23.868\sym{**} &    6.107\sym{*}  &   17.761\sym{**} \\
                &  (0.029)         & (11.517)         &  (3.298)         &  (8.835)         \\
[1em]
K-H Only $\times$ Share EOR Norway $\times$ Post&    0.050\sym{**} &   48.537\sym{***}&   11.312\sym{**} &   37.225\sym{***}\\
                &  (0.023)         & (10.443)         &  (4.564)         &  (6.358)         \\
[1em]
Share EOR Norway $\times$ Post&    0.015         &   -2.162         &   -1.434         &   -0.728         \\
                &  (0.011)         &  (6.591)         &  (3.072)         &  (3.867)         \\
[1em]
Share EOR $\times$ Post&    0.000         &   -5.994\sym{***}&   -0.916\sym{*}  &   -5.078\sym{***}\\
                &  (0.002)         &  (1.389)         &  (0.473)         &  (1.113)         \\
[1em]
Share Norway $\times$ Post    &   -0.005         &    5.974         &    4.432\sym{**} &    1.542         \\
                &  (0.008)         &  (5.201)         &  (2.198)         &  (3.261)         \\
    \midrule
Year FE         &      Yes         &      Yes         &      Yes         &      Yes         \\
Firm FE         &      Yes         &      Yes         &      Yes         &      Yes         \\
Observations    &     1377         &     1377         &     1377         &     1377         \\
R-squared       &     0.88         &     0.83         &     0.82         &     0.81         \\
\bottomrule\bottomrule
    \end{tabular}
    \vspace{1em} \\
    
    \vspace{0.5em}
    \begin{tabular}{l*{8}{c}}
    \multicolumn{9}{c}{\textbf{Panel (b): Replication of Appendix Table \ref{tab:know-risk-oil}}}\\
\toprule\toprule
                &\multicolumn{4}{c}{Number of EOR Eligible Fields:}&\multicolumn{4}{c}{Number of Non-EOR Eligible Fields:}\\ \cmidrule(lr){2-5}\cmidrule(lr){6-9}
                Dependent variable &\multicolumn{1}{c}{Heavy}&\multicolumn{1}{c}{Sour}&\multicolumn{1}{c}{Deep}&\multicolumn{1}{c}{Other}&\multicolumn{1}{c}{Heavy}&\multicolumn{1}{c}{Sour}&\multicolumn{1}{c}{Deep}&\multicolumn{1}{c}{Other}\\
                &\multicolumn{1}{c}{Oil}&\multicolumn{1}{c}{Oil}&\multicolumn{1}{c}{Well}&\multicolumn{1}{c}{(Lower Risk)}&\multicolumn{1}{c}{Oil}&\multicolumn{1}{c}{Oil}&\multicolumn{1}{c}{Well}&\multicolumn{1}{c}{(Lower Risk)}\\
                &\multicolumn{1}{c}{(1)}         &\multicolumn{1}{c}{(2)}         &\multicolumn{1}{c}{(3)}         &\multicolumn{1}{c}{(4)}           &\multicolumn{1}{c}{(5)}         &\multicolumn{1}{c}{(6)}         &\multicolumn{1}{c}{(7)}         &\multicolumn{1}{c}{(8)}                 \\
                \midrule
    K-H \& S-O $\times$ Sh. Eligible Norw. Fields $\times$ Post&    --         &    0.776         &    5.019\sym{*}  &    1.138\sym{*}  &    0.914\sym{***}&    8.292         &   14.071         &    1.487\sym{**} \\
                &      --         &  (0.670)         &  (2.955)         &  (0.639)         &  (0.114)         &  (7.002)         &  (8.632)         &  (0.592)         \\
[1em]
K-H Only $\times$ Sh. Eligible Norw. Fields $\times$ Post&    --         &    2.994\sym{***}&   11.319\sym{***}&    0.062         &    0.814         &   35.265\sym{***}&   29.898\sym{***}&    3.482         \\
                &      --         &  (0.796)         &  (4.045)         &  (0.635)         &  (0.790)         &  (5.737)         &  (6.950)         &  (2.229)         \\
[1em]
Sh. Eligible Norw. Fields $\times$ Post&    --         &    0.096         &   -1.525         &    0.108         &   -0.049         &    0.088         &   -0.315         &   -0.263         \\
                &      --         &  (0.463)         &  (2.983)         &  (0.141)         &  (0.098)         &  (2.265)         &  (3.035)         &  (0.364)         \\
[1em]
Sh. Eligible North Sea Fields $\times$ Post&    --         &   -0.162         &   -0.871\sym{*}  &   -0.039         &    0.055         &   -3.133\sym{***}&   -4.243\sym{***}&   -0.108\sym{*}  \\
                &      --         &  (0.211)         &  (0.467)         &  (0.033)         &  (0.053)         &  (0.871)         &  (1.006)         &  (0.063)         \\
[1em]
Share Norway $\times$ Post    &    --         &   -0.085         &    4.411\sym{**} &    0.003         &   -0.081         &   -0.234         &    0.512         &    0.485         \\
                &      --         &  (0.399)         &  (2.152)         &  (0.142)         &  (0.086)         &  (2.034)         &  (2.626)         &  (0.301)         \\
\midrule
Year FE         &      --         &      Yes         &      Yes         &      Yes         &      Yes         &      Yes         &      Yes         &      Yes         \\
Firm FE         &      --         &      Yes         &      Yes         &      Yes         &      Yes         &      Yes         &      Yes         &      Yes         \\
Observations    &     --         &     1377         &     1377         &     1377         &     1377         &     1377         &     1377         &     1377         \\
R-squared       &        --         &     0.77         &     0.82         &     0.78         &     0.33         &     0.78         &     0.77         &     0.71         \\
    \bottomrule\bottomrule
    \end{tabular}
    \vspace{1em} \\
    
    \vspace{0.5em}
    \begin{tabular}{l*{7}{c}}
    \multicolumn{7}{c}{\textbf{Panel (c): Replication of Panel (b) of Table \ref{tab:know-risk}}}\\
\toprule\toprule
                &\multicolumn{6}{c}{Diversification in Technology Portfolio (1-HHI)} \\ 
                \cmidrule(lr){2-7}
             Dependent variable   &\multicolumn{1}{c}{Exploration}&\multicolumn{1}{c}{Extraction}&\multicolumn{1}{c}{Well }&\multicolumn{1}{c}{Field} &\multicolumn{1}{c}{Tariff} &\multicolumn{1}{c}{CAPEX}\\
                &\multicolumn{1}{c}{\& Prod. Tech.}&\multicolumn{1}{c}{Technologies}&\multicolumn{1}{c}{ Depth}&\multicolumn{1}{c}{Sizes} &\multicolumn{1}{c}{Intensity} &\multicolumn{1}{c}{Intensity}\\                
                &\multicolumn{1}{c}{(1)}         &\multicolumn{1}{c}{(2)}         &\multicolumn{1}{c}{(3)}         &\multicolumn{1}{c}{(4)}         &\multicolumn{1}{c}{(5)}         &\multicolumn{1}{c}{(6)} \\ \midrule
                
    K-H \& S-O $\times$ Share EOR Norway $\times$ Post&    0.895\sym{**} &    0.426         &    0.621         &    0.918\sym{**} &    0.534         &    0.804\sym{*}  \\
                &  (0.393)         &  (0.439)         &  (0.457)         &  (0.380)         &  (0.390)         &  (0.451)         \\
[1em]
K-H Only $\times$ Share EOR Norway $\times$ Post&    1.249\sym{***}&    1.828\sym{***}&    1.592\sym{***}&    2.047\sym{***}&    1.558\sym{***}&    1.268\sym{***}\\
                &  (0.312)         &  (0.306)         &  (0.385)         &  (0.283)         &  (0.316)         &  (0.432)         \\
[1em]
Share EOR Norway $\times$ Post&   -0.402         &   -0.591         &   -0.558         &   -0.427         &   -0.567         &   -0.341         \\
                &  (0.497)         &  (0.507)         &  (0.537)         &  (0.490)         &  (0.486)         &  (0.493)         \\
[1em]
Share EOR $\times$ Post&   -0.262\sym{**} &   -0.112         &   -0.208\sym{**} &   -0.188\sym{*}  &   -0.162         &   -0.208\sym{**} \\
                &  (0.108)         &  (0.095)         &  (0.100)         &  (0.104)         &  (0.101)         &  (0.100)         \\
[1em]
Share Norway $\times$ Post    &   -0.103         &    0.114         &    0.100         &   -0.037         &    0.011         &   -0.092         \\
                &  (0.367)         &  (0.377)         &  (0.402)         &  (0.366)         &  (0.366)         &  (0.362)         \\
\midrule
Year FE         &      Yes         &      Yes         &      Yes         &      Yes         &      Yes         &      Yes         \\
Firm FE         &      Yes         &      Yes         &      Yes         &      Yes         &      Yes         &      Yes         \\
Observations    &     1377         &     1377         &     1377         &     1377         &     1377         &     1377         \\
R-squared       &     0.37         &     0.32         &     0.35         &     0.33         &     0.34         &     0.34         \\
    \bottomrule\bottomrule
    \end{tabular}
    \end{tabular}}

    \end{adjustbox}
		\begin{tablenotes}
			\footnotesize \vspace{-0.2em}
			\parbox{\textwidth}{\justifying  \item \hspace{-0.2em}\textbf{Notes:} Replications controlling for ``Share Norway $\cdot$ Post,'' which is excluded from our baseline analyses because of multicollinearity with ``Share EOR Norway $\cdot$ Post'' (correlation of 0.8). ``Share Norway'' is the share of Norwegian fields in which firm $i$ was active in the North Sea before 1985. Because of space constraints, Panel (c) replicates only Panel (b) of Table \ref{tab:know-risk}. Standard errors are clustered at the firm level.}
		\end{tablenotes}
	\end{table}

	\begin{table}[!htbp]
		\centering
		\caption{Analyses in Tables \ref{tab:know-firm}, \ref{tab:know-risk-oil}, and \ref{tab:know-risk} including ``S-O Only''}
		\label{tab:multiple_panels_SO}
		\centering
		\begin{adjustbox}{width=.92\textwidth,center}
{
\def\sym#1{\ifmmode^{#1}\else\(^{#1}\)\fi}
\begin{tabular}{c}

    \vspace{0.5em}
    \begin{tabular}{l*{4}{c}}
    \multicolumn{5}{c}{\textbf{Panel (a): Replication of Table \ref{tab:know-firm}}}\\
    \toprule\toprule
        &\multicolumn{1}{c}{Market}&\multicolumn{3}{c}{Number of Fields}\\ \cmidrule(lr){3-5}
Dependent variable        &   Share      & All  & EOR & Non-EOR   \\
        & \multicolumn{1}{c}{}   & Fields  & Eligible & Eligible \\
        &\multicolumn{1}{c}{(1)}         &\multicolumn{1}{c}{(2)}         &\multicolumn{1}{c}{(3)}         &\multicolumn{1}{c}{(4)}         \\
\midrule
K-H \& S-O $\times$ Share EOR Norway $\times$ Post&    0.132\sym{***}&   28.083\sym{***}&    9.253\sym{***}&   18.830\sym{**} \\
                &  (0.025)         &  (9.807)         &  (2.788)         &  (7.796)         \\
[1em]
K-H Only $\times$ Share EOR Norway $\times$ Post&    0.050\sym{**} &   47.330\sym{***}&   10.469\sym{**} &   36.861\sym{***}\\
                &  (0.023)         & (10.435)         &  (4.844)         &  (6.225)         \\
[1em]
S-O Only $\times$ Share EOR Norway $\times$ Post&    0.127\sym{***}&   10.272\sym{***}&    5.193\sym{**} &    5.078\sym{***}\\
                &  (0.004)         &  (3.297)         &  (2.172)         &  (1.272)         \\
[1em]
Share EOR Norway $\times$ Post&    0.008\sym{**} &    5.226         &    4.061\sym{*}  &    1.165         \\
                &  (0.004)         &  (3.254)         &  (2.127)         &  (1.343)         \\
\midrule
Share EOR $\times$ Post&      Yes         &      Yes         &      Yes         &      Yes         \\
Year FE         &      Yes         &      Yes         &      Yes         &      Yes         \\
Firm FE         &      Yes         &      Yes         &      Yes         &      Yes         \\
Observations    &     1406         &     1406         &     1406         &     1406         \\
R-squared        &     0.86         &     0.83         &     0.81         &     0.81         \\
\bottomrule\bottomrule
    \end{tabular}
    \vspace{1em} \\
    
    \vspace{0.5em}
    \begin{tabular}{l*{8}{c}}
    \multicolumn{9}{c}{\textbf{Panel (b): Replication of Appendix Table \ref{tab:know-risk-oil}}}\\
\toprule\toprule
                &\multicolumn{4}{c}{Number of EOR Eligible Fields}&\multicolumn{4}{c}{Number of Non-EOR Eligible Fields}\\ \cmidrule(lr){2-5}\cmidrule(lr){6-9}
                Dependent variable &\multicolumn{1}{c}{Heavy}&\multicolumn{1}{c}{Sour}&\multicolumn{1}{c}{Deep}&\multicolumn{1}{c}{Other}&\multicolumn{1}{c}{Heavy}&\multicolumn{1}{c}{Sour}&\multicolumn{1}{c}{Deep}&\multicolumn{1}{c}{Other}\\
                &\multicolumn{1}{c}{Oil}&\multicolumn{1}{c}{Oil}&\multicolumn{1}{c}{Well}&\multicolumn{1}{c}{(Lower Risk)}&\multicolumn{1}{c}{Oil}&\multicolumn{1}{c}{Oil}&\multicolumn{1}{c}{Well}&\multicolumn{1}{c}{(Lower Risk)}\\
                &\multicolumn{1}{c}{(1)}         &\multicolumn{1}{c}{(2)}         &\multicolumn{1}{c}{(3)}         &\multicolumn{1}{c}{(4)}           &\multicolumn{1}{c}{(5)}         &\multicolumn{1}{c}{(6)}         &\multicolumn{1}{c}{(7)}         &\multicolumn{1}{c}{(8)}                 \\           
\midrule 
K-H \& S-O $\times$ Share EOR Norway $\times$ Post&    --         &    0.704         &    8.150\sym{***}&    1.140\sym{*}  &    0.856\sym{***}&    8.078         &   14.399\sym{*}  &    1.835\sym{***}\\
                &      --         &  (0.559)         &  (2.485)         &  (0.577)         &  (0.093)         &  (6.383)         &  (7.793)         &  (0.488)         \\
[1em]
K-H Only $\times$ Share EOR Norway $\times$ Post&    --         &    2.988\sym{***}&   10.481\sym{**} &    0.060         &    0.829         &   35.194\sym{***}&   29.703\sym{***}&    3.400         \\
                &      --         &  (0.782)         &  (4.335)         &  (0.634)         &  (0.781)         &  (5.681)         &  (6.825)         &  (2.295)         \\
[1em]
S-O Only $\times$ Share EOR Norway $\times$ Post&    --         &    0.455         &    4.178\sym{**} &    1.035\sym{***}&   -0.027         &   -0.328         &    2.814\sym{***}&    0.763\sym{***}\\
                &      --         &  (0.281)         &  (2.102)         &  (0.116)         &  (0.047)         &  (0.662)         &  (0.954)         &  (0.261)         \\
[1em]
Share EOR Norway $\times$ Post&    --         &   -0.010         &    3.946\sym{*}  &    0.111         &   -0.150\sym{**} &   -0.221         &    0.298         &    0.340         \\
                &      --         &  (0.297)         &  (2.049)         &  (0.130)         &  (0.062)         &  (0.785)         &  (1.039)         &  (0.260)         \\             
\midrule
Share EOR $\times$ Post & -- & Yes & Yes & Yes & Yes & Yes & Yes & Yes \\
Year FE         & -- & Yes & Yes & Yes & Yes & Yes & Yes & Yes \\
Firm FE         & -- & Yes & Yes & Yes & Yes & Yes & Yes & Yes \\
Observations    &     --         &     1406         &     1406         &     1406         &     1406         &     1406         &     1406         &     1406         \\
R-squared       &        --         &     0.77         &     0.81         &     0.81         &     0.32         &     0.77         &     0.77         &     0.70         \\
\bottomrule\bottomrule
    \end{tabular}
    \vspace{1em} \\
    
    \vspace{0.5em}
    \begin{tabular}{l*{7}{c}}
    \multicolumn{7}{c}{\textbf{Panel (c): Replication of Table \ref{tab:know-risk}}}\\
\toprule\toprule
                &\multicolumn{6}{c}{Diversification in Technology Portfolio ($1-$HHI)} \\ 
                \cmidrule(lr){2-7}
                Dependent variable &\multicolumn{1}{c}{Exploration}&\multicolumn{1}{c}{Extraction}&\multicolumn{1}{c}{Well }&\multicolumn{1}{c}{Field} &\multicolumn{1}{c}{Tariff} &\multicolumn{1}{c}{CAPEX}\\
                &\multicolumn{1}{c}{\& Prod. Tech.}&\multicolumn{1}{c}{Technologies}&\multicolumn{1}{c}{ Depth}&\multicolumn{1}{c}{Sizes} &\multicolumn{1}{c}{Intensity} &\multicolumn{1}{c}{Intensity}\\                
                &\multicolumn{1}{c}{(1)}         &\multicolumn{1}{c}{(2)}         &\multicolumn{1}{c}{(3)}         &\multicolumn{1}{c}{(4)}         &\multicolumn{1}{c}{(5)}         &\multicolumn{1}{c}{(6)} \\ 

\multicolumn{7}{l}{\textit{Panel (a): Baseline}}  \\   
\midrule 
Share EOR Norway $\times$ Post&   -1.491\sym{**} &   -1.264\sym{**} &   -1.069\sym{*}  &   -1.251\sym{**} &   -1.500\sym{***}&   -1.134\sym{**} \\
                &  (0.581)         &  (0.530)         &  (0.575)         &  (0.585)         &  (0.466)         &  (0.542)         \\
[1em]
R-squared       &     0.37         &     0.31         &     0.35         &     0.32         &     0.34         &     0.33         \\
\midrule\midrule
\rule{0pt}{4ex}  \\[-3ex] 
\multicolumn{7}{l}{\textit{Panel (b): Know-How (K-H) \& State-Owned (S-O)} }       \\   
\midrule 
K-H \& S-O $\times$ Share EOR Norway $\times$ Post&    2.287\sym{***}&    1.391         &    1.906\sym{**} &    2.519\sym{***}&    1.483\sym{**} &    2.031\sym{**} \\
                &  (0.674)         &  (0.853)         &  (0.857)         &  (0.649)         &  (0.695)         &  (0.883)         \\
[1em]
K-H Only $\times$ Share EOR Norway $\times$ Post&    3.520\sym{***}&    4.996\sym{***}&    4.332\sym{***}&    5.809\sym{***}&    4.278\sym{***}&    3.523\sym{***}\\
                &  (0.848)         &  (0.916)         &  (1.133)         &  (0.812)         &  (0.888)         &  (1.174)         \\
[1em]
S-O Only $\times$ Share EOR Norway $\times$ Post&   -2.161\sym{***}&   -1.867\sym{***}&   -1.024         &   -2.009\sym{***}&   -1.464\sym{***}&   -1.001\sym{*}  \\
                &  (0.578)         &  (0.549)         &  (0.638)         &  (0.527)         &  (0.469)         &  (0.592)         \\                
[1em]
R-squared       &     0.39         &     0.33         &     0.39         &     0.33         &     0.39         &     0.33         \\
\midrule\midrule
Share EOR $\times$ Post&      Yes         &      Yes         &      Yes         &      Yes         &      Yes         &      Yes         \\
Year FE         &      Yes         &      Yes         &      Yes         &      Yes         &      Yes         &      Yes         \\
Firm FE         &      Yes         &      Yes         &      Yes         &      Yes         &      Yes         &      Yes         \\
Observations    &     1406         &     1406         &     1406         &     1406         &     1406         &     1406         \\
\bottomrule\bottomrule
    \end{tabular}
    \end{tabular}}

    \end{adjustbox}

		\begin{tablenotes}
			\footnotesize 
			\parbox{\textwidth}{\justifying  \item \hspace{-0.2em}\textbf{Notes:} Replications including the ``S-O Only'' treatment (Britoil and Norway State DFI) which is excluded from our baseline analyses. Standard errors are clustered at the firm level.}
		\end{tablenotes}
	\end{table}

	\begin{table}
		\centering
		\caption{Oil majors, market shares, and portfolio expansions} \label{tab:know-firm-om}
		\centering
		\begin{adjustbox}{max width={.9\textwidth},center}
{
\def\sym#1{\ifmmode^{#1}\else\(^{#1}\)\fi}
\begin{tabular}{l*{4}{c}}
\toprule
        &\multicolumn{1}{c}{Market}&\multicolumn{3}{c}{Number of Fields}\\ \cmidrule(lr){3-5}
      Dependent variable  &   Share      & All  & EOR & Non-EOR   \\
        & \multicolumn{1}{c}{}   & Fields  & Eligible & Eligible \\
        &\multicolumn{1}{c}{(1)}         &\multicolumn{1}{c}{(2)}         &\multicolumn{1}{c}{(3)}         &\multicolumn{1}{c}{(4)}         \\  \\

\multicolumn{5}{l}{\textit{Panel (a): Regression \eqref{eq:know-how-firm} with ``Oil Major'' instead of ``K-H'' and ``S-O''}} \\        
        \midrule
Oil Major $\times$ Share EOR Norway $\times$ Post&   -0.027         &    2.381         &   -3.163         &    5.544         \\
                &  (0.023)         & (12.024)         &  (4.303)         &  (8.109)         \\
[1em]
Share EOR Norway $\times$ Post&    0.031         &    9.031\sym{*}  &    6.209\sym{**} &    2.822         \\
                &  (0.021)         &  (5.250)         &  (2.880)         &  (2.597)         \\
[1em]
Share EOR $\times$ Post&   -0.000         &   -7.553\sym{***}&   -1.813\sym{***}&   -5.740\sym{***}\\
                &  (0.003)         &  (1.487)         &  (0.610)         &  (1.041)         \\
[1em]
Observations    &     1377         &     1377         &     1377         &     1377         \\
R-squared       &     0.87         &     0.81         &     0.80         &     0.79         \\
\midrule\midrule
\rule{0pt}{4ex}  \\[-3ex] 

\multicolumn{5}{l}{\textit{Panel (b): as in Panel (a) but controlling for ``Share Norway $\cdot$ Post''}} \\
\midrule
Oil Major $\times$ Share EOR Norway $\times$ Post&   -0.027         &    5.176         &   -1.444         &    6.620         \\
                &  (0.021)         & (11.384)         &  (3.769)         &  (7.883)         \\
[1em]
Share EOR Norway $\times$ Post&    0.029         &   -1.536         &   -0.292         &   -1.245         \\
                &  (0.021)         &  (8.520)         &  (3.868)         &  (4.919)         \\
[1em]
Share EOR $\times$ Post&    0.000         &   -6.208\sym{***}&   -0.986\sym{**} &   -5.222\sym{***}\\
                &  (0.002)         &  (1.436)         &  (0.488)         &  (1.138)         \\
[1em]
Share Norway $\times$ Post&    0.002         &    7.505         &    4.617\sym{**} &    2.888         \\
                &  (0.011)         &  (5.380)         &  (2.190)         &  (3.374)         \\
[1em]
Observations    &     1377         &     1377         &     1377         &     1377         \\
R-squared       &     0.87         &     0.82         &     0.82         &     0.79         \\
\midrule\midrule 
Average dependent variable & 0.012 &  5.216 & 2.116 & 3.100 \\
\midrule
Year FE         &      Yes         &      Yes         &      Yes         &      Yes         \\
Firm FE         &      Yes         &      Yes         &      Yes         &      Yes         \\
\bottomrule
\multicolumn{5}{l}{* -- $p < 0.1$; ** -- $p < 0.05$; *** -- $p < 0.01$}
\end{tabular}
}
\end{adjustbox}

		\begin{tablenotes}
			\footnotesize \vspace{1em}
			\parbox{\textwidth}{\justifying  \item \hspace{-0.2em}\textbf{Notes:} Panel (a) presents estimated coefficients from regression \eqref{eq:know-how-firm} where we use the indicator variable ``Oil Major'' to distinguish oil majors from the other firms. The oil majors include the so-called ``seven sisters'' (Exxon, Mobil, Chevron, Texaco, Gulf Oil, Royal Dutch Shell, and BP) as well as other very large players of the time (Total, Eni, Phillips Petroleum, Conoco, Amoco). Panel (b) additionally controls for ``Share Norway $\cdot$ Post,'' which is firm $i$'s share of EOR-eligible fields in the North Sea before 1985 times the post-1985 indicator. The dependent variables are firm $i$'s market share (Column 1), and the number of fields (Column 2), number of EOR-eligible fields (Column 3), and number of non-EOR-eligible fields (Column 4) in $i$'s portfolio in year $t$. All regressions control for firm and year fixed-effects. Standard errors are clustered at the firm level.}
		\end{tablenotes}
	\end{table}

	\begin{table}
		\centering
		\caption{Oil majors and portfolio diversification} \label{tab:know-risk-oil-om}
		\centering
		\begin{adjustbox}{width=1.05\textwidth,center}
{
\def\sym#1{\ifmmode^{#1}\else\(^{#1}\)\fi}
\begin{tabular}{l*{8}{c}}
\toprule
                &\multicolumn{4}{c}{Number of EOR Eligible Fields}&\multicolumn{4}{c}{Number of Non-EOR Eligible Fields}\\ \cmidrule(lr){2-5}\cmidrule(lr){6-9}
     Dependent variable           &\multicolumn{1}{c}{Heavy}&\multicolumn{1}{c}{Sour}&\multicolumn{1}{c}{Deep}&\multicolumn{1}{c}{Other}&\multicolumn{1}{c}{Heavy}&\multicolumn{1}{c}{Sour}&\multicolumn{1}{c}{Deep}&\multicolumn{1}{c}{Other}\\
                &\multicolumn{1}{c}{Oil}&\multicolumn{1}{c}{Oil}&\multicolumn{1}{c}{Well}&\multicolumn{1}{c}{(Lower Risk)}&\multicolumn{1}{c}{Oil}&\multicolumn{1}{c}{Oil}&\multicolumn{1}{c}{Well}&\multicolumn{1}{c}{(Lower Risk)}\\
                &\multicolumn{1}{c}{(1)}         &\multicolumn{1}{c}{(2)}         &\multicolumn{1}{c}{(3)}         &\multicolumn{1}{c}{(4)}           &\multicolumn{1}{c}{(5)}         &\multicolumn{1}{c}{(6)}         &\multicolumn{1}{c}{(7)}         &\multicolumn{1}{c}{(8)}                 \\      \\

\multicolumn{9}{l}{\textit{Panel (a): Regression \eqref{eq:know-how-firm} with ``Oil Major'' instead of ``K-H'' and ``S-O''}} \\        
        \midrule
Oil Major $\times$ Share EOR Norway $\times$ Post&    --         &    0.422         &   -3.104         &   -0.121         &    0.013         &    7.790         &    5.476         &   -0.042         \\
                &      --         &  (0.775)         &  (4.091)         &  (0.282)         &  (0.222)         &  (6.852)         &  (7.176)         &  (0.804)         \\
[1em]
Share EOR Norway $\times$ Post&    --         &    0.049         &    5.967\sym{**} &    0.261         &   -0.040         &   -0.307         &    1.291         &    0.649\sym{*}  \\
                &      --         &  (0.349)         &  (2.746)         &  (0.187)         &  (0.127)         &  (1.027)         &  (1.791)         &  (0.390)         \\
[1em]
Share EOR $\times$ Post&    --         &   -0.170         &   -1.754\sym{***}&   -0.050         &    0.056         &   -3.380\sym{***}&   -4.647\sym{***}&   -0.226\sym{***}\\
                &      --         &  (0.188)         &  (0.603)         &  (0.032)         &  (0.051)         &  (0.768)         &  (0.899)         &  (0.074)         \\
[1em]
Observations    &     --         &     1377         &     1377         &     1377         &     1377         &     1377         &     1377         &     1377         \\
R-squared       &        --         &     0.76         &     0.80         &     0.77         &     0.31         &     0.75         &     0.75         &     0.67         \\
\midrule\midrule
\rule{0pt}{4ex}  \\[-3ex] 

\multicolumn{9}{l}{\textit{Panel (b): as in Panel (a) but controlling for ``Share Norway $\cdot$ Post''}} \\
\midrule
Oil Major $\times$ Share EOR Norway $\times$ Post&    --         &    0.410         &   -1.420         &   -0.093         &    0.003         &    8.006         &    6.087         &    0.167         \\
                &      --         &  (0.781)         &  (3.571)         &  (0.260)         &  (0.220)         &  (6.782)         &  (7.041)         &  (0.748)         \\
[1em]
Share EOR Norway $\times$ Post&    --         &    0.095         &   -0.400         &    0.156         &   -0.003         &   -1.123         &   -1.020         &   -0.143         \\
                &      --         &  (0.563)         &  (3.707)         &  (0.202)         &  (0.142)         &  (2.640)         &  (3.732)         &  (0.486)         \\
[1em]
Share EOR $\times$ Post&    --         &   -0.176         &   -0.943\sym{*}  &   -0.037         &    0.052         &   -3.276\sym{***}&   -4.353\sym{***}&   -0.125\sym{**} \\
                &      --         &  (0.212)         &  (0.483)         &  (0.032)         &  (0.053)         &  (0.905)         &  (1.027)         &  (0.061)         \\
[1em]
Share Norway $\times$ Post&    --         &   -0.033         &    4.522\sym{**} &    0.075         &   -0.026         &    0.580         &    1.641         &    0.563\sym{*}  \\
                &      --         &  (0.392)         &  (2.129)         &  (0.117)         &  (0.101)         &  (1.987)         &  (2.656)         &  (0.292)         \\
[1em]
Observations    &     --         &     1377         &     1377         &     1377         &     1377         &     1377         &     1377         &     1377         \\
R-squared       &        --         &     0.76         &     0.81         &     0.77         &     0.31         &     0.75         &     0.75         &     0.68         \\
\midrule\midrule
Average dependent variable & -- & 0.580 & 2.015 & 0.097 & 0.060 & 1.864 & 2.356 & 0.156 \\
\midrule
Year FE         &      --         &      Yes         &      Yes         &      Yes         &      Yes         &      Yes         &      Yes         &      Yes         \\
Firm FE         &      --         &      Yes         &      Yes         &      Yes         &      Yes         &      Yes         &      Yes         &      Yes         \\
\bottomrule
\multicolumn{9}{l}{* -- $p < 0.1$; ** -- $p < 0.05$; *** -- $p < 0.01$}
\end{tabular}
}
\end{adjustbox}
		\begin{tablenotes}
			\footnotesize \vspace{1em}
			\parbox{\textwidth}{\justifying  \item \hspace{-0.2em}\textbf{Notes:} Panel (a) presents estimated coefficients from regression \eqref{eq:know-how-firm} where we use the indicator variable ``Oil Major'' to distinguish oil majors from the other firms. The oil majors include the so-called ``seven sisters'' (Exxon, Mobil, Chevron, Texaco, Gulf Oil, Royal Dutch Shell, and BP) as well as other very large players of the time (Total, Eni, Phillips Petroleum, Conoco, Amoco). Panel (b) additionally controls for ``Share Norway $\cdot$ Post,'' which is firm $i$'s share of EOR-eligible fields in the North Sea before 1985 times the post-1985 indicator. The dependent variables are the numbers of different types of EOR-eligible fields (Columns 1-4) and of different types of non-EOR-eligible fields (Columns 5-8) that were in firm $i$'s portfolio in year $t$. Each column focuses on different field types: heavy, sour, deep well, or other non-heavy, non-sour, non-deep. EOR cannot be implemented with heavy oil, hence Column 1 is empty. All regressions control for firm and year fixed-effects. Standard errors are clustered at the firm level. }
		\end{tablenotes}
	\end{table}

	\begin{table}
		\centering
		\caption{Oil majors and technological diversification \label{tab:know-risk-om}}
		\centering
		\begin{adjustbox}{width=1.05\textwidth,center}
{
\def\sym#1{\ifmmode^{#1}\else\(^{#1}\)\fi}
\begin{tabular}{l*{7}{c}}
\toprule
                &\multicolumn{6}{c}{Diversification in Technology Portfolio ($1-$HHI)} \\ 
                \cmidrule(lr){2-7}
             Dependent variable &\multicolumn{1}{c}{Exploration}&\multicolumn{1}{c}{Extraction}&\multicolumn{1}{c}{Well }&\multicolumn{1}{c}{Field} &\multicolumn{1}{c}{Tariff} &\multicolumn{1}{c}{CAPEX}\\
                &\multicolumn{1}{c}{\& Prod. Tech.}&\multicolumn{1}{c}{Technologies}&\multicolumn{1}{c}{Depth}&\multicolumn{1}{c}{Sizes} &\multicolumn{1}{c}{Intensity} &\multicolumn{1}{c}{Intensity}\\                
                &\multicolumn{1}{c}{(1)}         &\multicolumn{1}{c}{(2)}         &\multicolumn{1}{c}{(3)}         &\multicolumn{1}{c}{(4)}         &\multicolumn{1}{c}{(5)}         &\multicolumn{1}{c}{(6)} \\ 

\multicolumn{7}{l}{\textit{Panel (a): Regression \eqref{eq:know-how-firm} with ``Oil Major'' instead of ``K-H'' and ``S-O''}}  \\   
\midrule 
Oil Major $\times$ Share EOR Norway $\times$ Post&    0.094         &    0.561         &    0.171         &    0.120         &    0.569         &    0.178         \\
                &  (1.068)         &  (1.413)         &  (1.304)         &  (1.369)         &  (1.142)         &  (1.268)         \\
[1em]
Share EOR Norway $\times$ Post&   -1.156         &   -1.096         &   -0.911         &   -0.927         &   -1.394\sym{**} &   -0.985         \\
                &  (0.819)         &  (0.686)         &  (0.841)         &  (0.768)         &  (0.619)         &  (0.779)         \\
[1em]
Share EOR $\times$ Post&   -0.723\sym{**} &   -0.410         &   -0.667\sym{**} &   -0.571\sym{*}  &   -0.494\sym{*}  &   -0.570\sym{**} \\
                &  (0.297)         &  (0.275)         &  (0.281)         &  (0.298)         &  (0.280)         &  (0.283)         \\
[1em] 
Observations    &     1377         &     1377         &     1377         &     1377         &     1377         &     1377         \\
R-squared       &     0.36         &     0.31         &     0.34         &     0.32         &     0.34         &     0.33         \\
\midrule\midrule
\rule{0pt}{4ex}  \\[-1ex] 

\multicolumn{7}{l}{\textit{Panel (b): as in Panel (a) but controlling for ``Share Norway $\cdot$ Post''}} \\
\midrule
Oil Major $\times$ Share EOR Norway $\times$ Post&   -0.001         &    0.243         &    0.257         &    0.092         &    0.573         &    0.089         \\
                &  (0.410)         &  (0.490)         &  (1.276)         &  (1.386)         &  (1.146)         &  (1.314)         \\
[1em]
Share EOR Norway $\times$ Post&   -0.298         &   -0.586         &   -1.326         &   -0.888         &   -1.485         &   -0.703         \\
                &  (0.623)         &  (0.586)         &  (1.766)         &  (1.716)         &  (1.573)         &  (1.675)         \\
[1em]
Share EOR $\times$ Post&   -0.328\sym{***}&   -0.170\sym{*}  &   -0.756\sym{***}&   -0.726\sym{**} &   -0.631\sym{**} &   -0.757\sym{***}\\
                &  (0.103)         &  (0.090)         &  (0.259)         &  (0.278)         &  (0.263)         &  (0.258)         \\
[1em]
Share Norway $\times$ Post&   -0.099         &    0.120         &    0.243         &   -0.077         &    0.013         &   -0.248         \\
                &  (0.358)         &  (0.364)         &  (1.060)         &  (1.022)         &  (0.977)         &  (0.971)         \\
[1em]
Observations    &     1353         &     1353         &     1353         &     1353         &     1353         &     1353         \\
R-squared       &     0.38         &     0.32         &     0.38         &     0.32         &     0.38         &     0.32         \\
\midrule\midrule
Year FE         &      Yes         &      Yes         &      Yes         &      Yes         &      Yes         &      Yes         \\
Firm FE         &      Yes         &      Yes         &      Yes         &      Yes         &      Yes         &      Yes         \\
\midrule
\end{tabular}
}
\end{adjustbox}
		\begin{tablenotes}
			\footnotesize \vspace{1em}
			\parbox{\textwidth}{\justifying  \item \hspace{-0.2em}\textbf{Notes:} Panel (a) presents estimated coefficients from regression \eqref{eq:know-how-firm} where we use the indicator variable ``Oil Major'' to distinguish oil majors from the other firms. The oil majors include the so-called ``seven sisters'' (Exxon, Mobil, Chevron, Texaco, Gulf Oil, Royal Dutch Shell, and BP) as well as other very large players of the time (Total, Eni, Phillips Petroleum, Conoco, Amoco). Panel (b) additionally controls for ``Share Norway $\cdot$ Post,'' which is firm $i$'s share of EOR-eligible fields in the North Sea before 1985 times the post-1985 indicator. The dependent variables evaluate technological diversification according to \eqref{eq:diversification} across technology types (Columns 1-2), field characteristics (Columns 3-4), and technology-related expenditures (Column 5-6). See main text for more details. Each dependent variable is standardized. All regressions control for firm and year fixed effects. Standard errors are clustered at the firm level.}
		\end{tablenotes}
	\end{table}

	\newpage
	\clearpage
	
	\setcounter{table}{0} \renewcommand{\thetable}{F\arabic{table}} %
	\setcounter{figure}{0} \renewcommand{\thefigure}{F\arabic{figure}}
	\setcounter{equation}{0} \renewcommand{\theequation}{F\arabic{equation}}
	\setcounter{section}{0} \renewcommand{\thesection}{F}
	
	\section{Geographic Proximity to Fields Adopting EOR\label{apndx:dist}}
	We investigate whether geographic proximity to a field that already adopted EOR before 1985 increased the likelihood of EOR adoption after 1985. To do this, we extend field-level regression \eqref{eq:did-field} by interacting the treatment with G-P$_f$, defined as the inverse distance to the nearest pre-1985 EOR-adopting field:
	\begin{equation}\label{eq:know-field}
		\begin{aligned}
			\mathbbm{1}\{\text{Adopted EOR}_{ft}\} &= \beta_{\text{treat}} \cdot \text{G-P}_{f} \cdot \text{Norway}_f \cdot \text{EOR}_f \cdot \text{Post}_t  \\
			&+ \beta_{1} \cdot \text{G-P}_f  \cdot \text{EOR}_{f} \cdot \text{Post}_t \\ 
			&+ \beta_{2} \cdot \text{G-P}_f \cdot \text{Post}_t \\
			&+ \beta_{3} \cdot \text{Norway}_{f}  \cdot \text{EOR}_{f} \cdot \text{Post}_t  \\
			&+ \beta_{4} \cdot\text{Post}_t + \tau_{f} + \omega_{o(f)} + \iota_{c(f)t} + \epsilon_{ft},   
		\end{aligned}
	\end{equation}
	where $\omega_{o(f)}$ is an operator $o$-specific fixed effect which controls for the possibility that the same operator $o$ was active in multiple fields and therefore that EOR adoption in nearby fields was driven by the same operator rather than by geographical proximity. Table \ref{tab:know-field} reports the estimation results, which show that geographical proximity to EOR-adopting fields played no role in subsequent EOR adoption. Though less precisely estimated, $\widehat{\beta}_3$ and $\widehat{\beta}_4$ are similar to the corresponding estimates in Appendix Table \ref{tab:did-field-appendix}, Panel (a).

	\begin{table}[!htb]
		\centering
		\caption{EOR adoption and geographic proximity to adopting fields \label{tab:know-field}}
		\centering
		{
\def\sym#1{\ifmmode^{#1}\else\(^{#1}\)\fi}
\begin{adjustbox}{width=.5\textwidth, center}
\begin{tabular}{l*{3}{c}}
\toprule        
Dept. var. & \multicolumn{2}{c}{$\mathbbm{1}\{\text{Adopted EOR}_{ft}\}$}\\ 
\cmidrule(lr){2-3}
                 &\multicolumn{1}{c}{(1)}         &\multicolumn{1}{c}{(2)}         &           \\
\midrule
\\
\text{G-P} $\times$ Norway $\times$ EOR $\times$ Post ($\beta_{\text{treat}}$) 
&    0.394         &   -0.091         \\
                &  (1.383)         &  (1.546)         \\
[1em]
\text{G-P} $\times$ EOR $\times$ Post ($\beta_{1}$) 
&    0.665         &    0.667         \\
                &  (0.432)         &  (0.454)         \\
[1em]
\text{G-P} $\times$ Post ($\beta_{2}$) 
 & 0.044         &    0.043         \\
                &  (0.032)         &  (0.033)         \\
[1em]
Norway $\times$ EOR $\times$ Post ($\beta_{3}$) 
&    0.222         &    0.222         \\
                &  (0.224)         &  (0.235)         \\
[1em]
EOR $\times$ Post ($\beta_{4}$) 
&    0.009         &    0.021         \\
                &  (0.021)         &  (0.024)         \\

\midrule

Country-Year FE &       Yes         &      Yes         \\
Field FE        &       Yes         &      Yes         \\
Operator FE     &       No         &      Yes         \\
Observations    &     1989         &     1808         \\
R-squared       &     0.85         &     0.86         \\
\bottomrule
\multicolumn{3}{l}{* -- $p < 0.1$; ** -- $p < 0.05$; *** -- $p < 0.01$}
\end{tabular}
\end{adjustbox}
}

		\begin{tablenotes}
			\footnotesize \vspace{1em}
			\parbox{\textwidth}{\justifying  \item \hspace{-0.2em}\textbf{Notes:} Estimated coefficients from field-level regression \eqref{eq:did-field}. The dependent variable is an indicator equal to 1 if field $f$ has adopted EOR by year $t$ and 0 otherwise. Variable G-P$_f$ is field $f$'s inverse distance to the nearest pre-1985 EOR-adopting field. The indicator variables ``Norway,'' ``EOR,'' and ``Post'' equal 1 for Norwegian fields, for EOR-eligible fields, and for the years after 1985. Each regression includes country-by-year and field fixed effects. Column 2 additionally controls for operator fixed effects. Standard errors are clustered at the field level.}
		\end{tablenotes}
	\end{table} 
	

	\setcounter{table}{0} \renewcommand{\thetable}{G\arabic{table}} %
	\setcounter{figure}{0} \renewcommand{\thefigure}{G\arabic{figure}}
	\setcounter{section}{0} \renewcommand{\thesection}{G}
	
	\section{Classification of Technologies}\label{apndx:technologies}

	\begin{table}[ht!]
		\centering
		\caption{Classification of Exploration, Production, and Extraction Technologies}
		\label{tab:tech_classification}
		\small
		\begin{tabularx}{\textwidth}{llX}
			\toprule
			\textbf{Technology} & \textbf{Category / Type} & \textbf{Primary Function} \\ \midrule \\
			\multicolumn{3}{l}{\textit{Panel A: Exploration \& Production Technologies (Field Development)}} \\ \midrule
			ERD & E\&P Technology & Drilling wells that reach farther from the platform to access remote or hard-to-reach reservoirs. \\
			FPSO & Production (Floating) & Floating unit for offshore production, storage, and offloading; ideal for deepwater without fixed infrastructure. \\
			Fixed Platform & Production (Fixed) & Stationary platform supporting drilling and production in relatively shallow waters. \\
			Onshore & Production (Location) & Facilities located on land rather than offshore. \\
			Jack-up & Production (Mobile) & Mobile rig with legs lowered to the seabed; provides a stable base for drilling and production. \\
			Semi-sub & Production (Floating) & Anchored offshore platform used for deepwater production in harsh environments. \\
			Subsea & Production (Sub-sea) & Wells located on the seabed and connected back to a host platform or onshore facility. \\
			TLP & Production (Floating) & Floating platform moored to the seabed by tensioned tendons; used in ultra-deepwater. \\
			Undeveloped & Exploration Stage & Fields that have not yet reached production; associated with exploration and early development. \\ 
			\midrule \\
			\multicolumn{3}{l}{\textit{Panel B: Extraction Technologies (Hydrocarbon Recovery)}} \\ \midrule
			Artificial Lift & Primary Recovery & Mechanical methods (e.g., beam pumps) to bring oil to the surface when natural reservoir pressure is low. \\
			Gas Injection & Secondary Recovery & Injecting gas into the reservoir to maintain pressure and sweep oil toward production wells. \\
			Water Injection & Secondary Recovery & Injecting water into the reservoir to provide pressure support and displace hydrocarbons. \\
			Gas \& Water Inj. & Secondary Recovery & Simultaneous or alternating injection of gas and water to maximize displacement efficiency. \\
			Lift \& Gas Inj. & Secondary Recovery & Combination of mechanical artificial lift and gas injection for enhanced pressure support. \\
			Lift \& Water Inj. & Secondary Recovery & Combination of mechanical artificial lift and water flooding. \\
			Lift, Gas \& Water & Secondary Recovery & Integrated recovery strategy using lift systems alongside both gas and water injection. \\
			Unknown & Unclassified & Method used for extraction or enhancement is not specified in the underlying dataset. \\ \bottomrule
		\end{tabularx}
	\end{table}

	
	\setcounter{table}{0} \renewcommand{\thetable}{H\arabic{table}} %
	\setcounter{figure}{0} \renewcommand{\thefigure}{H\arabic{figure}}
	\setcounter{section}{0} \renewcommand{\thesection}{H}
	
	\section{Data, Extraction, and Coding of Constitutional Non-Retroactivity Provisions}\label{apndx:retroactivity}
	
	This appendix describes the automated procedure used to code whether a country’s constitutional text contains an explicit prohibition on retroactive laws that extends beyond criminal punishment. The pipeline is reproducible and consists of four steps: (i) retrieve constitutional texts from the \textit{Comparative Constitution Project} (CCP) API, (ii) extract a small set of candidate provisions using hybrid semantic and lexical retrieval, (iii) classify the candidate excerpts using an LLM constrained to excerpt-only evidence and a structured output schema, and (iv) apply deterministic guardrails and targeted fallback searches when evidence is weak. The script outputs a country-level binary indicator, \texttt{has\_noncriminal\_prohibition}, alongside domain tags, exception flags, evidence quotes, and a confidence score for auditability.
	
	\subsection{Data Source}\label{s:retro_data}
	
	Constitutional texts are obtained from the CCP API \citepsec{ElkinsGinsburg2025CCP}, which provides English translations and constitution-level metadata. The script restricts attention to constitutions that are in force and excludes draft records (identified by Constitute identifiers with draft suffixes). When multiple in-force constitutions are returned for the same country, the script retains a single document per country by selecting the most recently revised text, based on revision-year metadata (falling back to enactment year if needed).
	
	\subsection{Text Processing and Segmentation}\label{s:retro_segmentation}
	
	For each retained constitution, the script downloads the \texttt{HTML} representation and converts it to plain text by stripping markup and normalizing whitespace. The text is segmented into quasi-article units using common heading markers (e.g., ``Article \#,'' ``Section \#,'' and related variants); when headings are not detected, the script falls back to paragraph-based segmentation. Segmentation defines the retrieval unit and ensures that downstream decisions can be traced to identifiable constitutional provisions.

	\subsection{Hybrid Retrieval of Candidate Provisions}\label{s:retro_retrieval}
	
	Retroactivity provisions are rare relative to the length of a typical constitution. The pipeline therefore uses a hybrid retrieval strategy to identify a small set of candidate excerpts before invoking the LLM.
	
	First, it applies \textit{semantic retrieval}. Each segment is embedded using the OpenAI \texttt{text-embedding-3-large} model and scored by cosine similarity against a fixed set of query prompts spanning three families: criminal-only formulations, fiscal formulations (especially taxation), and general or mixed-domain formulations. For each family, the script retains the top-$k$ segments (default $k=12$) and deduplicates across queries and families by keeping the highest similarity score per segment.\footnote{We experiment with different values of $k$ with similar results. } Query embeddings are pre-computed once and reused across countries.
	
	Second, it adds a \textit{lexical supplement}: any segment containing explicit retroactivity cue terms---such as \emph{retroactive}, \emph{retrospective}, \emph{ex post facto},\footnote{An \emph{ex post facto} clause is a constitutional prohibition on retroactive \emph{criminal} lawmaking—typically barring (i) criminalizing conduct after the fact or (ii) increasing punishment retroactively. In many constitutions it appears as the formula ``No \emph{ex post facto} law shall be passed,'' often alongside a ban on bills of attainder. The phrase is used verbatim in some constitutional texts (e.g., the U.S. Constitution and the Philippines Constitution). \emph{Ex post facto} language also appears in some constitutional rights catalogues/headings (e.g., ``protection from ex post facto laws'' in Nigeria and South Africa). These examples are illustrative: many constitutions express the same idea without using the Latin phrase (e.g., ``no penalty without prior law'').}
	\emph{non-retroactivity}, \emph{irretroactivity}, \emph{acquired rights}, \emph{vested rights}, \emph{res judicata}, \emph{only for the future}, or \emph{regressively}---is included if not already retrieved semantically. This hybrid design addresses cases in which embedding similarity fails to surface provisions using paraphrastic or indirect formulations that nonetheless contain clear retroactivity language.
	
	The top candidates are then selected and expanded with immediate neighbors (one segment before and after each candidate) to preserve local context such as carve-outs and exceptions. Individual excerpts are truncated to 2{,}000 characters. The expanded set---which may exceed 20 articles when the top candidates are scattered across distinct sections of the constitution---is forwarded to the LLM.
	
	
	\subsection{LLM-based Classification with Structured Outputs}\label{s:retro_llm}
	
	The substantive coding is performed by an LLM that receives only the retrieved excerpts. The prompt requires the model (i) to base its decision exclusively on the excerpts, (ii) to return short verbatim evidence quotes supporting the decision, and (iii) not to treat general legality/authorization language (e.g., ``in accordance with the law'' or ``no tax without law'') as retroactivity prohibitions. The model is instructed to set \texttt{has\_noncriminal\_prohibition} to \texttt{True} only when the excerpts contain explicit temporal non-retroactivity language outside criminal punishment (e.g., ``retroactive/retrospective,'' ``shall not apply to acts occurring before,'' ``in force at the time'').
	
	The model returns a JSON object validated against a strict schema containing: \texttt{has\_noncriminal\_prohibition}; a domain label (criminal, fiscal, civil, administrative, general, or unclear); an exception flag and exception-type label; evidence quotes; notes; and a confidence score on $[0,1]$.
	
	\subsection{Deterministic Guardrails and Fallback Searches}\label{s:retro_guardrails}
	
	The script post-processes the LLM output with deterministic guardrails designed to prevent systematic misclassification. The guardrails (a) exclude patterns that are commonly mistaken for non-criminal retroactivity prohibitions (e.g., savings/transitional provisions, commencement-only language, constitutional-court remedy clauses, and tax-authorization clauses without explicit retroactivity language); (b) enforce scope restrictions (e.g., treat \emph{ex post facto} language as criminal-only unless the text explicitly extends beyond penal matters); and (c) apply positive overrides when explicit general or fiscal non-retroactivity language is present. A key design feature is a temporal-marker check: positive classifications are intended to rest on explicit temporal or retroactivity language in the cited evidence.
	
	Two targeted fallback mechanisms address retrieval failures. If the excerpts supporting a negative classification appear exclusively criminal in character, the script performs a targeted lexical search for non-criminal retroactivity language and re-runs the LLM on any matches. Separately, if the initial run yields weak support (missing evidence quotes or low confidence), the script performs a broader lexical search for retroactivity-related terms and re-runs the LLM. A fallback verdict replaces the initial verdict only when it improves evidentiary support under the same guardrails.
	
	\subsection{Output}\label{s:retro_output}
	
	The script writes one row per country to a CSV file containing the constitution identifier, \texttt{has\_noncriminal\_prohibition}, domain and exception fields, evidence quotes, notes, and the model confidence score. These fields support replication and manual auditing, and the map in Figure~\ref{fig:retroactivity-map} is produced by aggregating this country-level output.
	
	The resulting dataset covers 193 countries.\footnote{This count matches the CCP coverage at the time of data retrieval, which may differ slightly across API versions or update cycles.} The classification pipeline processes all 193 constitutions. Following \citesec{ginsburg2015does}, we treat the United Kingdom, Israel, and New Zealand as countries without a codified constitution; the map construction step reassigns these three countries to a separate category (shown in white in Figure \ref{fig:retroactivity-map}), leaving an empirical universe of 190 codified constitutions for the substantive analysis.
	
	As an additional validation step, we verify the country-level output in two complementary ways. First, two researchers independently audited 50\% of the country codings by inspecting the extracted evidence quotes and the underlying CCP excerpts; there were no discrepancies in the \texttt{has\_noncriminal\_prohibition} indicator. Second, following \citesec{asirvatham2026GPTMeasurement}, we re-ran the full pipeline after replacing the direct LLM call with a GABRIEL-based call that executes the same structured-prompt classification on the retrieved excerpts and enforces the identical strict JSON schema (i.e., it returns the same set of fields, including \texttt{has\_noncriminal\_prohibition}, domain labels, exceptions, evidence quotes, notes, and confidence). The resulting country classifications were similar, providing an external check that the core findings are not sensitive to the specific LLM-call used for the final structured-output step.

	\subsection{Discussion of design choices}\label{s:retro_design}
	
	Several design choices shape interpretation. The measure relies on CCP's English translations and metadata, ensuring cross-country comparability while inheriting translation and documentation error \citepsec{ElkinsGinsburg2025CCP}. Candidate provisions are identified using hybrid retrieval---embedding similarity combined with lexical cues and neighbor expansion---to reduce missed clauses when drafting is indirect or exceptions appear nearby. The LLM is constrained to these excerpts and must return short verbatim evidence quotes, which makes codings auditable but ties performance to retrieval quality. Deterministic guardrails then remove systematic false positives (e.g., savings/commencement provisions and tax-authorization clauses lacking explicit retroactivity language) and enforce conservative scope (e.g., treating \emph{ex post facto} as criminal-only unless explicitly extended). Accordingly, \texttt{has\_noncriminal\_prohibition} should be interpreted as a conservative measure of explicit non-criminal non-retroactivity.

	\singlespacing
	\bibliographystylesec{ecca-mod.bst}
	\bibliographysec{bibliography.bib}

\end{document}